\UseRawInputEncoding

\documentclass[nofootinbib,prx,aps,twocolumn,floatfix,10pt,longbibliography,superscriptaddress,notitlepage]{revtex4-1}
\usepackage[english]{babel}
\usepackage{breakcites}
\usepackage{listings}

\usepackage[T1]{fontenc}
\usepackage{textcomp}
\usepackage{gensymb,braket}

\usepackage{graphicx}
\usepackage{braket}
\usepackage{color}
\usepackage{epstopdf}
\usepackage{mathtools}
\usepackage{amsmath}
\usepackage{amssymb}
\usepackage{float}
\usepackage{comment}
\usepackage{amsmath}
\usepackage{multirow}

\usepackage{amsfonts}
\usepackage{bm}
\usepackage{bbold}

\usepackage{color}

\graphicspath{{./figs/}}
\usepackage{gensymb}
\usepackage[colorinlistoftodos,prependcaption]{todonotes}
\usepackage{xargs}  
\newcommandx{\unsure}[2][1=]{\todo[linecolor=red,backgroundcolor=red!25,bordercolor=red,#1]{#2}}
\newcommandx{\change}[2][1=]{\todo[linecolor=blue,backgroundcolor=blue!25,bordercolor=blue,#1]{#2}}
\newcommandx{\info}[2][1=]{\todo[linecolor=OliveGreen,backgroundcolor=OliveGreen!25,bordercolor=OliveGreen,#1]{#2}}
\newcommandx{\improvement}[2][1=]{\todo[linecolor=Plum,backgroundcolor=Plum!25,bordercolor=Plum,#1]{#2}}
\newcommandx{\thiswillnotshow}[2][1=]{\todo[disable,#1]{#2}}
\newcommandx{\greencom}[2][1=]
{\todo[inline, color=green!40,#1]{#2}}
\newcommandx{\bluecom}[2][1=]
{\todo[inline, color=blue!40,#1]{#2}}

\usepackage[colorlinks]{hyperref}
\definecolor{winered}{rgb}{0.5,0,0}
\hypersetup
{
colorlinks=true,
linkcolor=winered,
urlcolor={winered},
filecolor={winered},
citecolor={winered},
allcolors={winered}
}

\usepackage{letltxmacro}
\LetLtxMacro{\ORIGselectlanguage}{\selectlanguage}
\makeatletter
\DeclareRobustCommand{\selectlanguage}[1]{%
  \@ifundefined{alias@\string#1}
    {\ORIGselectlanguage{#1}}
    {\begingroup\edef\x{\endgroup
       \noexpand\ORIGselectlanguage{\@nameuse{alias@#1}}}\x}%
}
\newcommand{\definelanguagealias}[2]{%
  \@namedef{alias@#1}{#2}%
}
\makeatother

\definelanguagealias{en}{english}
\definelanguagealias{EN}{english}

\graphicspath{{figure/}}

\begin{document}

\title{Quasinormal modes
 and  Purcell factors of coupled loss-gain resonators and index-modulated ring resonators near exceptional points}

\author{Juanjuan Ren}
\email{jr180@queensu.ca}
\affiliation{Department of Physics, Engineering Physics, and Astronomy, Queen's
University, Kingston, Ontario K7L 3N6, Canada}
\author{Sebastian Franke} 
\email{sebastian.franke@tu-berlin.de}
\affiliation{Technische Universit\"at Berlin, Institut f\"ur Theoretische Physik,
Nichtlineare Optik und Quantenelektronik, Hardenbergstra{\ss}e 36, 10623 Berlin, Germany}
\affiliation{Department of Physics, Engineering Physics, and Astronomy, Queen's University, Kingston, Ontario K7L 3N6, Canada}
  \author{Stephen Hughes}
  \email{shughes@queensu.ca}
 \affiliation{Department of Physics, Engineering Physics, and Astronomy, Queen's University, Kingston, Ontario K7L 3N6, Canada}

\begin{abstract}

We first present 
a quasinormal mode (QNM) theory for  coupled loss and gain resonators, working in the vicinity of an exceptional point.
Assuming linear media, which can be fully quantified using the complex pole properties of the QNMs, we show how 
the QNMs yield a quantitatively 
good model to a full dipole spontaneous emission response in Maxwell's equations at a variety of spatial positions and frequencies (linear response).
We also develop a highly accurate and intuitive QNM coupled-mode theory, which can be used to rigorously model such systems using only the QNMs of the bare resonators, where the hybrid QNMs of the complete system are automatically obtained. 
Near a lossy exceptional point, 
we analytically show how the QNMs yield a 
Lorentzian-like  and  a Lorentzian-squared-like response
for the spontaneous emission lineshape, consistent with other works. However, using rigorous analytical and numerical solutions
for microdisk resonators,  we demonstrate that the general lineshapes are far richer than what has been previously predicted. Indeed, the classical picture of spontaneous emission can take on a wide range of positive and negative Purcell factors from the hybrid modes of the coupled loss-gain system. 
These negative Purcell factors are unphysical and signal a clear breakdown of the
classical dipole picture of spontaneous emission in such media, though the negative local density of states is correct.
We also show the rich spectral features of the Green function propagators,
which can be used to model various  physical observables.
Second, we  present a QNM approach to model index-modulated ring resonators working near an exceptional point
and show  unusual chiral power flow from linearly polarized emitters, in agreement with recent experiments, which is quantitatively explained without invoking the interpretation of a  missing dimension (the Jordan vector) and a decoupling from the cavity eigenmodes.
\end{abstract}
\maketitle 

\section{Introduction}

Lossless photonic systems (such as closed resonators with no material absorption) can be formulated as  
a Hermitian eigenvalue problem, which yields real eigenfrequencies
from the source-free Helmholz equation, and corresponding normal modes (NMs). This is also true for
periodic systems with bound modes, e.g.,  lossless waveguide modes.
However, real cavity structures (resonators) with open boundary conditions yield finite loss (or gain) and  produce complex eigenfrequencies.
Thus most optical systems are naturally dissipative via material absorption and/or radiation.

A common design approach to improving resonators is
to increase the photonic local density of states (LDOS), by
reducing radiation losses and the effective mode volume, thus increasing the
Purcell factor for enhanced spontaneous emission (SE).
An alternative approach to reducing loss is through gain compensation,
where a gain medium 
can be  introduced and controlled through stimulated emission or parametric processes.
Not necessarily related to lasing, a gain medium can be introduced
to the system and treated as a {\it linear amplifying medium}~\cite{raabe_qed_2008,PhysRevA.64.033812}, which must satisfy strict criteria that relates to the complex poles
of the Green function.
Aided by the rapid development of  optical nanotechnologies, coupled loss and gain structures have been under intense investigation recently, especially after the demonstrations of parity-time (PT) symmetry~\cite{bender_real_1998,bender_pt-symmetric_1999,levai_systematic_2000,bender_generalized_2002,bender_complex_2002,mostafazadeh_pseudo-hermiticity_2002,bender_must_2003,bender_making_2007} in optical 
systems~\cite{el-ganainy_theory_2007,makris_beam_2008,klaiman_visualization_2008,guo_observation_2009,longhi_bloch_2009,mostafazadeh_spectral_2009,ruter_observation_2010,kottos_broken_2010,longhi_optical_2010,PhysRevA.91.053825,konotop_nonlinear_2016,feng_non-hermitian_2017,longhi_parity-time_2017,lupu_optimal_2017,el-ganainy_non-hermitian_2018,jin_parity-time-symmetric_2018,morozko_modal_2020}, which support so-called exceptional points (EPs)~\cite{berry_physics_2004,heiss_exceptional_2004,heiss_physics_2012,PhysRevX.6.021007,miri_exceptional_2019,chen_sensitivity_2019,jin_hybrid_2020} (two isolated eigenfrequencies and eigenmodes coalesce), along with many interesting and counter-intuitive phenomena, such as non-reciprocal propagation (isolators)~\cite{lin_unidirectional_2011,feng_experimental_2013,peng_paritytime-symmetric_2014,chang_paritytime_2014,jin_incident_2018}, mode switching~\cite{zyablovsky_pt-symmetry_2014,doppler_dynamically_2016,xu_topological_2016,heiss_circling_2016}, efficient sensing~\cite{chen_exceptional_2017,chen_parity-time-symmetric_2018}, absorber~\cite{longhi_pt_2010,chong_p_2011,sun_experimental_2014,jin_unidirectional_2016}, and lasing~\cite{brandstetter_reversing_2014,feng_single-mode_2014,hodaei_parity-time-symmetric_2014,peng_loss-induced_2014}.

To have a detailed understanding of such systems, and also for connecting
to new applications in quantum optics,
it is desirable to have an accurate  model of coupled
loss-gain resonators at the level of a rigorous and intuitive mode theory, which can allow one to describe light-matter interactions at various spatial points and frequencies.
From a theoretical perspective, 
temporal coupled mode theory (CMT) has proved to be an efficient approach to investigate coupled resonator systems~\cite{haus_coupled-mode_1991,fan_temporal_2003}, where only the solution from the bare systems are required,
and one assumes that the coupled modes can be represented by a linear superposition of the bare modes (weak coupling regime). However, the coupling coefficients are typically used as heuristic parameters in that they are usually extracted from fitting the full solution of the coupled system properties~\cite{artar_directional_2011,peng_paritytime-symmetric_2014,yang_active_2019,park_symmetry-breaking-induced_2020}, or they are mainly used to explain the basic physics of coupling.

Another potential problem with such approaches is that
the underlying modes of the bare resonators are assumed to be NMs (Hermitian system), and a finite decay rate, to account for real losses, is added phenomenologically.
This finite decay rate already relates to a non-Hermitian problem, so it is natural to model the bare resonators also with a non-Hermitian theory, where the
correct eigenvalues and modes would then be obtained for a more general CMT for loss-gain systems.
{\it This is not only a more correct approach, but it changes some of the fundamental coupling regimes and constraints significantly, and open up much richer light-matter interaction  regimes.}

In recent years,
  the theory of open cavity modes has been shown to be 
 accurately described in terms of  quasinormal modes (QNMs)~\cite{lai_time-independent_1990,leung_completeness_1994,leung_time-independent_1994,leung_completeness_1996,lee_dyadic_1999,kristensen_generalized_2012,sauvan_theory_2013,kristensen_modes_2014,bai_efficient_2013-1,PhysRevA.98.043806,PhysRevX.7.021035,lalanne_light_2018,kristensen_modeling_2020}, which are open cavity modes with complex eigenfrequencies and spatially diverging modes (with finite loss). These open cavity modes naturally include the effect of losses and can also be used to
 construct the photon 
   Green function,  which  describes a wide range of light-matter interactions~\cite{lee_dyadic_1999,muljarov_brillouin-wigner_2010,kristensen_generalized_2012,sauvan_theory_2013,kristensen_modes_2014,muljarov_exact_2016,lalanne_light_2018,kristensen_modeling_2020}.
Moreover, as shown recently,  QNMs can be fully quantized  and used to show  departures from the usual NM quantum optics theories~\cite{franke_quantization_2019,PhysRevResearch.2.033332,PhysRevResearch.2.033456}.
Several CMT approaches based on QNMs have also been successfully developed~\cite{vial_coupling_2016,Kristensen_coupled_modes_2017,cognee_hybridization_2020,tao_coupling_2020} for coupled passive resonator systems.

In this work, we first
present a QNM theory for 
general media containing both lossy resonators and gain resonators.
We  describe a rigorous and intuitive CMT based on the Green function solution for the coupled loss-gain QNMs, where  we analytically obtain the hybrid modes from
{\it only} the QNMs of the bare resonators (gain or loss). We demonstrate the extremely high accuracy of the 
analytical 
theory 
by comparing with full numerical dipole simulations for whispering-gallery modes (WGMs) of microdisk resonators, and show excellent agreement for various designs
and spatial positions without using any fitting parameters.
A QNM approach also allows one to justify the the underlying assumptions of a linear gain medium, which requires an analysis of the complex poles. Without such an analysis, then 
composite systems may not even constitute a physically meaningful solution with gain treated at the level of a material response in Maxwell's equations. Indeed, as far as we are aware, this is likely the only approach to confirm this directly (in the absence of having an analytical expression for the medium Green function).

\begin{figure}[th]
    \centering
    \includegraphics[width=0.95\columnwidth]{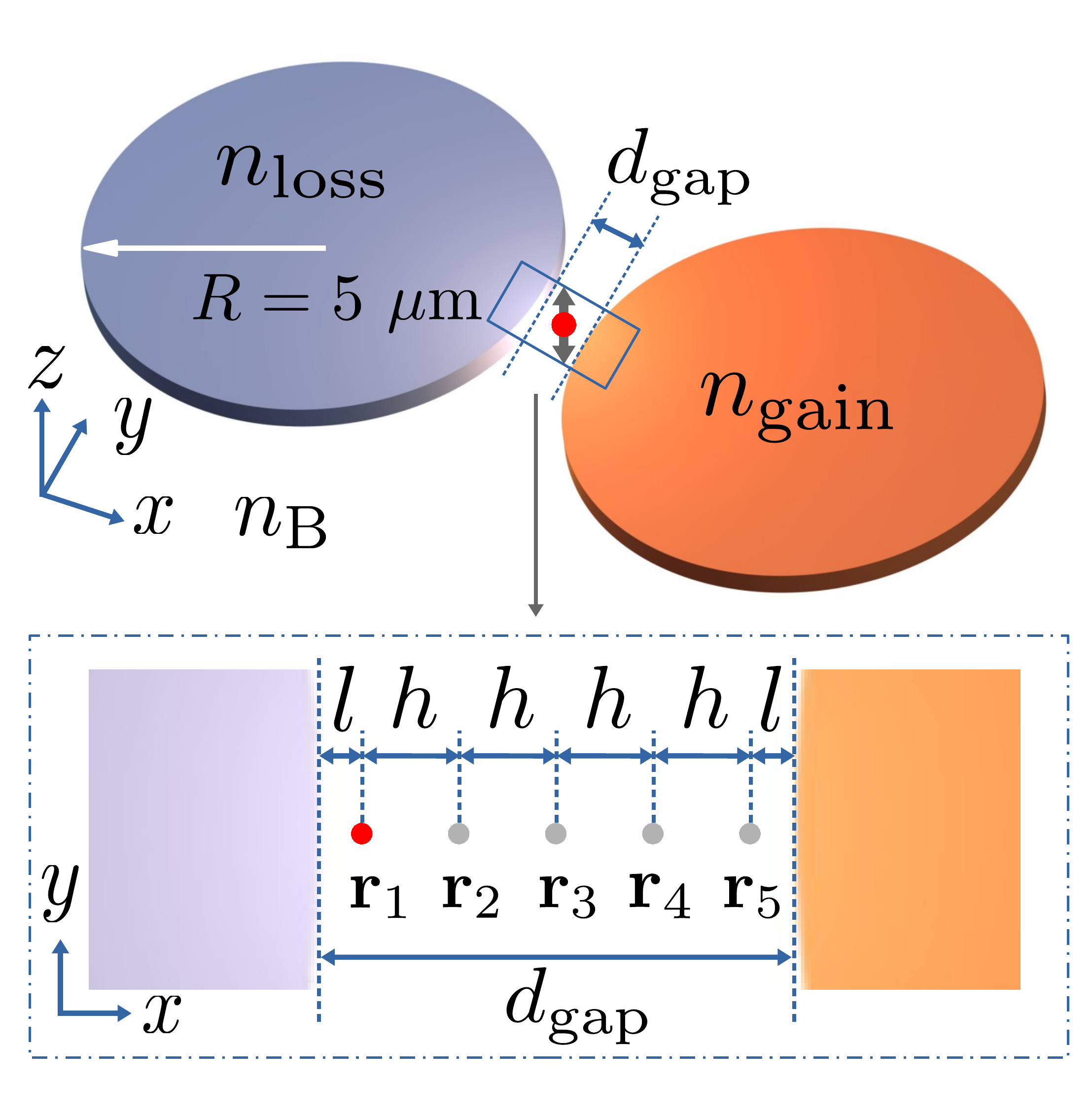}
        \caption{Schematic diagram of a coupled loss and gain microdisk  resonator system. The radius of both resonators is $R=5\,\mu$m. The refractive index of the lossy (gain) resonator is $n_{\rm loss}$ ($n_{\rm gain}$), in a  background medium with  $n_{\rm B}=1$ (free space). The gap distance between the resonators is $d_{\rm gap}$, and the  dipole (emitter) is placed within the gap, at several possible positions:  $\mathbf{r}_1$, $\mathbf{r}_2$, $\mathbf{r}_3$, $\mathbf{r}_{4}$, and $\mathbf{r}_{5}$ (see text). The origin of the coordinate system is at the gap center.}\label{disk_sche}
\end{figure}

After obtaining the physically meaningful QNMs
for loss-gain resonator systems,
we apply our theory to study the
unusual Purcell factors and Green function propagators at various spatial positions for eigenfrequencies close to an EP. 
For our numerical example, 
the considered coupled loss-gain disk resonators are shown in 
Fig.~\ref{disk_sche}(a), where one microdisk has material loss and the other has material gain. 
Such systems with balanced gain and loss form a general optical system to investigate PT symmetry and EP physics~\cite{chang_paritytime_2014,peng_paritytime-symmetric_2014,chen_parity-time-symmetric_2018}. 
In our model, we choose a gain coefficient that is slightly less than the loss coefficient, which can support 
hybridized QNMs with finite loss, which is also a 
requirement for assuming linear gain media~\cite{raabe_qed_2008}.

Next, we apply our general QNM approach to model emitters
coupled to index-modulated ring resonators, which were recently
experimentally studied~\cite{chen_revealing_2020};
in this work, the authors found the following interesting observation, quoting from their abstract:
``We find a chirality-reversal
phenomenon in a ring cavity where the radiation field reveals the missing dimension of the Hilbert space, known as the Jordan
vector. This phenomenon demonstrates that the radiation field of an emitter can become fully decoupled from the eigenstates
of its environment.'' In contrast to this interpretation, using similar
resonator designs, we show that the
chiral reversal behaviour is quantitatively explained by the emitter coupling to the two fundamental QNMs of the resonator structure. We explain this net chiral power flow analytically in terms of only the QNM properties, and show excellent agreement with full dipole simulations from Maxwell's equations.

The rest of our paper is organized as follows:
In Sec.~\ref{sec:sec2}, we introduce optical QNMs, Green functions in terms of QNMs, and show how these relate to SE decay and Purcell factors.
In Sec.~\ref{sec:sec3},
we present a detailed CMT, using both NMs and QNMs,
to obtain analytical insight into coupled loss-gain resonators. We subsequently use these to explain when an EP may form, and how things change when one uses a QNM theory. We obtain explicit expressions for the hybrid modes using only the QNMs of the bare loss or gain resonators. We explain the limits and failures of the usual CMT for such systems.\
Section~\ref{sec:sec4} discusses Green functions and Purcell factors
at the EP, and shows how a Lorentzian-like or Lorentzian-squared-like lineshape forms~\cite{PhysRevA.84.063833,Pick2017,PhysRevResearch.2.023375}.
Section \ref{sec:sec5} presents detailed numerical
results for coupled microdisk resonators, and  confirms the excellent agreement with our analytical CMT and full dipole solutions, for various gap distances between the resonators. We then study various Purcell factor regimes at different dipole positions as a function of frequency, and show highly unusual and rich spectral lineshapes, including negative Purcell factors, and discuss the essential role of the QNM phase; again, all of these show quantitatively good agreement with full dipole calculations.
Negative total purcell factors are not physical and motivate the need for a corrected derivation of the accepted Fermi's golden rule
for such media, which is described elsewhere~\cite{EPQuantumPaper}.
We also show the Green function propagators, which connects to various observables outside the resonators, which also yield  rich non-Lorentzian lineshapes.
Finally, in
Sec.~\ref{sec:final}, we 
study EP physics for emitters coupled
to index-modulated ring resonators,
and connect to recent experiments
on chiral emission for linear dipoles~\cite{chen_revealing_2020} using the QNM theory.
We give our conclusions in Sec.~\ref{sec:sec6}.

In addition to the main text, we also present several  Appendices.
Appendix~\ref{sec:numerics} discusses the 
numerical QNM normalization approaches using COMSOL, where we show three different approaches yielding the same normalized QNMs within numerical precision.
Appendix \ref{sec:degenerate_WGMs} discusses why we only need to consider one  QNM of the miscrodisk resonator for the dipole locations we study, which is constructed from a symmetric linear combination of clockwise and counter clockwise  WGMs.
Full dipole calculations in COMSOL are also discussed in Appendix \ref{sec:full_simulation}, which are used to check the validity of the QNM results.
To compare with the coupled cavity systems, the results for single loss and single gain cavities are shown in Appendix \ref{sec:single_modes}, which also confirms the extremely high accuracy of the single QNM approximation for these resonators. 
Naturally,  one can also solve the coupled system with a QNM approach directly, instead of using CMT from the bare solutions; thus
Appendix~\ref{sec:directQNMs} shows the direct QNM approach for a coupled resonators, where quantitatively good agreement with our analytical CMT results and full dipole results are obtained. 
In addition to  the loss-gain cavities shown in the main text, we also show two more loss-gain examples in Appendix \ref{sec:morelossgain}, with different gain coefficients.

\section{Quasinormal modes and semiclassical theory of spontaneous emission  and  Purcell factors}
\label{sec:sec2}

We first introduce the electric-field QNMs, $\tilde{\mathbf{f}}_{{\mu}}\left(\mathbf{r}\right)$, which are solutions to the 
Helmholtz equation,
\begin{equation}\label{smallf}
\boldsymbol{\nabla}\times\boldsymbol{\nabla}\times\tilde{\mathbf{f}}_{{\mu}}\left(\mathbf{r}\right)-\left(\dfrac{\tilde{\omega}_{{\mu}}}{c}\right)^{2}
\epsilon(\mathbf{r},\tilde{\omega}_{\mu})\,\tilde{\mathbf{f}}_{{\mu}}\left(\mathbf{r}\right)=0,
\end{equation}
where $c$ is the vacuum speed of light,
$\tilde{\omega}_{{\mu}}= \omega_{{\mu}}-i\gamma_{{\mu}}$ is the complex eigenfrequency of each QNM,
and $\epsilon(\mathbf{r},\tilde{\omega}_{\mu})$ 
is the dielectric function, which is in general complex and dispersive,
though for our numerical examples below, we will assume this is a constant complex value in the frequency regime of interest (this is not a model restriction). 
The open boundary conditions ensure the Silver-M\"uller radiation condition~\cite{Kristensen2015}.
It is also worth noting that this boundary condition leads to quite different  asymptotic behaviour of gain QNMs and loss QNMs due to the change of sign for $\gamma_\mu$. Namely,
the lossy QNMs diverge in space but converge in time, while the gain QNMs converge in space, but diverge in time. For the composite system, then the hybrid modes must converge in time, forcing the complex poles to have loss.
These subtleties are clearly missing and overlooked in heuristic theories of coupled loss and gain resonators, but are essential to get correct for a physically meaningful model.

Using $n$ to represent the refractive index and $\alpha$ the loss or gain coefficient, for a lossy resonator with permittivity $\epsilon_{\rm L}=(n_{\rm L}+i \alpha_{\rm L})^2$, 
   we assume  a dominant QNM resonance 
   $\tilde{\omega}_{\rm L}\approx \omega_{\rm L}-i\gamma_{\rm L}$,
   where $\alpha_{\rm L},\gamma_{\rm L}>0$.
  Similarly, for the gain resonator,
$\epsilon_{\rm G}=(n_{\rm G}+i \alpha_{\rm G})^2$,
    we have a dominant QNM resonance
   $\tilde{\omega}_{\rm G}\approx \omega_{\rm L}-i\gamma_{\rm G}$,  
   where now 
   $\alpha_{\rm G},\gamma_{\rm G}<0$.
The quality factor is defined from $Q_{\mu}=\omega_{\mu}/(2|\gamma_{\mu}|)$.
Since we treat the gain amplifier in terms of a linear amplifying medium (e.g., we neglect saturation effects in the gain medium), the composite system must have $\gamma_\mu^{+/-}{>}0$ for the hybrid modes~\cite{raabe_qed_2008}.
We will refer to the gain  and loss QNMs
as $\tilde{\bf f}^{\rm G}$
 and    $\tilde{\bf f}^{\rm L}$, respectively,
 and the hybrid modes (i.e., in the presence of coupling)
 as $\tilde{\bf f}^{\pm}$.
   
To connect to
a general definition of the
SE in an arbitrary medium, we seek to obtain the 
Green function, defined through
\begin{equation}
\boldsymbol{\nabla}\times\boldsymbol{\nabla}\times{
{\bf G}}({\bf r},{\bf r}_{0},\omega)-\frac{\omega^2}{c^2}\epsilon({\bf r},\omega){{\bf G}}({\bf r},{\bf r}_{0},\omega)=\frac{\omega^2}{c^2}{\bf 1}\delta({\bf r}-{\bf r}_{0}),\label{eq:GreenHelmholtz}
\end{equation}
with corresponding radiation conditions, and $\mathbf{1}$ is the unit tensor.
The normalized QNMs
can be used to define the 
 Green function for locations near (or within) the scattering geometry through~\cite{leung_completeness_1994,ge_quasinormal_2014}
\begin{equation}
\mathbf{G}\left(\mathbf{r},\mathbf{r}_{0},\omega\right)= \sum_{\mu} A_{\mu}\left(\omega\right)\,\tilde{\mathbf{f}}_{\mu}\left({\bf r}\right)\tilde{\mathbf{f}}_{\mu}\left({\bf r}_{0}\right),\label{eq:GFwithSUM}
\end{equation}
with 
    $A_{\mu}(\omega)={\omega}/{2(\tilde{\omega}_{\mu}-\omega)}$.
    Expanding the Green function with the QNMs can easily be used to compute the SE
rate and the Purcell factor.
We stress again that the medium must meet the condition
for  linear amplifying media, which coincides with a causal Green function in the sense of linear response theory (Kramers-Kr\"onig relations). Here, the Green function must be analytic in the upper half complex plane to fulfill the Kramers-Kr\"onig relations, and this can be rigorously justified by using a QNM approach. Indeed, without such an approach, it is not known whether the model represents a physically meaningful solution for Maxwell's equations.

Considering a 
dipole emitter, $\mathbf{d} =d_{0}\mathbf{n}_{\rm d}$, at location ${\bf r}_{\rm d}$, then the classical SE rate is
~\cite{kristensen_modes_2014}
 \begin{align}
 \begin{split}
 \Gamma_{\rm }(\mathbf{r}_{\rm d},\omega)&=\frac{2}{\hbar\epsilon_{0}}\mathbf{d}\cdot{\rm Im}\{\mathbf{G}(\mathbf{r}_{\rm d},\mathbf{r}_{\rm d},\omega)\}\cdot\mathbf{d},
 \end{split}  
 \end{align}
 and the 
 generalized Purcell factor reads \cite{Anger2006,kristensen_modes_2014} 
 \begin{align}\label{QNMpurcell}
 \begin{split}
     &F_{{\rm }}^{\rm }(\mathbf{r}_{\rm d},\omega) =1+\frac{\Gamma(\mathbf{r}_{\rm d},\omega)}{\Gamma_{0}(\mathbf{r}_{\rm d},\omega)},
 \end{split}
 \end{align}
 where
  $\Gamma_{0}(\mathbf{r}_{\rm d},\omega)=2\mathbf{d}\cdot{\rm Im}\{\mathbf{G}_{\rm B}(\mathbf{r}_{\rm d},\mathbf{r}_{\rm d},\omega)\}\cdot\mathbf{d}/(\hbar\epsilon_0)$,
  and ${\bf G}_{\rm B}$
  is the Green function for a homogeneous medium (known analytically). For a 2D TM (TE) dipole,  ${\rm Im}\{\mathbf{G}_{\rm B}(\mathbf{r}_{\rm d},\mathbf{r}_{\rm d},\omega)\}={\omega^2}/{4c^2}$ (${\omega^2}/{8c^2}$).  The factor of 1 appears naturally for dipole positions outside the resonator~\cite{ge_design_2014}.

For an arbitrary photonic cavity medium, the  QNMs for both the bare resonators (i.e., without coupling),  and also for the coupled system, can be obtained
from an efficient dipole scattering approach 
in complex frequency~\cite{bai_efficient_2013-1}, 
described in more detail in Appendix~\ref{sec:numerics}.
 The total Green function
can also be obtained numerically
from the full dipole response (i.e., without any modal approximations),
which we carry out in COMSOL to check the accuracy of the QNM expansion form. Although the hybrid QNMs can  be obtained numerically as well,
it is far more insightful to develop 
a coupled-mode formalism to describe the coupling geometry.

\section{Coupled Mode Theory with an intuitive Green function expansion}\label{sec:sec3}

\subsection{Wave equation and normal modes}

Before developing a QNM CMT, here  we  present a  NM approach
and also connect to the common literature for describing
when EPs can occur
for coupled loss-gain cavity modes.

To simplify the equations and terminology,
we introduce shorthand notation, and  define the wave equation:
\begin{equation}
\label{eq:WE1}
{\cal L} \ket{\bf E}=
\omega^2 \hat \epsilon_{\rm t} \ket{\bf E},
\end{equation}
where the fields are assumed to have a harmonic frequency dependence, $e^{-i\omega t}$,
${\cal L} = c^2 \boldsymbol{\nabla} \times \boldsymbol{\nabla} \times $, and $\epsilon_{\rm t}({\bf r})$
is the total dielectric constant
that we will assume is nondispersive.
The operator $\hat \epsilon_{\rm t}$
is defined as $\braket{{\bf r}|\hat \epsilon_{\rm t}|{\bf r}'}=\epsilon_{\rm t}({\bf r})
{\bf \delta}({\bf r}-{\bf r}')$,
and the electric field
is given by a projection onto space, $\braket{{\bf r}|{\bf E}}={\bf E}({\bf r},\omega)$.

To construct a Green function solution, we consider a situation  where we start with cavity 1, and then add  cavity 2. The  dielectric
constant defining  cavity 1 is
$\hat \epsilon_1 = \hat \epsilon_{\rm b} + \hat V_1$, so we can also write the wave equation as
\begin{equation}
\label{eq:WE2}
({\cal L}-\omega^2 \hat \epsilon_1) \ket{\bf E}=
\omega^2 \hat V_2 \ket{\bf E},
\end{equation}
where $\hat V_2$
defines the dielectric constant change after
adding in cavity 2, and $\hat \epsilon_{\rm b}$
defines the entire background without either cavity. Naturally, we can also
start from cavity 2 and add in cavity 1, and the end Green function that includes both cavities must be the same.

\subsection{Coupled mode theory and lossless exceptional points using normal modes}

Exploiting the fact that
${\cal L}$ is a linear 
self-adjoint operator over space,
 the homogeneous part
of Eq.~\eqref{eq:WE2} defines an orthogonal set
of eigenstates on a single cavity. It follows that
\begin{equation}
    {\cal L} \ket{{\bf f}_k}
    = \omega_k^2 \hat \epsilon_1
    \ket{{\bf f}_k},
\end{equation}
where $\omega_k$ are the eignfrequencies of the eigenstates
${\bf f}_k$.
These states are also complete and orthogonal~\cite{PhysRevE.68.046606}, so
\begin{align}
\sum_k \hat \epsilon_1 \ket{{\bf f}_k}\bra{{\bf f}_k}&=
{\bf 1}, \nonumber \\
\braket{{\bf f}_i|\hat \epsilon_1|{\bf f}_j} &= \delta_{ij},
\end{align}
and the sum includes all modes, physical and unphysical.

Next, we can formulate a scattering problem,
based on some (arbitrary) reference input field, so that
\begin{equation}
    \ket{\bf E}
    = \ket{{\bf E}^{\rm 0}}
    + \hat {\bf G} \hat V_2 \ket{{\bf E}^0},
\end{equation}
where ${\bf G}({\bf r},{\bf r}')=
\braket{{\bf r}|\hat {\bf G}|{\bf r}'}$
is the total Green function of the system (including both cavities), and is defined from
\begin{equation}
\label{eq:GWE}
    ({\cal L}-\omega^2 \hat \epsilon_1
    - \omega^2 \hat V_2) \hat {\bf G}
    = ({\cal L}-\omega^2 \hat \epsilon_{\rm t}) \hat {\bf G} = \omega^2{\bf 1}.
\end{equation}

For {\it weakly coupled} resonators, we can expand $\hat {\bf G}$ in terms of a restricted set of
carefully chosen basis states, which will be the dominant modes
of the individual cavity systems~\cite{PhysRevE.68.046606}. Thus one obtains the Green function expansion
\begin{equation}
\hat {\bf G} = \sum_{\alpha,\beta} 
B_{\alpha,\beta}
\ket{{\bf f}_\alpha} \bra{{\bf f}_\beta},
\end{equation}
where both sums extend over all states of interest. If we obtain a solution
for $B_{\alpha,\beta}$, then the
scattering problem is solved as we have the total Green function, including the new poles of the coupled resonator system.

In the absence of cavity 2, 
we define the
 cavity mode of the bare cavity 1 (i.e., no cavity mode 2 yet), from
\begin{equation}
    {\cal L} \ket{{\bf f}_1}
    =\omega_1^2 \hat \epsilon_1 
    \ket{{\bf f}_1},
\end{equation}
with the normalization
$\braket{{\bf f}_1|\hat \epsilon_1|{\bf f}_1}
=1$. For simplicity, we assume the cavity supports a dominant single mode in the frequency of interest, but this can easily be generalized to allow for $N$ modes per cavity.
Similarly, we can define the solution of cavity 2, from
\begin{equation}
    {\cal L} \ket{{\bf f}_2}
    =\omega_2^2 \hat \epsilon_2 
    \ket{{\bf f}_2},
\end{equation}
with $\braket{{\bf f}_2|\hat \epsilon_2|{\bf f}_2}
=1$.

Subsequently, we substitute the mode expansions into the
main Green function Eq.~\eqref{eq:GWE},
to obtain
\begin{align}
\label{eq:MGF}
    \sum_{\alpha,\beta}
    [\omega_\alpha^2\braket{{\bf f}_i|\hat \epsilon_\alpha|
    {\bf f}_\alpha}&
    - \omega^2
    \braket{{\bf f}_i|\hat \epsilon_{\rm t}|
    {\bf f}_\alpha}
    ]
    B_{\alpha,\beta}\braket{{\bf f}_\beta|\hat \epsilon_{\rm t}|
    {\bf f}_j}  \nonumber \\
    &=\omega^2\braket{{\bf f}_i|\hat \epsilon_{\rm t}|
    {\bf f}_j},
\end{align}
where $i,j$ refer to any basis states,
and $\hat \epsilon_\alpha$ can refer
to $\hat \epsilon_1$ or $\hat \epsilon_2$.
Equation \eqref{eq:MGF}
defines a matrix equation whose  poles correspond to the new eigenfrequencies
of the composite system. To proceed,
we exploit the fact that the modes are only weakly coupled to each other, and so 
\begin{equation}
\label{eq:totalorth}
    \braket{{\bf f}_\alpha|\hat \epsilon_{\rm t}|{\bf f}_{\beta}} = \delta_{\alpha\beta},
\end{equation}
which can typically be easily checked numerically (else this approximation can also be relaxed, which just yields a more complex matrix to be solved).

The matrix defined from
Eq.~\eqref{eq:MGF}, namely 
$MBT = T$, has
the solution
$B_{\alpha,\beta}=[M^{-1}]_{\alpha,\beta}$,
with elements:
\begin{align}
\label{eq:FullM}
M_{\alpha,\alpha} &=
\frac{1}{\omega^2}(\omega_\alpha^2 -\omega^2),
\nonumber \\
M_{\alpha,\beta\neq \alpha} &=
\frac{1}{\omega^2}(-\omega_\beta^2 
\braket{\braket{{\bf f}_\alpha
|\hat V_\alpha|{\bf f}_\beta}}). 
\end{align}
Thus, the  matrix $M$ is 
\begin{equation}
M=
\frac{1}{\omega^2}
\begin{pmatrix} 
\omega_1^2 -\omega^2 & -{\omega_2^2} \braket{{\bf f}_1
|\hat V_1|{\bf f}_2} \\
-{\omega_1}^2 \braket{{\bf f}_2
|\hat V_2|{\bf f}_1} & \omega_2^2 -\omega^2
\end{pmatrix},
\end{equation}
and  we  define the inter-mode coupling rate:
\begin{equation}
\kappa_{12}=\frac{\omega_2}{2} \braket{{\bf f}_1
|\hat V_1|{\bf f}_2},
\end{equation}
so that
\begin{equation}
M=
\frac{1}{\omega^2}
\begin{pmatrix} 
\omega_1^2 -\omega^2 & -2{\omega_2}
\kappa_{12} \\
-2{\omega_1} \kappa_{21}  & \omega_2^2 -\omega^2
\end{pmatrix}.
\end{equation}

Matrix inversion   can be solved without approximations,
however it is appropriate to obtain 
an easier form within a rotating-wave approximation. Using
$\omega_\alpha^2-\omega^2
\approx (\omega_\alpha-\omega)2\omega_\alpha
\approx(\omega_\alpha-\omega)2\omega$, 
then
\begin{equation}
M=
\frac{2}{\omega}
\begin{pmatrix} 
\omega_1 -\omega & -\kappa_{12} \\
-\kappa_{21} & \omega_2 -\omega
\end{pmatrix},
\end{equation}
and we obtain
an explicit solution for the Green function
expansion coefficients
\begin{align}
\label{eq:roots1}
B_{\alpha,\beta}
    = 
    \frac{\omega/2}{(\omega-\omega_+)(\omega-\omega_-)}
    \begin{pmatrix} 
\omega_2 -\omega & \kappa_{12} \\
\kappa_{21} & \omega_1 -\omega
\end{pmatrix},
\end{align}
where the pole frequencies are 
\begin{equation}
\omega_{\pm}=
\frac{\omega_1+\omega_2}{2}
\pm \frac{\sqrt{4\kappa_{12}\kappa_{21} + (\omega_1-\omega_2)^2}}{2}.
\end{equation}
%
Finally, we note that
for closed cavity systems,
unitarity of 
    a Hermitian system
    also {\rm requires} that
    $\kappa_{12}=\kappa_{21}^*$, 
    and the pole frequencies are simply
\begin{equation}
\omega_{\pm}=
\frac{\omega_1+\omega_2}{2}
\pm \frac{\sqrt{4|\kappa_{12}|^2 + (\omega_1-\omega_2)^2}}{2}.
\end{equation}

This concludes the derivation of the NM Green function with weakly coupled cavities.
With regards to EPs, 
if we now consider the case with two cavity systems, one with loss:
$-\gamma$ ($\omega_1 =
\omega_0-i \gamma $),
and one with a loss compensating
gain: $+\gamma$ ($\omega_2 =\omega_0 + i \gamma $),
then one might be tempted to predict a situation where $\omega_{\pm} \rightarrow \omega_0$, if 
$|\kappa_{12}|=\pm\gamma$.
The problem with this argument is that the
original cavity 
modes here do not satisfy a Hermitian eigenvalue problem (assuming they are open and/or contain some loss), and thus the above coupled mode Green function solutions are not valid. Strictly, they are only valid for real eigenfrequency cavity modes.
For very high $Q$ cavities, however, the theory may be approximately correct, but the question and definition of a true  EP still becomes questionable.


\subsection{Coupled mode theory and lossy exceptional points  using quasinormal modes}\label{sec:IIIC}

Since we are interested in open cavities with loss and gain, we now adopt a more rigorous and appropriate resonator approach, using QNMs. One form of the QNM normalization can be defined from:
\begin{equation}
\braket{\braket{\tilde {\bf f}_1|\hat \epsilon_1|\tilde {\bf f}_1}}
\rightarrow 
\int d{\bf r} \epsilon_1({\bf r}) \tilde {\bf f}_1({\bf r}) \tilde {\bf f}_1({\bf r}) =1,
\end{equation}
where some coordinate transform has been applied to regularize the outgoing surface fields, e.g., through
 perfectly matched layers~\cite{sauvan_theory_2013},
but such terms will not be needed in the region for CMT overlap integrals as discussed below.
Alternative QNM normalizations 
 are 
discussed in Appendix~\ref{sec:numerics}.

These fields are now solutions to the eigenvalue problem with complex
frequencies, and  virtually all of the previous
equations apply, with some simple replacements:

(i) The eigenfrequencies become complex
and formally discrete\footnote{Although the NMs are also assumed to be discrete for resonator problems, formally they yield continuous eigenfrequencies.}:
\begin{equation}
    \omega_k \rightarrow \tilde \omega_\mu,
\end{equation}
with $\tilde \omega_\mu = \omega_\mu - i \gamma_\mu$.

(ii)
The QNM  Green function expansion is
\begin{equation}\label{eq:QNMGreen}
\hat {\bf G} = \sum_{\alpha,\beta} 
B_{\alpha,\beta}
\ket{\tilde{\bf f}_\alpha} \bra{\tilde{\bf f}_\beta^*}.
\end{equation}


(iii)
The completeness relation becomes
\begin{align}
\frac{1}{2} \sum_{\mu=\pm 1,2,...} \hat \epsilon_1 \ket{\tilde{\bf f}_\mu} 
\bra{\tilde {\bf f}_\mu^*}&=
{\bf 1}, \nonumber \\
\braket{\braket{\tilde{\bf f}_i|\hat \epsilon_1|{\tilde{\bf f}}_j}} &= \delta_{ij},
\end{align}
which is assumed to be valid  for spatial regions near or inside the scattering geometry.

(iv)
For regions far outside the scattering geometry,
one can use regularized QNMs (non divergent), such that
\begin{equation}
\hat {\bf G} = \sum_{\alpha,\beta} 
\tilde B_{\alpha,\beta}
\ket{\tilde{\bf F}_\alpha}
\bra{\tilde{\bf F}_\beta^*},
\end{equation}
where $\ket{\tilde{\bf F}_\alpha}$
is obtained from a Dyson solution using the
original QNM~\cite{ge_quasinormal_2014}, or using near-field to far-field transformations~\cite{ren_near-field_2020}.
For high $Q$ cavities however,
using the QNMs in the perturbative cavity region
is in practice extremely accurate, as we  also confirm later.

With these replacements, we can use  the same approach as before.
Assuming again that the solution is first solved for cavity 1, and then we add in cavity 2, we
derive:
\begin{align}
\label{eq:BRWA}
\tilde B_{\alpha,\beta}
    = 
    \frac{\omega/2}{(\omega-\tilde \omega_+)(\omega-\tilde \omega_-)}
    \begin{pmatrix} 
\tilde \omega_2 -\omega & \tilde \kappa_{12} \\
\tilde \kappa_{21} & \tilde \omega_1 -\omega
\end{pmatrix},
\end{align}
where
%
\begin{equation}\label{QNMCMT_coup}
\tilde \kappa_{12}=
\frac{\tilde \omega_2}{2} \braket{\braket{\tilde{\bf f}_1
|\hat V_1|\tilde{\bf f}_2}},
\end{equation}
which notably
now uses an {\it unconjugated norm}
in the QNM overlap integrals.
%
For the QNM formalism, note $\tilde \kappa_{12} \neq \tilde \kappa_{21}^*$, in contrast to NM theory.
In QNM theory, 
these
off-diagonal terms are not the complex conjugates of
each other, since the open cavity system does not obey Hermiticity.
Indeed, using such a relation with open cavities is simply ill-defined.

The full QNM Green function solution (using Eq.~\eqref{eq:BRWA} and
Eq.~\eqref{eq:QNMGreen})
can be written as follows:
\begin{align}
\label{eq:Gsol1}
\hat {\bf G}
&= \frac{\frac{\omega}{2}(\tilde \omega_2-\omega) \ket{\tilde {\bf f}_1}\bra{\tilde{\bf f}_1^*}}{(\omega-\tilde \omega_+)(\omega-\tilde \omega_-)} 
+ \frac{\frac{\omega}{2}\tilde \kappa_{12} \ket{\tilde{\bf f}_1}\bra{\tilde{\bf f}_2^*}}{(\omega-\tilde \omega_+)(\omega-\tilde \omega_-)}
\nonumber \\
&+\frac{\frac{\omega}{2}\tilde \kappa_{21} \ket{\tilde{\bf f}_2}\bra{\tilde{\bf f}_1^*}}{(\omega-\tilde \omega_+)(\omega-\tilde \omega_-)}
+
\frac{\frac{\omega}{2}(\tilde \omega_1-\omega) \ket{\tilde{\bf f}_2}\bra{\tilde{\bf f}_2^*}}{(\omega-\tilde \omega_+)(\omega-\tilde \omega_-)},
\end{align}
with the two new QNM pole frequencies for the composite cavity system,
\begin{equation}
\label{eq:rootsRWA}
\tilde \omega_{\pm}=
\frac{\tilde\omega_1+\tilde\omega_2}{2}
\pm \frac{\sqrt{4\tilde \kappa_{12}\tilde \kappa_{21} + (\tilde\omega_1-\tilde\omega_2)^2}}{2}.
\end{equation}
In the limit of no coupling, 
$\tilde \kappa_{12}=\tilde \kappa_{21}=0$, and
$\tilde \omega_{\pm} \rightarrow \tilde \omega_{1,2}$, and we recover the 
original Green function expansion for two separated resonators.

Finally, we briefly discuss the condition for EPs.
Equation~\eqref{eq:rootsRWA} shows that the exceptional point occurs when
\begin{align}
\label{eq:EP1}
2\sqrt{\tilde \kappa_{21} \tilde \kappa_{12}}|_{\rm EP}\equiv A+iB
= \pm i (\tilde\omega_1-\tilde\omega_2),
\end{align}
and in general one might obtain a lossy EP (at best), since the resonators are open, and the coupled system must also yield
a finite loss for any linear amplifying medium~\cite{raabe_qed_2008}.
The influence of dissipation on EPs and CMT is often argued 
heuristically. For example (defining $\tilde{\omega}_{1}=\omega_{1}-i\gamma_{1}$ and $\tilde{\omega}_{2}=\omega_{2}-i\gamma_{2}$), in Ref.~\onlinecite{miri_exceptional_2019}, Miri and Al\`u
discussed two limits 
where (i) the coupling is real, i.e., $B=0$,  so the EP condition would become $A=\pm 2\gamma_{0}$, with
$\gamma_1=\gamma_0$
and $\gamma_2=-\gamma_0$ ($\omega_{1}=\omega_{2}=\omega_{0}$); and (ii)
assume the coupling is purely imaginary, i.e., $A=0$; then the EPs condition would become $B=\pm \Delta$,
where $\omega_1=\omega_0+\Delta$
and
$\omega_2=\omega_0$ ($\gamma_{1}=\gamma_{2}$). 
In practise it would be very difficult to reach exactly this regime (Eq.\eqref{eq:EP1}), though one can likely come close, and certainly find signatures of EP behavior.

Experimentally,
in Refs.~\onlinecite{chang_paritytime_2014,peng_paritytime-symmetric_2014,peng_chiral_2016}, the spectra below and above EPs are observed, as well as the associated unidirectional transmission, but an exact EP was not demonstrated. Note, however, that
these systems involve waveguides, so 
the CMT  theory should be modified to connect more closely to such works. The 
 theory in Ref.~\onlinecite{chen_parity-time-symmetric_2018}
 also finds non-ideal EPs if one shows a zoom-in of the EP region in the complex
 frequency plane.

For a high $Q$ resonator, one can assume that an approximate EP may be obtained when
${\rm Re} (\tilde \kappa_{12})
=  \gamma_0$,
but in general this condition is an approximate one for open resonators (and QNMs).
Moreover, the lineshapes associated with the QNMs
are far richer than with two coupled
Lorentzian lineshapes, as we will demonstrate  in more detail below, even for very high $Q$  modes ($Q\approx 10^5$).

 \subsection{Hybrid quasinormal modes using the couple
 mode theory}

Next, we 
introduce a model for obtaining the hybridized QNMs, which can be obtained analytically  in terms of 
 the uncoupled
QNMs; this  considerably simplifies the numerical solutions and helps to identify the underlying physics of how the hybridized modes are formed.

In terms of the hybrized eigenfrequencies
$\tilde{\omega}_\pm$,
defined in Eq.~\eqref{eq:rootsRWA}, we obtain:
\begin{align}\label{eq:QNMs_pm}
\ket{\tilde{\bf f}^\pm}&=
\frac{\tilde\omega_\pm-\tilde\omega_2}{\sqrt{(\tilde\omega_\pm-\tilde\omega_2)^2+ \tilde \kappa_{21}^2}}
\ket{\tilde{\bf f}_1}
\nonumber \\
&
+ \frac{-\tilde\kappa_{21}}{\sqrt{(\tilde\omega_\pm-\tilde\omega_2)^2+ \tilde \kappa_{21}^2}} 
\ket{\tilde{\bf f}_2},
\end{align}
or
\begin{align}
\ket{\tilde{\bf f}^\pm}&=
\frac{-\tilde\kappa_{12}}{\sqrt{(\tilde\omega_\pm-\tilde\omega_1)^2+ \tilde \kappa_{12}^2}}
\ket{\tilde{\bf f}_1}\nonumber \\
&+ \frac{\tilde\omega_\pm-\tilde\omega_1}{\sqrt{(\tilde\omega_\pm-\tilde\omega_1)^2+ \tilde \kappa_{12}^2}} 
\ket{\tilde{\bf f}_2}.
\end{align}
Assuming $\tilde\kappa=\tilde\kappa_{12} \approx \tilde\kappa_{21}$, then
\begin{align}
\ket{\tilde{\bf f}^\pm}&=
\frac{-\tilde\kappa_{}}{\sqrt{(\tilde\omega_\pm-\tilde\omega_1)^2+ \tilde \kappa_{}^2}}
\ket{\tilde{\bf f}_1}\nonumber \\
&+ \frac{\tilde\omega_\pm-\tilde\omega_1}{\sqrt{(\tilde\omega_\pm-\tilde\omega_1)^2+ \tilde \kappa_{}^2}} 
\ket{\tilde{\bf f}_2}.
\label{eq:QNMHybrid}
\end{align}

Subsequently, we also obtain the
new Green function:
\begin{align}
\label{eq:Gsol2}
\hat {\bf G}
&= \frac{\omega \ket{\tilde {\bf f}^+}\bra{\tilde{\bf f}^{+*}}}
{2(\omega-\tilde \omega_+)} 
 + \frac{\omega  \ket{\tilde{\bf f}^-}\bra{\tilde{\bf f}^{-*}}}{2(\omega-\tilde \omega_-)},
\end{align}
which gives the same results as Eq.~\eqref{eq:Gsol1}\footnote{Apart from at an EP, which is discussed later, as the hybrid modes become self-orthogonal.}, but is now in diagonalized form.

\section{Green functions and Purcell factors at the exceptional point}
\label{sec:sec4}

The SE rates (and generalized Purcell factors) can be significantly modified close to EPs~\cite{PhysRevA.84.063833,sunada_enhanced_2018,lin_enhanced_2016,Pick2017,pick_enhanced_2017,PhysRevResearch.2.023375}, where a squared Lorentzian contribution has been emphasized as well as signatures of linewidth narrowing.  With higher-order EPs, these linewidths may be reduced further~\cite{Pick2017}, e.g., a cubic Lorentzian lineshape
has been predicted with third-order EPs~\cite{lin_enhanced_2016}.

Below, we 
focus on the more general second-order EPs and first
briefly comment on previous theoretical predictions
about the modified lineshapes.
In Ref.~\onlinecite{PhysRevResearch.2.023375}, the frequency dependent response of SE has the following form:
\begin{equation}
    \tilde S(\omega)
= \frac{A}{(\omega-\omega_k)^2}-\frac{B}{(\omega-\omega_k)},
\end{equation}
which is a complex Lorentzian squared and a single Lorentzian, if very near the EP. 

A similar regime was predicted and shown numerically 
in Ref.~\onlinecite{Pick2017}. The main Green function
response was predicted to have the following form:
\begin{equation}
G(\omega)
= \frac{C}{\omega^2-\tilde \omega^2_{\rm EP}}
+ \frac{D}{(\omega^2-\tilde \omega^2_{\rm EP})^2},
\end{equation}
where $C$ and $D$ are connected with  the Jordan vectors.
They also give an approximate coupled mode theory expression for the  Green function expansion.


Next, we will first 
 show how our
QNM Green function is fully consistent with the above predictions, and then show why 
one can 
find a much richer range of complex lineshapes,
which we also demonstrate explicitly in the numerical results section.

We define the EP  resonance frequency 
from $\tilde \omega_{\pm}= \tilde \omega_{\rm EP}=(\tilde\omega_1+\tilde\omega_2)/2$,
which occurs when 
 $\tilde \Delta_{12}^2\equiv
(\tilde \omega_1-\tilde\omega_2)^2= - 4\tilde \kappa_{12}\tilde \kappa_{21}$. Thus we can write
\begin{equation}
M =
\frac{2}{\omega}
\begin{pmatrix} 
\tilde \omega_{\rm EP} -\omega
+ \tilde \omega_{\rm EP}-\tilde \omega_2 & -\tilde\kappa_{12} \\
-\tilde\kappa_{21} & 
\tilde \omega_{\rm EP} -\omega
+ \tilde \omega_{\rm EP}-\tilde \omega_1
\end{pmatrix},
\end{equation}
which is identical to what we have done already
but this form allows us to expand the
Green function near the EP resonances.
For example, the
$\hat {\bf G}_{1,1}$ term (expanded in terms of the bare mode from resonator 1) is:
\begin{equation}
   \hat {\bf G}_{1,1}
    = \frac{\omega}{2}
    \left (
\frac{\tilde \omega_{\rm EP}-\tilde\omega_{1}}{(\omega - \tilde
\omega_{\rm EP})^2}
- \frac{1}{\omega - \tilde 
\omega_{\rm EP}}
\right ) 
\ket{\tilde {\bf f}_1}
\bra{\tilde {\bf f}_1^*},
\end{equation}
where we  clearly see the separation
of a Lorentzian and a Lorentzian squared contribution,
in agreement with Refs.~\onlinecite{PhysRevA.84.063833,Pick2017,PhysRevResearch.2.023375}. 

We stress that our spectral forms are not actually Lorentzian or Lorentzian-squared  because of the QNM phase. Indeed, given the appropriate QNM phase, these terms can contribute negatively, a feature that is already known with coupled QNMs yielding a Fano resonance~\cite{RosenkrantzdeLasson2015,2017PRA_hybrid,ren_near-field_2020,el-sayed_quasinormal-mode_2020}.
It is also useful to  compare with the response of the single cavity solution, which is  simply 
\begin{equation}
\hat {\bf G}^{\rm cav1}
    = 
    \left (
\frac{\omega}{2( \tilde 
\omega_{1}-\omega)} \right ) 
\ket{\tilde {\bf f}_1}
\bra{\tilde {\bf f}_1^*},
\end{equation}
which has a single Lorentzian-like feature, again modified by the QNM phase. Thus the Lorentzian-squared feature is caused by the EP coupling regime.

Since $\tilde \omega_{\rm EP}= (\tilde \omega_1+\tilde \omega_2)/2$, we can also write the QNM Green function solution
to the coupled resonator EP regime as 
%
\begin{subequations}
\begin{align}
\label{eq:GEP3}
\hat {\bf G}_{1,1}
    &= 
    \frac{\omega}{2} \left (
\frac{-\tilde\Delta_{12}/2} {(\tilde 
\omega_{\rm EP}-\omega)^2}
+ \frac{1}{(\tilde 
\omega_{\rm EP}-\omega)}
\right ) 
\ket{\tilde {\bf f}_1}
\bra{\tilde {\bf f}_1^*},\\
\hat {\bf G}_{1,2}
    &= 
    \frac{\omega}{2}\left (
\frac{\tilde{\kappa}_{12}}{(\tilde 
\omega_{\rm EP}-\omega)^2}
\right ) 
\ket{\tilde {\bf f}_1}
\bra{\tilde {\bf f}_2^*},\\
\hat {\bf G}_{2,1}
    &= 
   \frac{\omega}{2} \left (
\frac{\tilde{\kappa}_{21}}{(\tilde 
\omega_{\rm EP}-\omega)^2}
\right ) 
\ket{\tilde {\bf f}_2}
\bra{\tilde {\bf f}_1^*},\\
\hat {\bf G}_{2,2}
    &= 
    \frac{\omega}{2} \left (
\frac{\tilde\Delta_{12}/2}{2( \tilde 
\omega_{\rm EP}-\omega)^2}
+ \frac{1}{ (\tilde 
\omega_{\rm EP}-\omega)}
\right ) 
\ket{\tilde {\bf f}_2}
\bra{\tilde {\bf f}_2^*},
\end{align}
\end{subequations}
with $\tilde\Delta_{12}=\tilde\omega_1-\tilde\omega_2$. Note that any divergences
from the hybrid modes at the EP are avoided here, since we  use an expansion in terms of the bare modes and CMT.

Interestingly, we have obtained this known (and highly unusual) form without having to perform any Jordan expansion around the EP pole~\cite{Pick2017,sunada_enhanced_2018,PhysRevResearch.2.023375}. 
Assuming the CMT is accurate (and we show later that it can be quantitatively accurate), this is clearly a more convenient form to work with.
It is also important to note that
the hybrid QNM modes, e.g., in Eq.~\eqref{eq:QNMHybrid},
diverge at the EP, since the QNMs at exactly this point are ill-defined and self-orthogonal. In practise, however, this is not a restriction, as most solutions will deviate from precisely this point, where the two Green function responses become identical and well defined; thus one can use either the bare modes (non-diagonal form) or the hybrid mode solutions (diagonal form). This is a significant advantage that benefits from
the constructed CMT.

In the next section,
 we will highlight much more general forms of the spectral lineshapes near the EP, which are fully verified by numerically exact solutions (within numerical precision). Our calculations also point out a fundamental problem with defining
a SE rate in media with coupled
loss and gain resonators, even when the
total Green function is analytic in the upper complex half plane~\cite{raabe_qed_2008}.

\section{Numerical Results for coupled loss-gain microdisk resonators}
\label{sec:sec5}

 We consider two coupled loss-gain microdisk resonators, both with a disk radius of $R=5~\mu$m (ct.~Fig.~\ref{disk_sche}). The refractive index of the lossy (gain) resonator is $n_{\rm loss}=2+10^{-5}i$ ($n_{\rm gain}=2-5\times10^{-6}i$), unless stated otherwise. The background medium is free space with  $n_{\rm B}=1$. The gap distance between the resonators is $d_{\rm gap}$ (around $1120\sim1200$~nm). The dipole (out-of-plane line current, a point in 2D, shown as red dot in Fig.~\ref{disk_sche}) is placed within the gap, at several possible positions:  $\mathbf{r}_{\rm d}=\mathbf{r}_1$, $\mathbf{r}_2$, $\mathbf{r}_3$, $\mathbf{r}_{4}$, and $\mathbf{r}_{5}$, where $l=10~$nm and $h=(d-2l)/4$ ($\mathbf{r}_3$ is at gap center).

\subsection{Quasinormal modes for single loss and single gain resonators}

\begin{figure*}[t]
    \centering
    \includegraphics[trim=0 -0.8cm 0 0,width=0.88\textwidth]{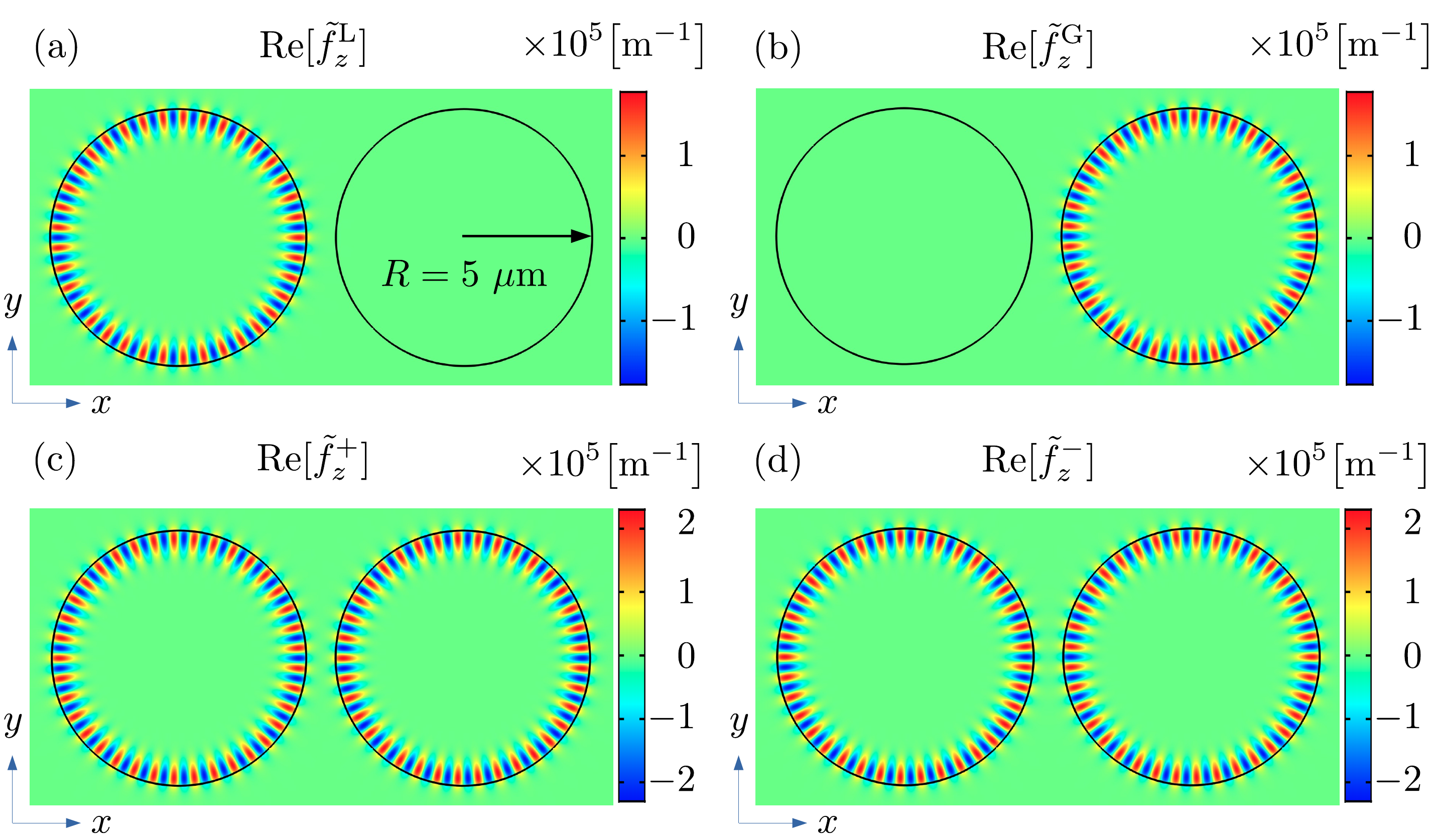}
    \centering
    \includegraphics[width=0.95\columnwidth]{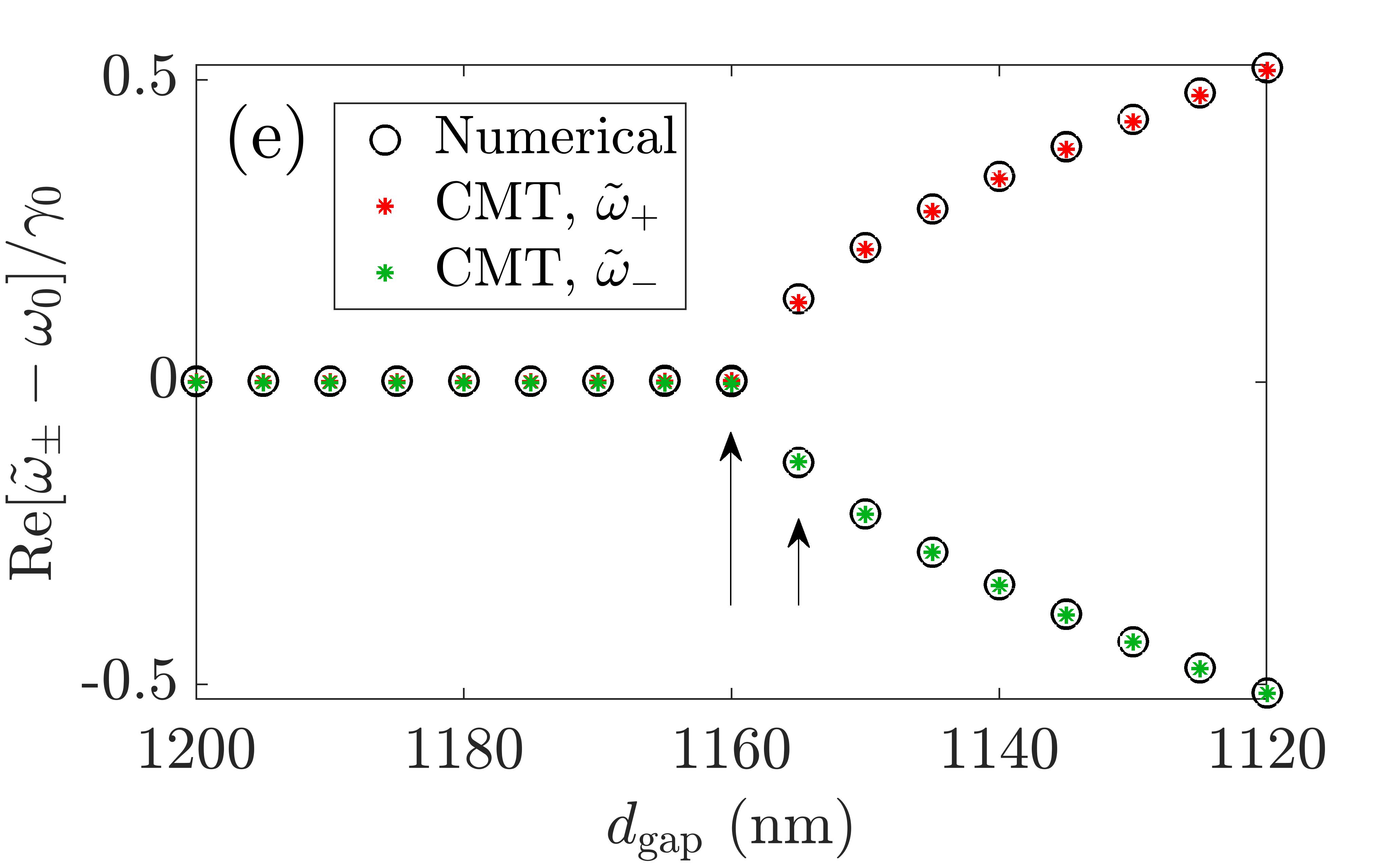}
    \includegraphics[width=0.95\columnwidth]{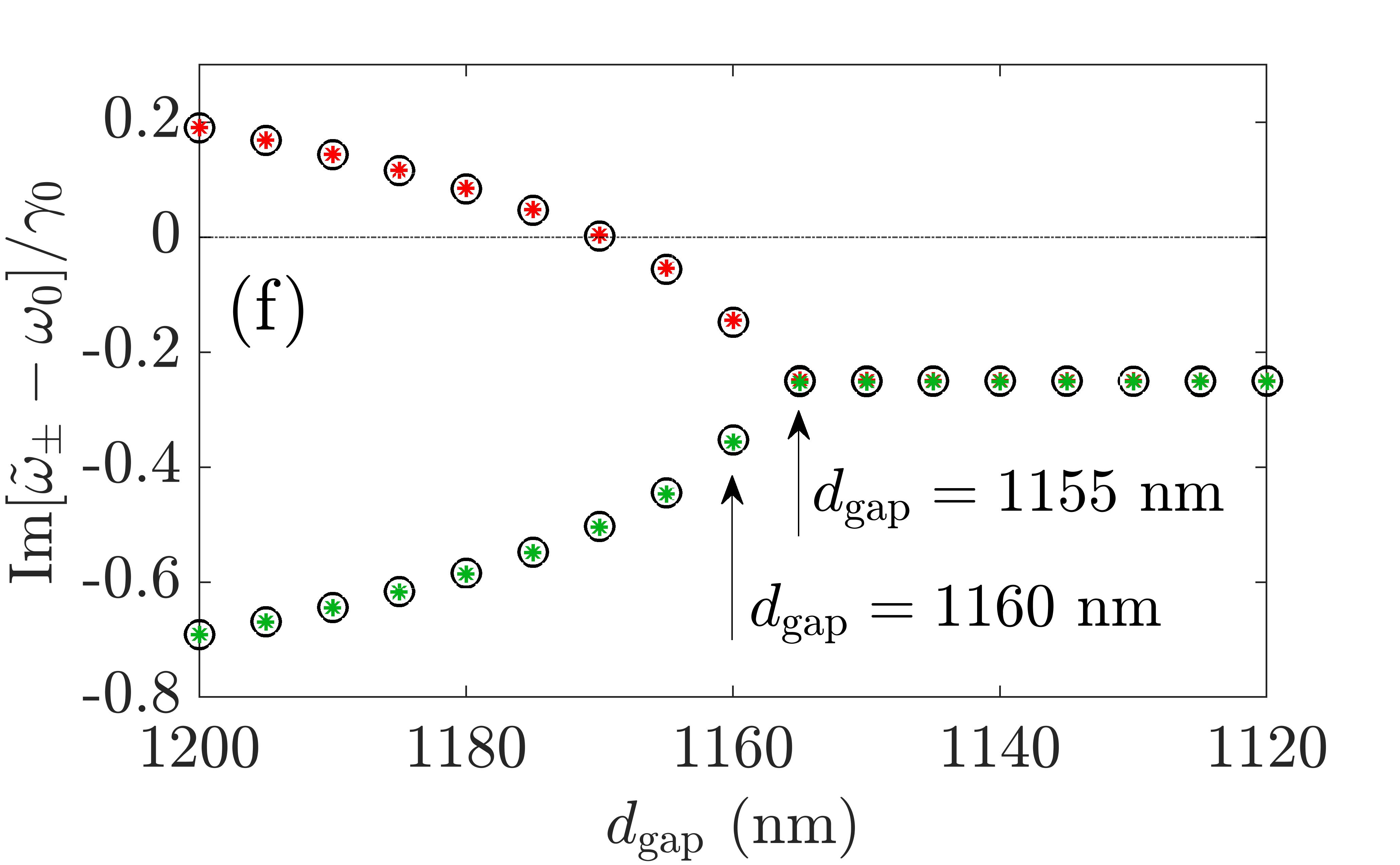}
    \caption{Spatial profile of the QNMs for the following cavity configurations: (a) lossy resonator only (${\rm Re}[\tilde{f}^{\rm L}_{z}]$\footnote{The cavity at the right is shown for clarity and for use in the CMT, but of course is not there for the single resonator calculations.}, Eq.~\eqref{2DQNM_norm}), (b) gain resonator only (${\rm Re}[\tilde{f}^{\rm G}_{z}]$, Eq.\eqref{2DQNM_norm}), and (c-d) Hybrid modes with coupling (${\rm Re}[\tilde{f}^{\pm}_{z}]$, Eq.\eqref{eq:QNMs_pm}), for a separation distance $d_{\rm gap}=1155$~nm; this regime is close to the EP  as shown in (e) and (f). (e-f) Complex QNM eigenfrequencies of the coupled microdisk resonators from a direct numerical eigenfrequency solver in COMSOL (numerical, black circles) and the analytical CMT (Eq.~\eqref{eq:rootsRWA}, red stars for $\tilde{\omega}_{+}$, green  stars for $\tilde{\omega}_{-}$), as a function of gap separation $d_{\rm gap}$. 
    Note using the eigenfrequency solver, there are two bare degenerate modes per resonator, and four new eigenfrequencies for coupled resonators (four black circles for each gap distance), but with two pair of degenerate modes. For the dipole QNM technique, only one of degenerate standing-wave QNMs  are obtained and used for single resonator (see text and Appendix~\ref{sec:numerics}); thus,   the analytical CMT produced two new eigenfrequencies in the presence of finite coupling (one red star and one green star for each gap distance).
    The arrows show two solutions near an EP. Note that $\omega_{0}$ and $\gamma_{0}$ are for uncoupled single lossy resonator with $n_{\rm loss}=2+10^{-5}i$ ($\tilde{\omega}_{\rm L}=\omega_{\rm L}-i\gamma_{\rm L}=\omega_{0}-i\gamma_{0})$; while for the uncoupled single gain cavity, with $n_{\rm loss}=2-5\times10^{-6}i$, the resonance is around at $\tilde{\omega}_{\rm G} \approx \omega_0+i0.5\gamma_{0}$. 
    }\label{QNM_distribution}
\end{figure*}

Before investigating the coupled resonator regime, we first show the QNMs for the single loss or gain  resonators, which will be used as input for the CMT in the next subsection.

A single 2D microdisk resonator supports WGMs~\cite{righini_whispering_2011,schunk_identifying_2014}, which generally can be described by three integers: radial mode number $q$~(=1,2,3...), azimuthal mode number $m$, and polarization $p$ (TM or TE).
Note that for the same $q$, $m$, and $p$, there are two degenerate counterpropagating modes~\cite{leung_completeness_1994,mazzei_controlled_2007,teraoka_resonance_2009,cognee_cooperative_2019}:
${E}_{\rm cw}(\mathbf{r},\phi)=E(\mathbf{r})e^{(-im \phi)}$ with clockwise (cw) direction and ${E}_{\rm ccw}(\mathbf{r},\phi)=E(\mathbf{r})e^{(im \phi)}$ with counter clockwise (ccw) direction.  for a TM mode, there is only the $z$ component for electric fields), which share the same eigenvalues 
( for a TE mode, the $z$ component  of the magnetic field has similar properties). 
A linear combinations of these modes result in degenerate standing waves~\cite{mazzei_controlled_2007,teraoka_resonance_2009,cognee_cooperative_2019}, such as a symmetric standing mode ${E}_{\rm s}={E}_{\rm cw}+ {E}_{\rm ccw} \propto \cos{(m \phi)}$, and an antisymmetric standing wave mode ${E}_{\rm as}={E}_{\rm cw}- {E}_{\rm ccw} \propto \sin{(m \phi)}$.

In general, the mode with $q=1$ and $m\gg1$ has a high $Q$ and strong field confinement.
In this work, we focus on a TM mode (${H}_{x},{H}_{y},{E}_{z}$) with $q=1$, and $m=37$ (yielding a resonant wavelength near the telecommunication band, around $1487~$nm).
To compute the normalized QNMs, we employ an efficient dipole scattering approach to obtain the QNMs in complex frequency~\cite{bai_efficient_2013-1}, where an out-of-plane line current (a point in 2D, $z$-polarized) is placed at $10$~nm away from the 2D microdisk (the dipole is at $\mathbf{r}_{\rm d}=\mathbf{r}_{1}$ for the single lossy cavity and at $\mathbf{r}_{\rm d}=\mathbf{r}_{5}$ for the single gain cavity). For more details, see Appendix~\ref{AppendixA_sub1}.


Conveniently, for our TM dipole location along  the $x$-axis,  we excite only one of the 
 two degenerate standing wave modes, as shown in Fig.~\ref{QNM_distribution}(a) for the single lossy cavity, which has the form $\cos(m\phi)$ if one considers $\phi=0$ at the positive $x$-axis. 
The orthogonal and generate QNM has the form $\sin(m\phi)$, constructed   from an asymmetric linear combination of the clockwise and counter clockwise modes, though we only need to consider the symmetric QNM for our chosen dipole locations below.
More details can be found in Appendix \ref{sec:degenerate_WGMs}.
Also note that this working QNM  dominates in the frequency region of interest below, because the two closest modes are $q=4$, $m=25$ and $q=3$, $m=29$, and the angular frequency spacing between them and the working mode are $\Delta\omega=3.43{~\rm THz}$ and $\Delta\omega=19.6{~\rm THz}$, which are much larger than the FWHM ($\sim12.52{~\rm GHz}$ for the single lossy cavity, and $\sim6.26{~\rm GHz}$ for the  single gain cavity). Also note,  the free spectral range for modes 
with $q=1$ is also much larger than these FWHM values, e.g., the angular frequency spacing between $q=1$, $m=36$ (or $q=1$, $m=38$) with the working mode ($q=1$, $m=37$), would be around $\Delta\omega=31.6{~\rm THz}$. Thus, we can adopt a single QNM approximation for the mode of interest for each resonator, which we will also verify below by performing  full dipole calculations with no mode approximations.

Numerically, the complex angular eigenfrequency for single lossy resonator is found to be $\tilde{\omega}_{\rm L}=\omega_{\rm L}-i\gamma_{\rm L}\equiv \omega_{0}-i\gamma_{0}=1.266666\times10^{15} - 6.260269\times10^{9}i$ rad/s, with a quality factor $Q\sim10^{5}$;
here  $\omega_{0}$ ($\gamma_{0}$) is the real part (opposite imaginary part) of the complex eigenfrequency.
The corresponding QNM field distribution (real part ${\rm Re}[\tilde{f}_{z}^{\rm L}]$), 
is shown in Fig.~\ref{QNM_distribution} (a); the imaginary part is much smaller than the real part, and thus is not shown.

To better understand the overlap integrals for use in CMT, the QNMs at the second resonator region are also shown ($1155~$ nm away, i.e., $d_{\rm gap}=1155~$nm), which is hardly seen on a linear scale because they are very small compared with fields close to the resonator.
This also indicates that the lossy QNMs show no sign of a spatial divergence in this region and thus they can be accurately used for input to CMT, which   is a consequence of the high
$Q$.

Similarly, the complex angular eigenfrequency for the single gain resonator is found at $\tilde{\omega}_{\rm G}=\omega_{\rm G}-i\gamma_{\rm G}\sim\omega_{0}+i0.5\gamma_{0}$. 
The corresponding QNM field distribution (real part ${\rm Re}[\tilde{f}_{z}^{\rm G}]$) 
is shown in Fig.~\ref{QNM_distribution}(b).
In addition, the Purcell factors from the dipole location ($10~$nm away from the loss and gain  resonator), using a single QNM contribution, (Eq.~\eqref{QNMpurcell}) agree quantitatively well with full dipole results
(see Appendix \ref{sec:single_modes}).
We also highlight that the Purcell factors (as defined in a semi-classical model) are net negative for single gain cavity only (since the field is being amplified rather than dissipated), a regime that will also be shown below
for the coupled resonator system.


\subsection{Hybrid quasinormal modes for coupled loss-gain resonators}

We next study the hybrid QNMs formed from the two coupled microdisk resonators,
using the QNM CMT and also using full numerical solutions to confirm the accuracy of our semi-analytical results.

Using only the QNMs from single lossy/gain cavities as input, the properties for the coupled modes are obtained analytically. First, the new angular eigenfrequencies $\tilde{\omega}_{\pm}$ are computed from Eq.~\eqref{eq:rootsRWA}, where the coupling coefficients $\tilde{\kappa}_{12}$ and $\tilde{\kappa}_{21}$ (both complex, Eq.~\eqref{QNMCMT_coup}) are related to the overlap QNM integrals for different gap distances $d_{\rm gap}$. 
Note the input variables $\tilde{\omega}_{1/2}$ and $\tilde{\mathbf{f}}_{1/2}$ in theory parts are changed to $\tilde{\omega}_{\rm L/G}$ and $\tilde{\mathbf{f}}^{\rm L/G}$ in the numerical parts accordingly.
As shown in Figs.~\ref{QNM_distribution}(e) and (f), the analytical eigenfrequencies (Eq.~\eqref{eq:rootsRWA}) 
versus 
full numerical solution in COMSOL (eigenfrequency solver)
show excellent agreement, and at all gap separations.
This quantitative  level of agreement is also obtained for the QNMs and the QNM Green functions as we will show in more detail below.
%

The complex coupling coefficients for $d_{\rm gap}=1155~$nm ($d_{\rm gap}=1160~$nm) are $\tilde{\kappa}_{12}/\gamma_{0}=-0.7614-5.745\times 10^{-6}i$ ($-0.7425-5.375\times10^{-6}i$) and $\tilde{\kappa}_{21}/\gamma_{0}=-0.7425+4.153\times10^{-5}i$ ($-0.7425+4.087\times10^{-6}i$), where the small imaginary part is mainly due to the high quality factor of the bare resonator and one can find they do not satisfy $\tilde{\kappa}_{12}=\tilde{\kappa}_{21}^{*}$ in general, as mentioned in Sec.~\ref{sec:IIIC}. This is also true for balanced loss-gain systems,  and these inconspicuous imaginary parts would affect the condition for finding a perfect EP, e.g., as  in Ref.~\cite{chen_parity-time-symmetric_2018}.



Note, with a direct QNM eigenfrequency solver, there would be four new eigenfrequencies for coupled resonators, because there are two degenerate standing modes per resonator (see Appendix~\ref{sec:degenerate_WGMs}); thus, there are four black circles for each gap distance in Fig.~\ref{QNM_distribution} (e) and (f), but with two pairs of degenerate modes. In contrast, with QNM dipole technique (see Appendix~\ref{sec:numerics}), only one of the degenerate standing modes are used. Hence,  when combining with the analytical CMT approach,  there are only two new eigenfrequencies for coupled resonators -- labelled by the red star and the green star for each gap distance in Fig.~\ref{QNM_distribution}(e) and (f).

Equation~\eqref{eq:QNMs_pm} gives the two new coupled QNMs $\tilde{\mathbf{f}}^{\pm}$ (hybrid QNMs) corresponding to the two new eigenfrequencies $\tilde{\omega}_{\pm}$, where the input fields $\tilde{\mathbf{f}}_{1}$ and $\tilde{\mathbf{f}}_{2}$ are now $\tilde{\mathbf{f}}^{\rm L}$ and $\tilde{\mathbf{f}}^{\rm G}$.
The spatial profile of the coupled QNMs (real part ${\rm Re}[\tilde{f}^{\pm}_{z}]$) for $d_{\rm gap}=1155~$nm are shown in Figs.~\ref{QNM_distribution} (c) and (d), where
the fields now extend over both resonators.
Below  we study two example gap cases close to the lossy EP,
namely $d_{\rm gap}=1155~$nm and $d_{\rm gap}=1160~$nm, indicated by the arrows  in Figs.~\ref{QNM_distribution}(e) and (f).

\subsection{Non-Lorentzian Purcell factors close to a lossy exceptional point: Quasinormal mode Green function solution versus full dipole simulations}

Next, we focus on the Purcell factors close to the lossy EPs in the coupled loss-gain resonators. 
Once the coupled QNMs (Eq.~\eqref{eq:QNMs_pm}) are obtained from CMT (the input fields $\tilde{\mathbf{f}}_{1}$ and $\tilde{\mathbf{f}}_{2}$ are now $\tilde{\mathbf{f}}^{\rm L}$ and $\tilde{\mathbf{f}}^{\rm G}$), the generalized Purcell factors are obtained analytically from Eq.~\eqref{QNMpurcell}, using the QNM Green function (Eq.~\eqref{eq:Gsol2}).
As shown in Fig.~\ref{disk_sche}, we 
consider five potential dipole positions along 
the $x$-axis, and study the Purcell factors as a function of frequency in each case.

\begin{figure}[t]
    \centering
    \includegraphics[width=0.99\columnwidth]{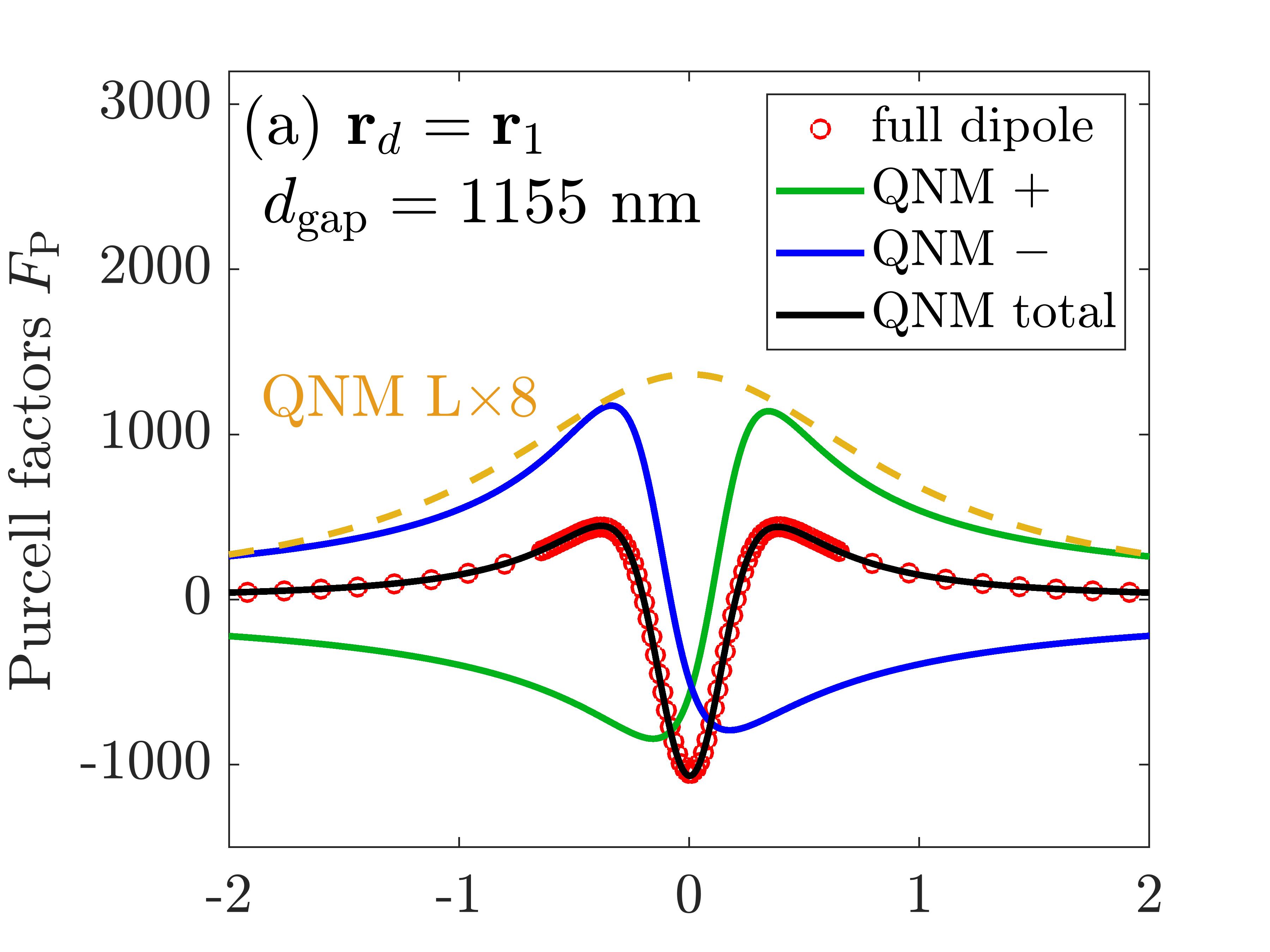}
    \includegraphics[width=0.99\columnwidth]{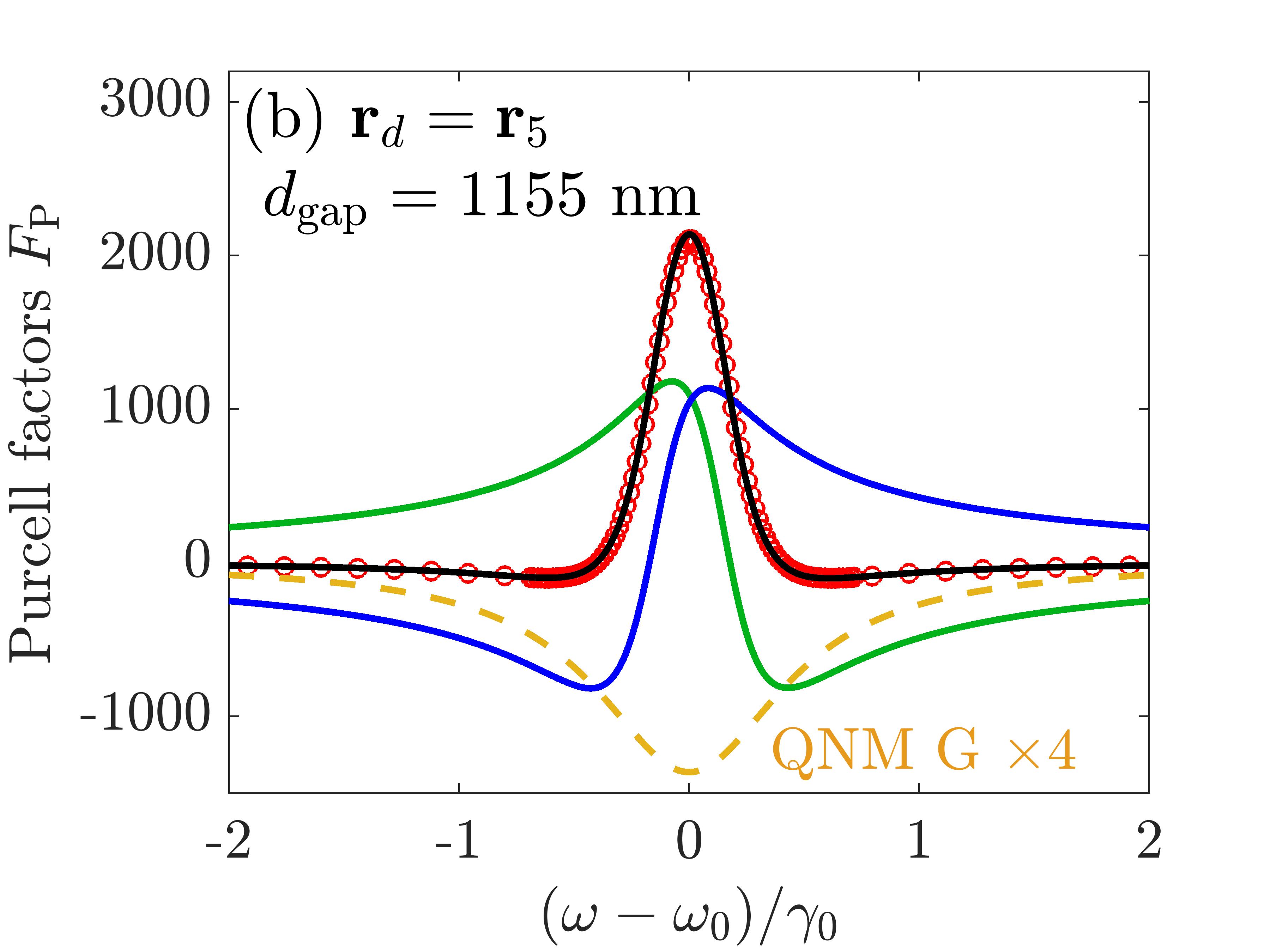}
    \caption{(a) Purcell factors for the coupled resonators with gap distance $d_{\rm gap}=1155~$nm, showing the analytical CMT solutions ((Eq.~\eqref{eq:Gsol2} and Eq.~\eqref{QNMpurcell} or Eq.~\eqref{eq:2Dpurcell}))
    for the hybrid QNM contributions as well as their sum. Also shown is the full dipole simulation with no approximations (red marker, Eq.~\eqref{Purcellfulldipole}), and the QNM contribution from a single lossy resonator (multiplied by a constant for clarity). The dipole is placed at $\mathbf{r}_{\rm d}=\mathbf{r}_{1}$ ($10~$ nm away from the lossy resonator).
    (b) Similar to (a), but with a different dipole position $\mathbf{r}_{\rm d}=\mathbf{r}_{5}$, $10~$nm away from the gain resonator. 
}\label{fig4}
\end{figure}
\textbf{\begin{figure}
    \centering
    \includegraphics[width=0.99\columnwidth]{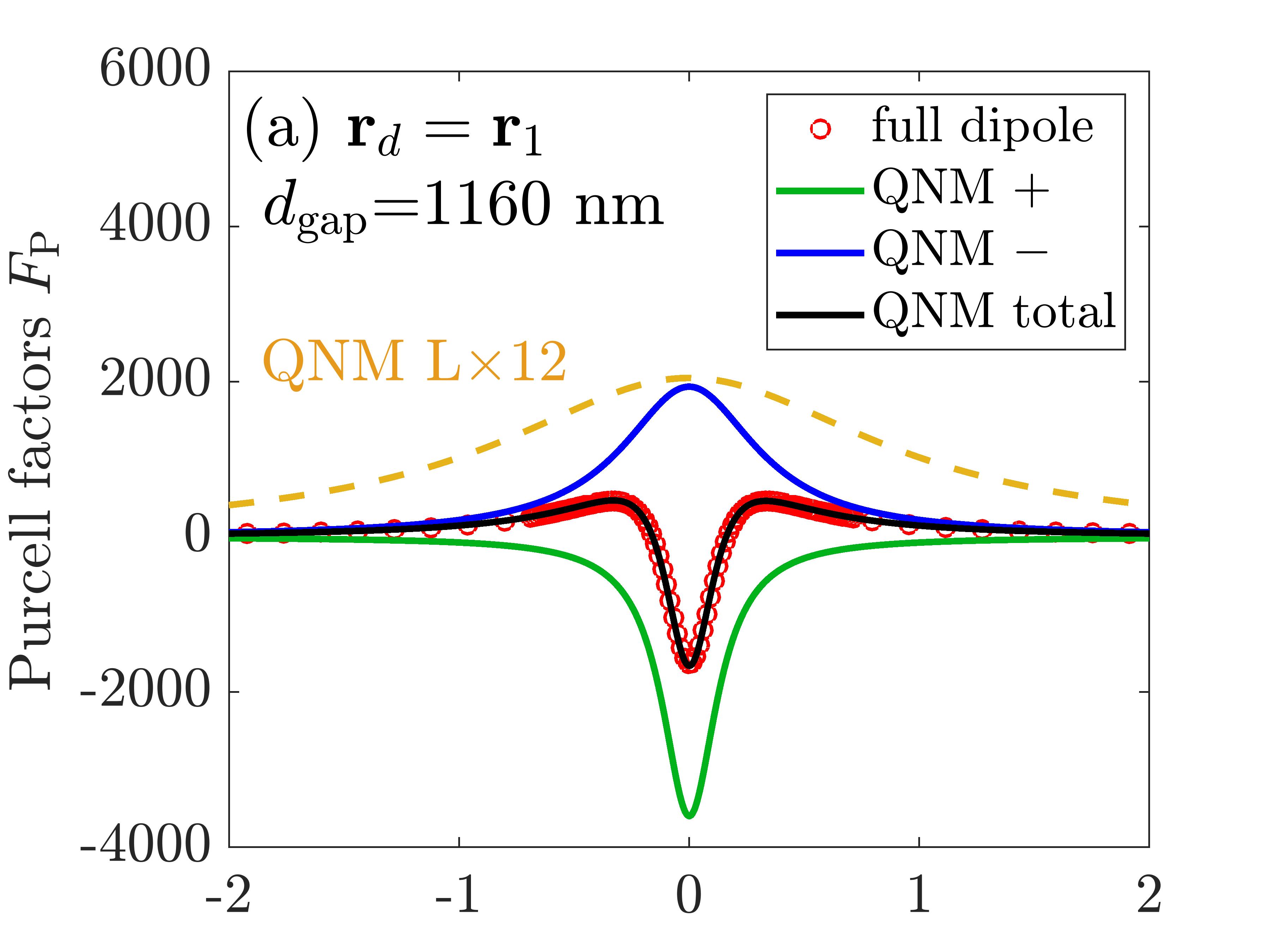}
    \includegraphics[width=0.99\columnwidth]{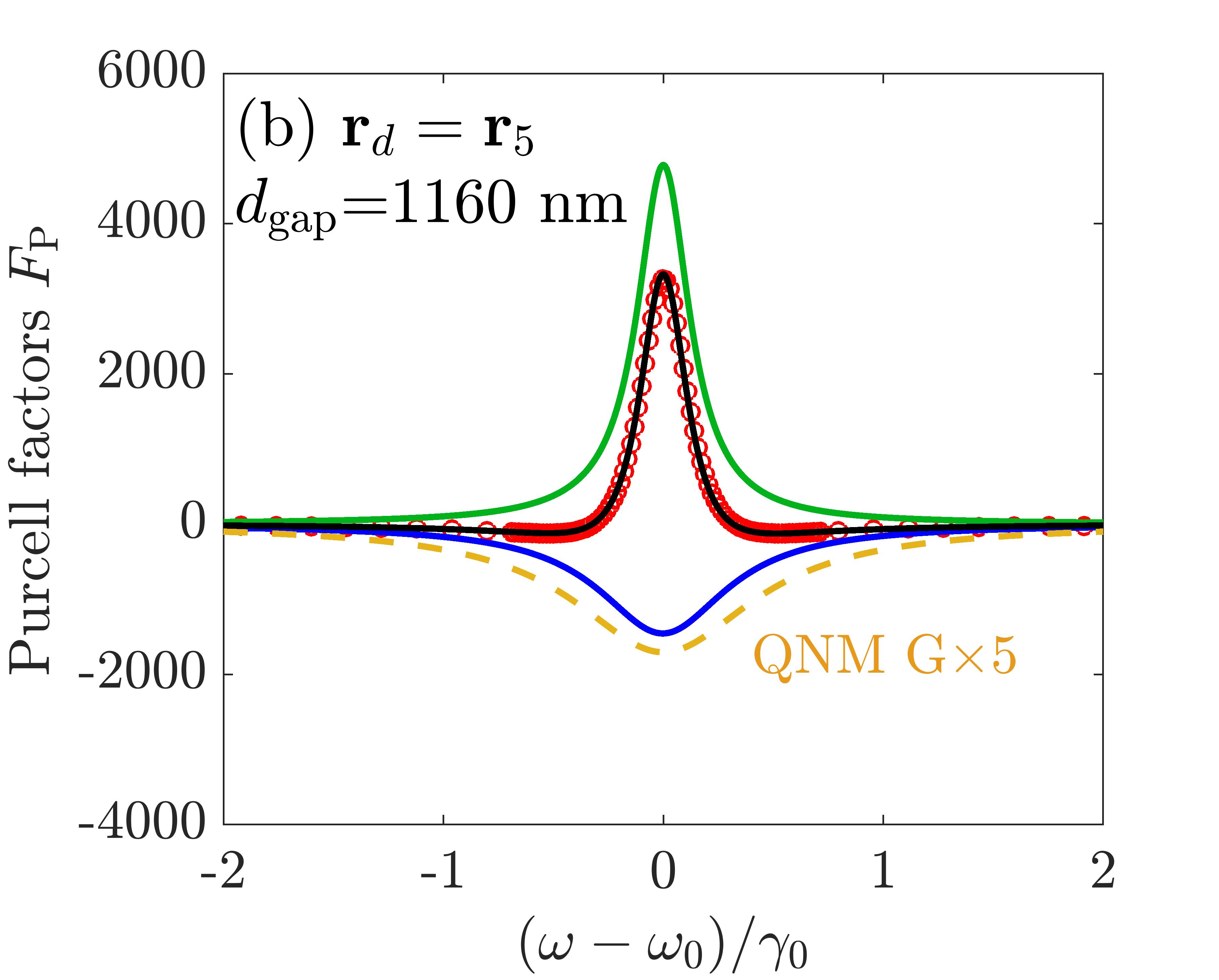}
    \caption{Similar to Fig.~\ref{fig4}, but with a different gap distance $d_{\rm gap}=1160~$nm. 
 }\label{fig4_prime}
\end{figure}}

For a dipole at the position  $\mathbf{r}_{\rm d}=\mathbf{r}_{1}$ ($10~$nm away from the lossy cavity), the Purcell factors are shown in Fig.~\ref{fig4}(a) with $d_{\rm gap}=1155~$nm, where  negative Purcell factors (black solid curve) are found in a wide range of frequencies.
The black solid line shows the analytical QNM Green function result using only the bare resonator parameters as input, and the red circles show full dipole solutions, showing quantitatively good agreement over all frequencies. We stress there are no fitting parameters in the QNM solutions. 
Moreover, from our theory, the contribution from
 the hybrid QNMs $\tilde{f}^{+}$ and $\tilde{f}^{-}$ can be shown separately, as indicated by the green and blue solid curves (Eq.~\eqref{eq:Gsol2}, the first and second term). 
As a reference, we also show the Purcell factors for a single lossy cavity (orange dashed curve), which is net positive and multiplied by $8$ for a better graphical comparison. The gain resonator clearly acts to suppress the
broadening and enhance the overall Purcell factors.

\begin{table*}[htb]
\caption {Phase contributions in the imaginary part of QNM Green function, which is used to explain the lineshapes of the corresponding Purcell factors shown in Fig.~\ref{fig4} and Fig.~\ref{fig4_prime}. 
} \label{phase} 
    \centering
    \begin{tabular}{|c|c|c|c|c|}
 \hline
   & \multicolumn{2}{|c|}{$\mathbf{r}_{\rm d}=\mathbf{r}_{1}$} & \multicolumn{2}{|c|}{$\mathbf{r}_{\rm d}=\mathbf{r}_{5}$} \\
 \hline
QNM L  & \multicolumn{2}{|c|}{$\cos{(2\phi^{\rm L})}=1.0000$,~$\sin{(2\phi^{\rm L})}=-0.0012$} & \multicolumn{2}{|c|}{}   \\
 \hline
QNM G  & \multicolumn{2}{|c|}{} & \multicolumn{2}{|c|}{$\cos{(2\phi^{\rm G})}=1.0000$,~$\sin{(2\phi^{\rm G})}=0.0005$}   \\
\hline
\multirow{2}{7em}{$d_{\rm gap}=1155~$nm}  & \multicolumn{2}{|c|}{$\cos{(2\phi^{+})}=0.1514$,~$\sin{(2\phi^{+})}=-0.9885$} & \multicolumn{2}{|c|}{$\cos{(2\phi^{+})}=0.1834$,~$\sin{(2\phi^{+})}=0.9830$}   \\
& \multicolumn{2}{|c|}{$\cos{(2\phi^{-})}=0.1939$,~$\sin{(2\phi^{-})}=0.9810$} & \multicolumn{2}{|c|}{$\cos{(2\phi^{-})}=0.1620$,~$\sin{(2\phi^{-})}=-0.9868$}\\
 \hline
\multirow{2}{7em}{$d_{\rm gap}=1160~$nm}  & \multicolumn{2}{|c|}{$\cos{(2\phi^{+})}=-1.0000$,~$\sin{(2\phi^{+})}=0.0009$} & \multicolumn{2}{|c|}{$\cos{(2\phi^{+})}=0.9999$,~$\sin{(2\phi^{+})}=0.0101$}   \\
& \multicolumn{2}{|c|}{$\cos{(2\phi^{-})}=1.0000$,~$\sin{(2\phi^{-})}=-0.0011$} & \multicolumn{2}{|c|}{$\cos{(2\phi^{-})}=-0.9999$,~$\sin{(2\phi^{-})}=-0.0129$}\\
 \hline
 \end{tabular}
\end{table*}

To help explain these unusual lineshapes, we can study the contributions to the QNM Green functions from the phases for the two hybrid QNMs.
Defining $\tilde{f}^{+}_{z}(\mathbf{r})=|\tilde{f}^{+}_{z}(\mathbf{r})|e^{i\phi^{+}(\mathbf{r})}$ and $\tilde{f}^{-}_{z}(\mathbf{r})=|\tilde{f}^{-}_{z}(\mathbf{r})|e^{i\phi^{-}(\mathbf{r})}$,  the QNM Green function
can be  expressed as
\begin{align}
&{G}_{zz}(\mathbf{r}_{\rm d},\mathbf{r}_{\rm d},\omega)= \nonumber 
\\
&\ =A^{+}(\omega)\tilde{f}^{+}_{z}(\mathbf{r}_{\rm d})\tilde{f}^{+}_{z}(\mathbf{r}_{\rm d})+A^{-}(\omega)\tilde{f}^{-}_{z}(\mathbf{r}_{\rm d})\tilde{f}^{-}_{z}(\mathbf{r}_{\rm d})\nonumber \\
&\ =A^{+}(\omega)e^{i2\phi^{+}(\mathbf{r}_{\rm d})}|\tilde{f}^{+}_{z}(\mathbf{r}_{\rm d})|^2
+A^{-}(\omega)e^{i2\phi^{-}(\mathbf{r}_{\rm d})}|\tilde{f}^{-}_{z}(\mathbf{r}_{\rm d})|^2,
\end{align}
from which we extract the imaginary part for use in Purcell's
formula~\cite{el-sayed_quasinormal-mode_2020}:
\begin{align}\label{purcellphase}
\begin{split}
&{\rm Im}[{G}_{zz}(\mathbf{r}_{\rm d},\mathbf{r}_{\rm d},\omega)]\\
=&\bigg[\cos{2\phi^{+}(\mathbf{r}_{\rm d})}+\frac{\omega_{+}-\omega}{\gamma_{+}}\sin{2\phi^{+}(\mathbf{r}_{\rm d})}\bigg]\Big|\tilde{f}^{+}_{z}(\mathbf{r}_{\rm d})\Big|^2 L^{+}(\omega)\\
+&\bigg[\cos{2\phi^{-}(\mathbf{r}_{\rm d})}+\frac{\omega_{-}-\omega}{\gamma_{-}}\sin{2\phi^{-}(\mathbf{r}_{\rm d})}\bigg]\Big|\tilde{f}^{-}_{z}(\mathbf{r}_{\rm d})\Big|^2 L^{-}(\omega),
\end{split}
\end{align}
where we introduced the normalized Lorentzian lineshapes,
\begin{align}
L^{+}(\omega)&=\frac{\omega}{2}\frac{\gamma_{+}}{(\omega_{+}-\omega)^2+\gamma_{+}^{2}}, \nonumber \\
L^{-}(\omega)&=\frac{\omega}{2}\frac{\gamma_{-}}{(\omega_{-}-\omega)^2+\gamma_{-}^{2}},
\end{align}
with
$\tilde{\omega}_{+}=\omega_{+}-i\gamma_{+}$,  $\tilde{\omega}_{-}=\omega_{-}-i\gamma_{-}$, $A^{+}(\omega)=\omega/[2(\tilde{\omega}_{+}-\omega)]$, and $A^{-}(\omega)=\omega/[2(\tilde{\omega}_{-}-\omega]$.  
Similar expressions have  been used to explain Fano resonances formed
by elastic QNMs in coupled cavity beams~\cite{el-sayed_quasinormal-mode_2020}.

For comparison, the Green functions for the single-mode bare QNMs are also given as 
\begin{align}
\begin{split}\label{Eq:G_loss_phase}
&{\rm Im}[{G}_{zz}^{\rm L}(\mathbf{r}_{\rm d},\mathbf{r}_{\rm d},\omega)]\\
=&\bigg[\cos{2\phi^{\rm L}(\mathbf{r}_{\rm d})}+\frac{\omega_{1}-\omega}{\gamma_{1}}\sin{2\phi^{\rm L}(\mathbf{r}_{\rm d})}\bigg]\Big|\tilde{f}^{\rm L}_{z}(\mathbf{r}_{\rm d})\Big|^2 L^{\rm L}(\omega),
\end{split}
\end{align}
\begin{align}
\begin{split}\label{Eq:G_gain_phase}
&{\rm Im}[{G}_{zz}^{\rm G}(\mathbf{r}_{\rm d},\mathbf{r}_{\rm d},\omega)]\\
=&\bigg[\cos{2\phi^{\rm G}(\mathbf{r}_{\rm d})}+\frac{\omega_{2}-\omega}{\gamma_{2}}\sin{2\phi^{\rm G}(\mathbf{r}_{\rm d})}\bigg]\Big|\tilde{f}^{\rm G}_{z}(\mathbf{r}_{\rm d})\Big|^2 L^{\rm G}(\omega),
\end{split}
\end{align}
with normalized Lorentzian lineshapes
for the loss and gain resonator modes,
and we have redefined the QNMs as
 $\tilde{f}^{\rm L}_{z}(\mathbf{r})=|\tilde{f}^{\rm L}_{z}(\mathbf{r})|e^{i\phi^{\rm L}(\mathbf{r})}$ and $\tilde{f}^{\rm G}_{z}(\mathbf{r})=|\tilde{f}^{\rm G}_{z}(\mathbf{r})|e^{i\phi^{\rm G}(\mathbf{r})}$.


In Fig.~\ref{fig4}(a), we show the
total Purcell factors for $d_{\rm gap}=1155~$nm,
as well as the hybrid mode contributions,
and the lossy mode result on its own for comparison.
First, the single mode case (orange dashed curve, single lossy resonator)
shows a typical Lorentzian peak, which can be explained by the QNM phase contributions: $\cos{[2\phi^{\rm L}(\mathbf{r}_{1})]}=0.999999$ and $\sin{[2\phi^{\rm L}(\mathbf{r}_{1})]}=-0.001244$ (also shown in Table ~\ref{phase}).
Applying these to Eq.~\eqref{Eq:G_loss_phase}  results in a typical Lorentzian lineshape.

Next we focus on the hybrid modes (coupled resonator case), where now: $\cos{[2\phi^{+}(\mathbf{r}_{1})]}=0.1514$, $\sin{[2\phi^{+}(\mathbf{r}_{1})]}=-0.9885$, and $\cos{[2\phi^{-}(\mathbf{r}_{1})]}=0.1939$, $\sin{[2\phi^{-}(\mathbf{r}_{1})]}=0.9810$ (also shown in Table~\ref{phase}), which explain the non-Lorentzian lineshapes for the separate contributions (two terms in Eq.~\eqref{purcellphase}, green solid curve and blue solid curve shown in Fig.~\ref{fig4} (a)). Then combining the weights for each term (from $|\tilde{f}^{+}_{z}(\mathbf{r}_{1})|^2$ and $|\tilde{f}^{-}_{z}(\mathbf{r}_{1})|^2$ as shown in Eq.~\eqref{purcellphase}), we obtain negative Purcell factors in a wide range of frequencies (black solid curve in Fig.~\ref{fig4} (a)).

Furthermore, these  QNM phases will result in 
position-dependent lineshapes for the Purcell factors.
Thus when we change the dipole position from $\mathbf{r}_{\rm d}=\mathbf{r}_{1}$ to $\mathbf{r}_{\rm d}=\mathbf{r}_{5}$ ($10~$nm away from the gain cavity),  the Purcell factor
lineshapes can change significantly, as shown in Fig.~\ref{fig4} (b) (again for $d_{\rm gap}=1155~$nm), where the total contribution (black solid curve) are found to be mainly net positive in a wide range of frequency, and also show excellent agreement with the full dipole method (red circles). Once again, the contribution from $\tilde{f}^{+}$ and $\tilde{f}^{-}$ can be given separately, as shown with the green and blue solid curves. In addition, the Purcell factors for the single gain cavity is shown as an orange dashed curve, which is net negative (and multiplied by $4$ for a better graphical comparison). 

As  before,  these lineshapes can also be well explained from the QNM phase terms.
For a single gain cavity, one finds that $\cos{[2\phi^{\rm G}(\mathbf{r}_{5})]}=0.999999$ and $\sin{[2\phi^{\rm G}(\mathbf{r}_{5})]}= 0.000527$ (also shown in Table ~\ref{phase}). However, with $\gamma_{\rm G}<0$, then $L^{\rm G}(\omega)$ shows a negative Lorentzian lineshape, yielding negative Purcell factors with the gain cavity only (orange dashed curve in Fig.~\ref{fig4}(b)). As for the coupled system, one can find $\cos{[2\phi^{+}(\mathbf{r}_{5})]}=0.1834$, $\sin{[2\phi^{+}(\mathbf{r}_{5})]}=0.9830$, and $\cos{[2\phi^{-}(\mathbf{r}_{5})]}=0.1620$, $\sin{[2\phi^{-}(\mathbf{r}_{5})]}=-0.9868$ (also shown in Table ~\ref{phase}), which explain the non-Lorentzian lineshapes for separate contribution (two terms in Eq.~\eqref{purcellphase}, green solid curve and blue solid curve shown in Fig.~\ref{fig4} (b)). Then combining the weights for each term (from $|\tilde{f}^{+}_{z}(\mathbf{r}_{5})|^2$ and $|\tilde{f}^{-}_{z}(\mathbf{r}_{5})|^2$ in Eq.~\eqref{purcellphase}), the mostly positive Purcell factors are obtained (black solid curve in Fig.~\ref{fig4}(b)).

Similar to case with a gap distance $d_{\rm gap}=1155~$nm, the Purcell factors with $d_{\rm gap}=1160~$nm for a dipole at $\mathbf{r}_{1}$ are shown in Fig.~\ref{fig4_prime}(a).
Their separate contributions from ${\rm QNM}~+$ (green solid curve) and ${\rm QNM}~-$ (blue solid curve) show Lorentzian lineshapes, but with different linewidth since $\gamma_{+}\neq\gamma_{-}$ (see Fig.~\ref{QNM_distribution} (f)) and different signs due to $\cos{[2\phi^{+}(\mathbf{r}_{1})]}=-1.0000$ and $\cos{[2\phi^{-}(\mathbf{r}_{1})]}=1.0000$.
Combing these contributions, the total contribution are found to be net negative in a wide range of frequencies.
When the dipole is at $\mathbf{r}_{5}$, the Purcell factors are shown in Fig.~\ref{fig4_prime} (b), where the Lorentzian lineshapes for separate contributions could also be explained by the QNM phases $\cos{[2\phi^{+}(\mathbf{r}_{5})]}=0.9999$ and $\cos{[2\phi^{-}(\mathbf{r}_{5})]}=-0.9999$. Considering their separate contributions, the total Purcell factors are now mostly positive. 
More details about the QNM phases are shown in Table~\ref{phase}.
\begin{figure}[ht]
    \centering
    \includegraphics[width=0.99\columnwidth]{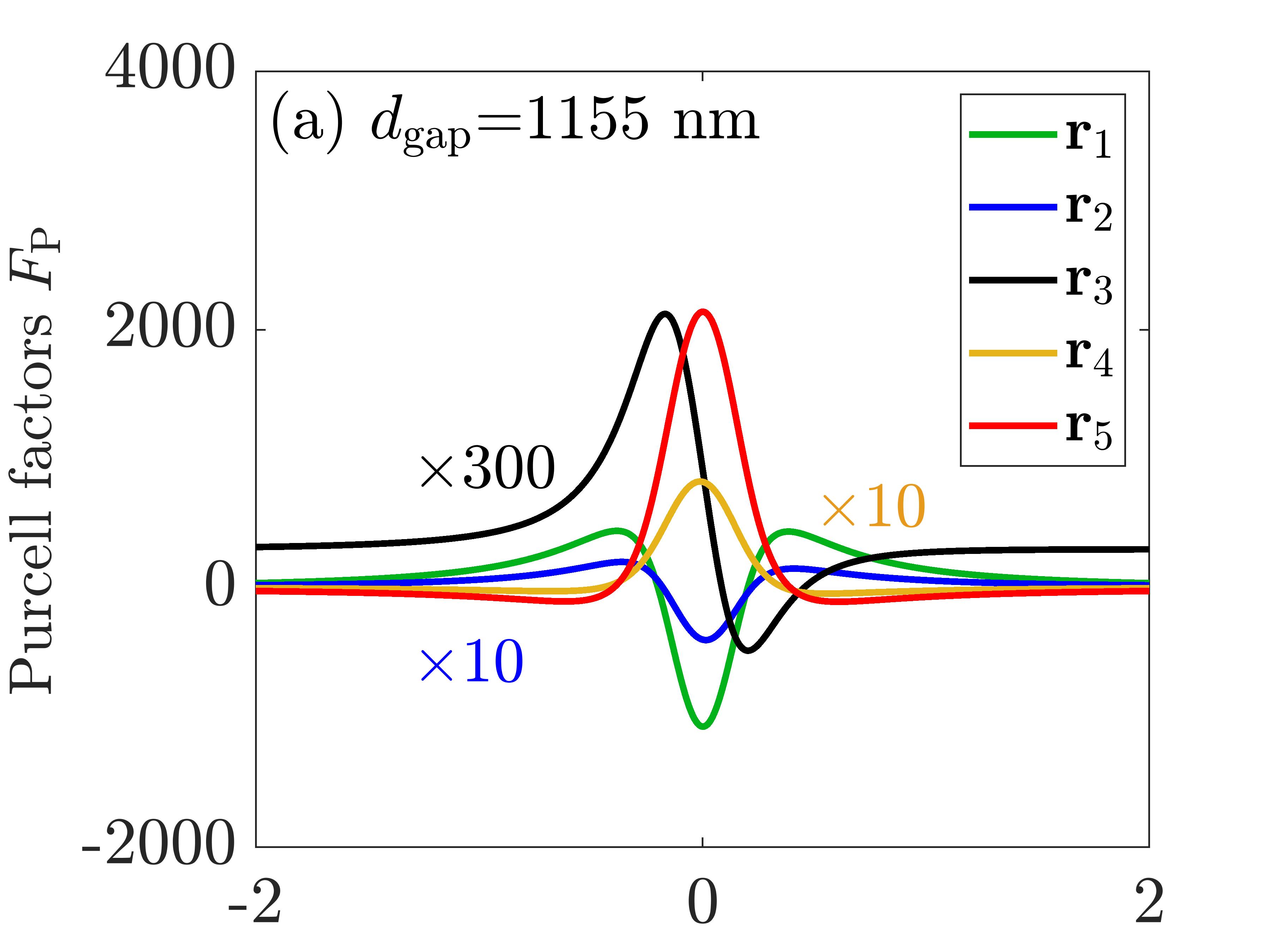}
    \includegraphics[width=0.99\columnwidth]{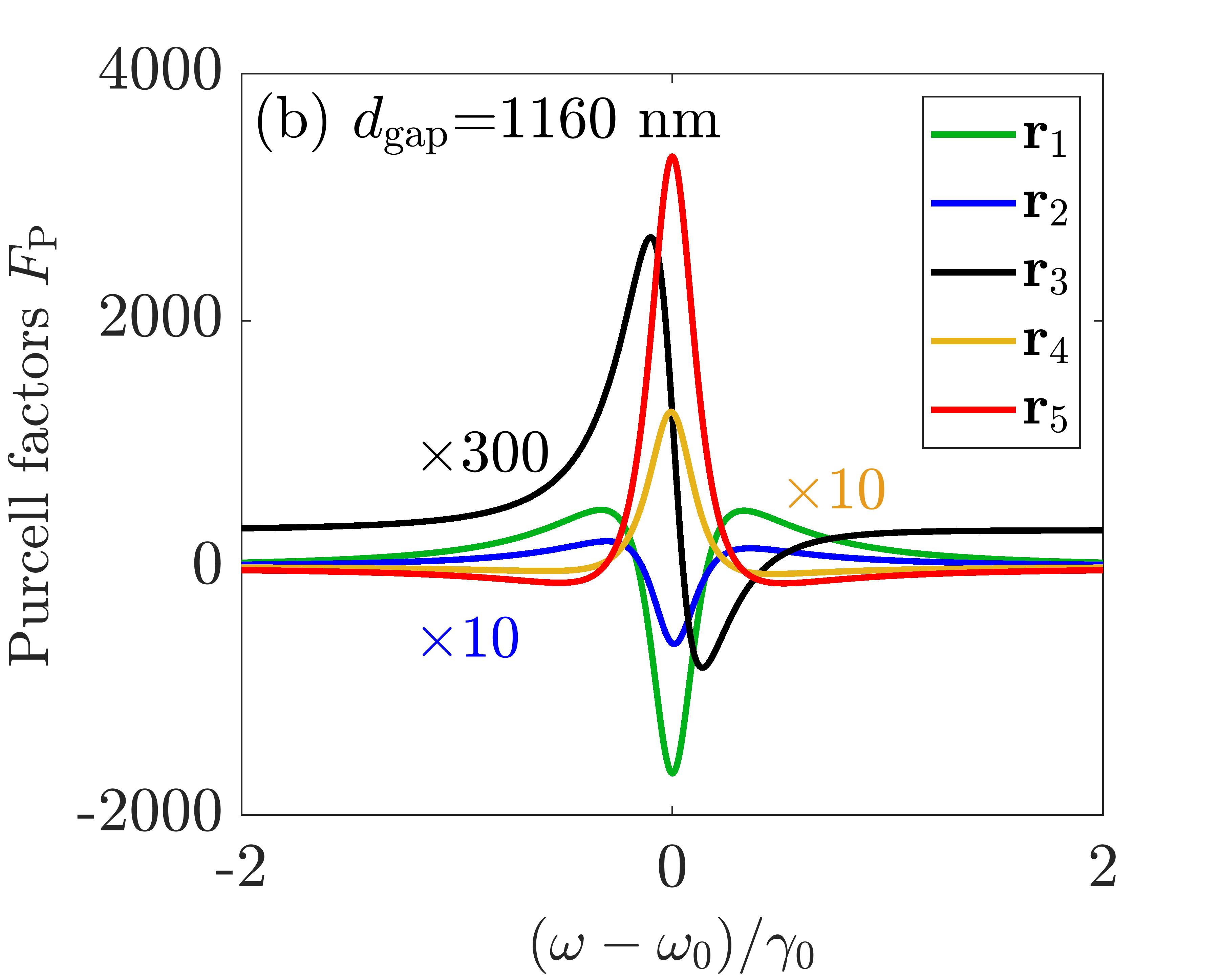}
    \caption{Purcell factors at different dipole positions $\mathbf{r}_{\rm d}=\mathbf{r}_{1}\sim{\bf r}_5$ (shown in Fig.~\ref{disk_sche}), for two different gap separations (a) $d_{\rm gap}=1155~$nm, and (b) $d_{\rm gap}=1160~$nm. Full dipole simulations are not shown for clarity, but yield the same quantitatively good fit as in the previous graphs. } \label{fig5}
\end{figure}

These  QNM phases have a significant effect on the spectral lineshapes of the Purcell factors, which can be further seen  in Fig.~\ref{fig5}(a) (for $d_{\rm gap}=1155~$nm) and (b) (for $d_{\rm gap}=1160~$nm), where the total Purcell factors are given for  five different dipole positions as shown in Fig.~\ref{disk_sche}.
For better comparison, the Purcell factors at $\mathbf{r}_{2}$, $\mathbf{r}_{3}$ and $\mathbf{r}_{4}$ are multiplied by $10$, $300$ and $10$. Note, we do not show the full dipole results here, as they are basically indistinguishable from the the two QNM solutions, similar to the previous comparisons.

We stress that these negative Purcell factors are unphysical because the classical Purcell factor formulas, and Fermi's golden rule, is no longer working with a gain medium~\cite{EPQuantumPaper}, though the local density of states (LDOS, proportional to ${\rm Im}[\mathbf{G}(\mathbf{r}_{\rm d},\mathbf{r}_{\rm d},\omega)]$) is correct. Reference ~\onlinecite{EPQuantumPaper} discusses the breakdown of classical formulas in more detail,
and shows that the correct SE rates must be described fully quantum mechanically, in which case the 
quantum  Purcell factors (modified SE decay rayes) are indeed net positive.

\subsection{Green function Propagators}

   \begin{figure}[t!]
     \centering
     \includegraphics[width=0.99\columnwidth]{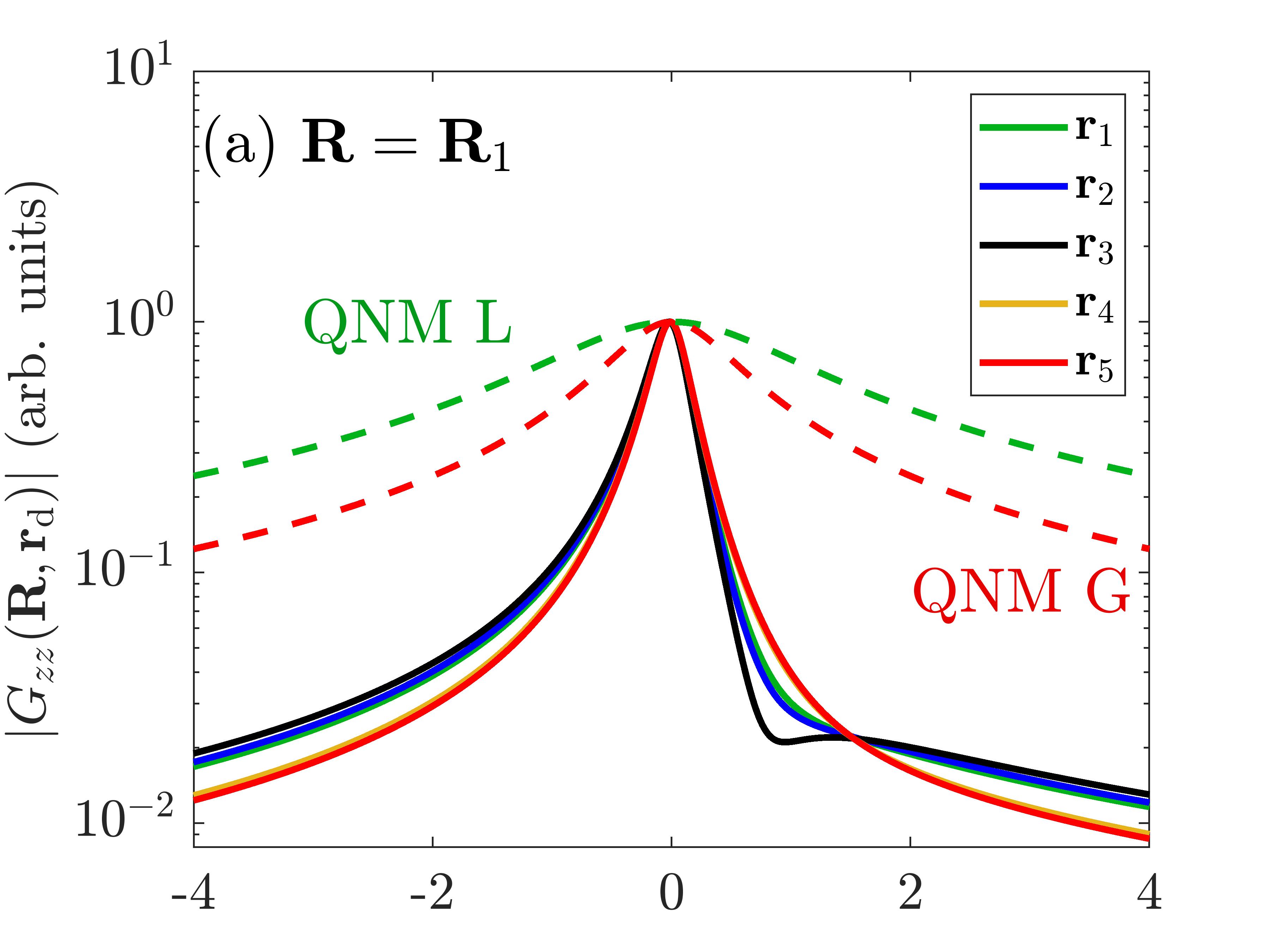}
     \includegraphics[width=0.99\columnwidth]{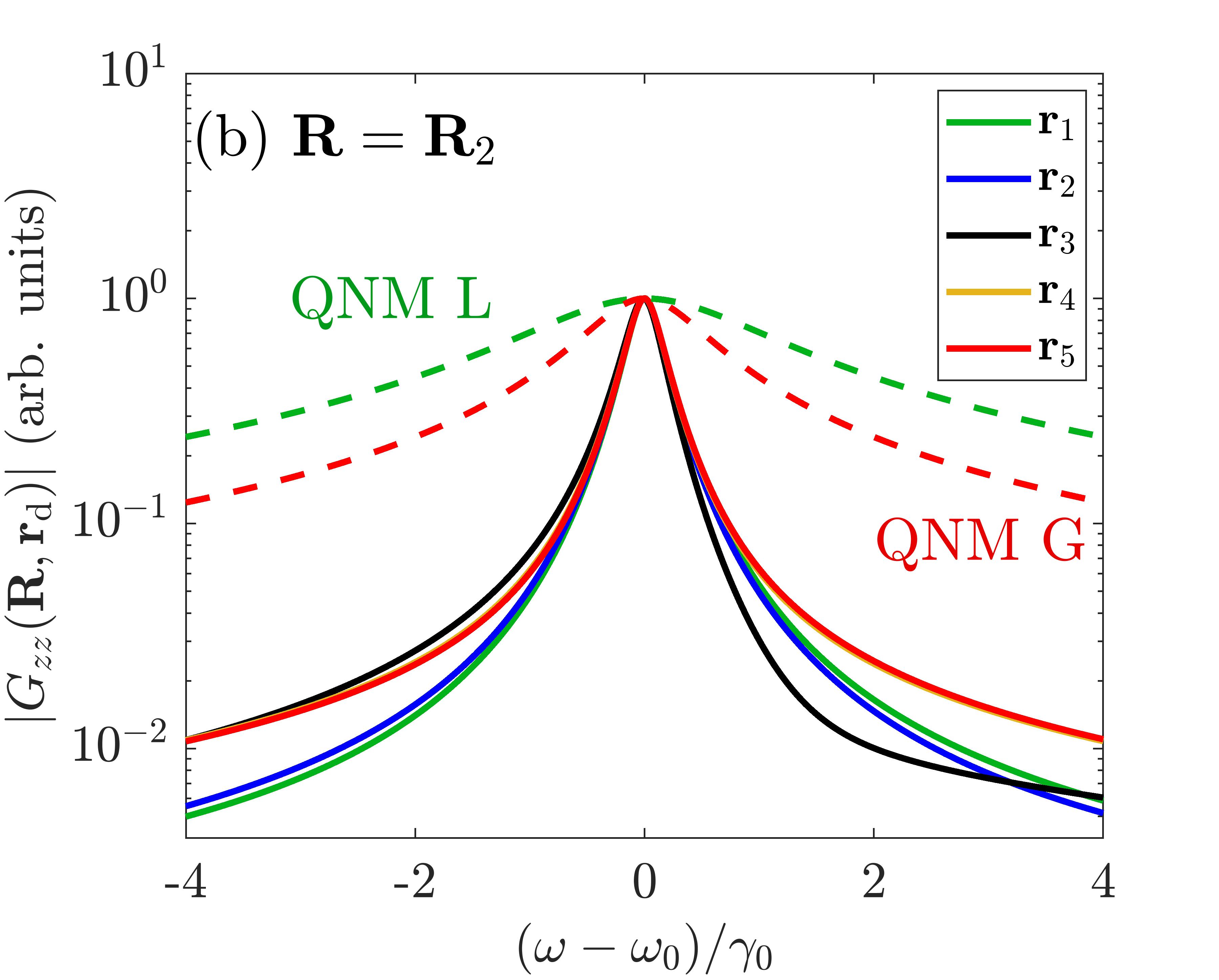}
     \caption{Green function propagator for the far field detected point is at (a) $\mathbf{R}=\mathbf{R}_{1}=$[0,10]~$\mu$m, and (b) $\mathbf{R}=\mathbf{R}_{2}=[14.09,0]~\mu$m (which is $3.51~\mu$m away from the gain cavity) with $d_{\rm gap}=1160~$nm. For QNM L (loss), the dipole is at $\mathbf{r}_{\rm d}=\mathbf{r}_{1}$, and for QNM G (gain), it is at $\mathbf{r}_{\rm d}=\mathbf{r}_{5}$. The results for $d_{\rm gap}=1155~$nm look similar and we do not show them.  }\label{fig5b_prime_semilog}
 \end{figure}

\begin{table}[htb]
\caption {Phase contributions from the QNMs in the propagator (Eq.~\eqref{eq:prop}). 
} \label{phasepropagator} 
    \centering
    \begin{tabular}{|c|c|c|}
 \hline
   & \multicolumn{1}{|c|}{$\mathbf{R}=\mathbf{R}_{1}$} & \multicolumn{1}{|c|}{$\mathbf{R}=\mathbf{R}_{2}$} \\
 \hline
QNM L  & \multicolumn{1}{|c|}{$\cos{[\phi^{\rm L}(\mathbf{R})+\phi^{\rm L}(\mathbf{r}_{\rm d})]}=-0.6905$} &\multicolumn{1}{|c|}{$-0.9989$}  \\
$\mathbf{r}_{\rm d}=\mathbf{r}_{1}$ &~$\sin{[\phi^{\rm L}(\mathbf{R})+\phi^{\rm L}(\mathbf{r}_{\rm d})]}=0.7234$ & $-0.0470$\\
 \hline
 QNM G  & \multicolumn{1}{|c|}{$\cos{[\phi^{\rm G}(\mathbf{R})+\phi^{\rm G}(\mathbf{r}_{\rm d})]}=-0.7036$} &\multicolumn{1}{|c|}{$0.2606$}  \\
$\mathbf{r}_{\rm d}=\mathbf{r}_{1}$ &~$\sin{[\phi^{\rm G}(\mathbf{R})+\phi^{\rm G}(\mathbf{r}_{\rm d})]}=0.7106$ & $-0.9655$\\
\hline
\multirow{2}{7em}{$d_{\rm gap}=1160~$nm} &$\cos{[\phi^{+}(\mathbf{R})+\phi^{+}(\mathbf{r}_{\rm d})]}=-0.0815$ & $0.9760$ \\
& $\sin{[\phi^{+}(\mathbf{R})+\phi^{+}(\mathbf{r}_{\rm d})]}=-0.9967$ & $0.2177$ \\
\multirow{2}{7em}{$\mathbf{r}_{\rm d}=\mathbf{r}_{1}$} & $\cos{[\phi^{-}(\mathbf{R})+\phi^{-}(\mathbf{r}_{\rm d})]}=-0.0575$ & $-0.9784$ \\
& $\sin{[\phi^{-}(\mathbf{R})+\phi^{-}(\mathbf{r}_{\rm d})]}=0.9983$ & $-0.2070$ \\
\hline
 \end{tabular}
\end{table}

This next section presents example Green function propagator results, namely $|\mathbf{G}(\mathbf{R},\mathbf{r}_{\rm d},\omega)|$, which can be useful to  relate to a range of experimental observables. 
For example, this function is required to model
the spectrum at a point
detector located at ${\bf R}$, emitted
from a dipole at ${\bf r}_{\rm d}$. Unlike the
(classical) Purcell factor, it is a well defined quantity for use in both classical and quantum field theory, and is used frequently in the exploration of light-matter coupling regimes
 \cite{PhysRevA.70.053823,PhysRevB.85.075303,PhysRevB.87.205425}. For example, 
 considering an excitation  dipole,  ${\bf d}$, then the  spectrum is 
 $S({\bf R},\omega) \propto |{\bf G}({\bf R},{\bf r}_{\rm d},\omega) \cdot{\bf d}|^2$.

A far-field detection point is first chosen at $\mathbf{R}=\mathbf{R}_{1}=[0,10]~\mu$m (the origin of the coordinate system is at the gap center). The corresponding propagators $|G_{zz}(\mathbf{R},\mathbf{r}_{\rm d},\omega)|$ (arbitrary units) for various dipole positions $\mathbf{r}_{\rm d}$ are shown in Fig.~\ref{fig5b_prime_semilog}(a), for  $d_{\rm gap}=1160~$nm. For comparison, the propagator for a single lossy (gain) cavity is also shown as green (red) dashed curve, where the dipole is placed at $\mathbf{r}_{\rm d}=\mathbf{r}_{1}$ ($\mathbf{r}_{5}$).
Furthermore, for the far field detected point at  $\mathbf{R}=\mathbf{R}_{2}=[14.09,0]~\mu$m, the propagators are shown in Fig.~\ref{fig5b_prime_semilog}(b).

To better explain the physics of the
propagator lineshape, we once more write out the two QNM expanded Green function in terms of the hybrid modes, 
\begin{align}
\begin{split}
&G_{zz}(\mathbf{R},\mathbf{r}_{\rm d},\omega)\\
&=A^{+}(\omega)\tilde{f}^{+}_{z}(\mathbf{R})\tilde{f}^{+}_{z}(\mathbf{r}_{\rm d})+A^{-}(\omega)\tilde{f}^{-}_{z}(\mathbf{R})\tilde{f}^{-}_{z}(\mathbf{r}_{\rm d})\\
&=A^{+}(\omega)e^{i(\phi^{+}(\mathbf{R})+\phi^{+}(\mathbf{r}_{\rm d}))}|\tilde{f}^{+}_{z}(\mathbf{R})| |\tilde{f}^{+}_{z}(\mathbf{r}_{\rm d})|\\
&+A^{-}(\omega)e^{i(\phi^{-}(\mathbf{R})+\phi^{-}(\mathbf{r}_{\rm d}))}|\tilde{f}^{-}_{z}(\mathbf{R})| |\tilde{f}^{-}_{z}(\mathbf{r}_{\rm d})|,
\end{split}
\label{eq:prop}
\end{align}
where again we can recognize
non-Lorentzian features from the QNM phase terms, now from both the dipole position and the detector position. The detailed phases are shown in Table~\ref{phasepropagator}.
Note, the two-space point Green function has a range of uses
for describing light-matter interactions, including the description of photon transport~\cite{PhysRevA.70.053823,PhysRevB.83.075305,PhysRevB.92.205420}, and can be used to model collective effects with multiple emitters and dipoles.

\section{Quasinormal modes for index-modulated ring resonators near exceptional points}
\label{sec:final}

\begin{figure*}
    \centering
    \includegraphics[width=1.85\columnwidth]{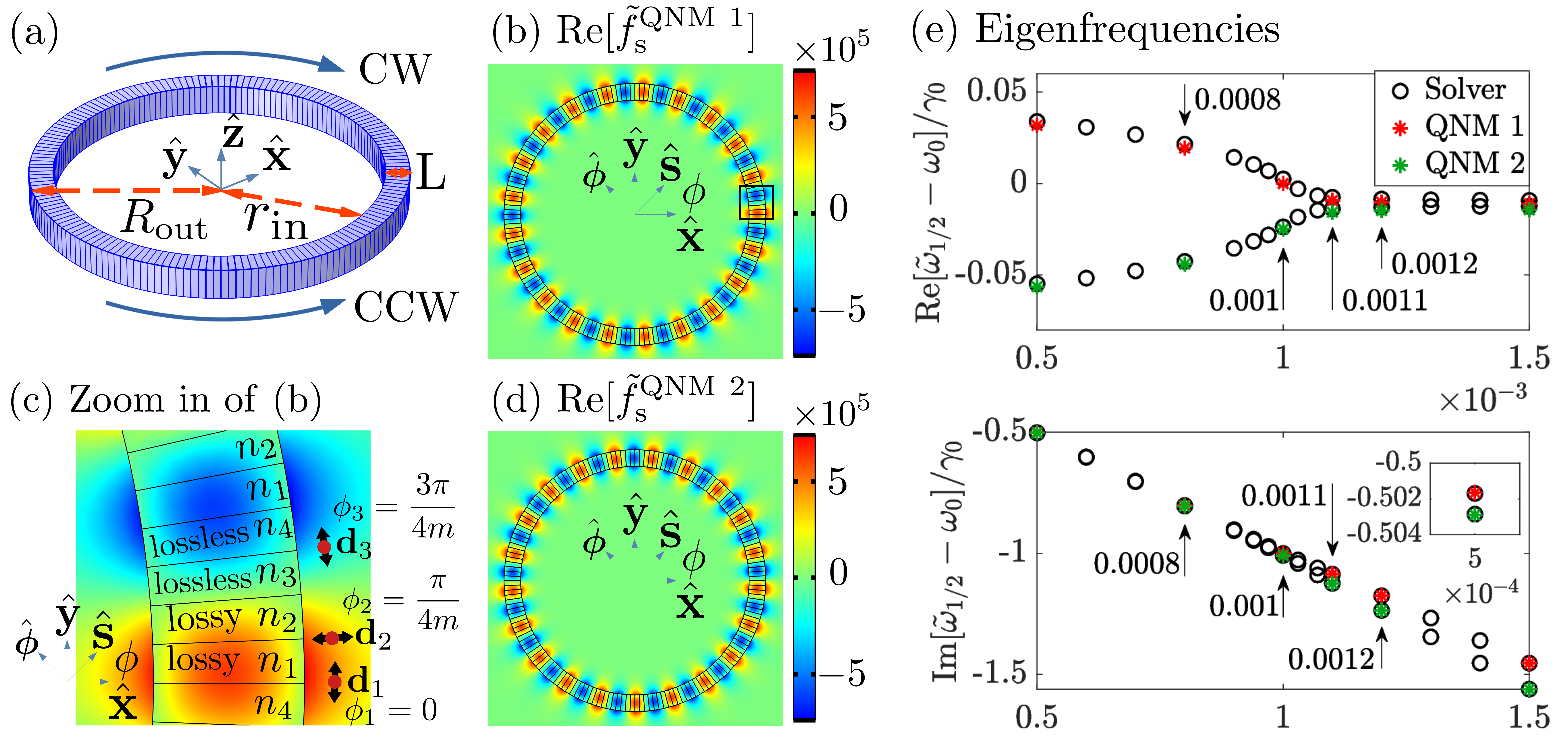}
       \caption{(a) Schematic  of the refractive-index-modulated ring resonator. The wo blue arrows label the direction of CW and CCW of the fields. The inner and outer radius (two red dashed arrows) of the ring are $r_{\rm in}=1109~$ nm and $R_{\rm out}=1271~$nm (width of the ring is ${\rm L}=R_{\rm out}-r_{\rm in}=162~$nm). In our calculations, we investigate a 2D ring resonator and couple it with an in-plane linear dipole. The ring is divided into $8m$~($m=21$) sections, where the refractive indices are $n_{1/2/3/4}$, which are distributed periodically along the CCW direction (see (c)). Field distribution ${\rm Re}[\tilde{f}_s]$ (with units $m^{-1}$) ($\tilde{\bf f}=\tilde{f}_{x}\hat{\bf x}+\tilde{f}_{y}\hat{\bf y}=\tilde{f}_{s}\hat{\bf s}+\tilde{f}_{\phi}\hat{\boldsymbol{\phi}}$) of (b) QNM~1 and (d) QNM~2 for ring structures with $\delta n_{\rm Re}=\delta n_{\rm Im}=0.001$. 
       (c) Zoom in of the black square in (b). In addition to general unit vectors $\hat{\bf x}$ and $\hat{\bf y}$, we also utilize $\hat{\bf s}=\hat{\bf x}\cos{\phi}+\hat{\bf y}\sin{\phi}$ and $\hat{\boldsymbol{\phi}}=-\hat{\bf x}\sin{\phi}+\hat{\bf y}\cos{\phi}$. Three example dipoles are shown, where one ($\mathbf{d}_{1}$) is a linear $\hat{\boldsymbol{\phi}}$ dipole at $\phi=\phi_{1}=0$ (i.e., $\hat{\bf y}$ dipole at $\phi=0$), one ($\mathbf{d}_{2}$) is a linear $\hat{\bf s}$ dipole at $\phi=\phi_{2}=\pi/4m$, and the left one ($\mathbf{d}_{3}$) is a linear $\hat{\boldsymbol{\phi}}$ dipole at $\phi=\phi_{2}=3\pi/4m$. 
       (e) Eigenfrequencies from QNMs ($\tilde{\omega}_{1}=\omega_{1}-i\gamma_{1}$ and $\tilde{\omega}_{2}=\omega_{2}-i\gamma_{2}$, red and green stars) and approximate COMSOL solver (black circles) for the ring resonator as a function of $\delta n_{\rm Im}$ when keeping $\delta n_{\rm Re}=0.001$. The insert in the bottom figure shows the zoom in region close to $\delta n_{\rm Im}=0.0005$. One will find that there are two QNMs when, e.g.,
       $\delta n_{\rm Re}=\delta n_{\rm Im}=0.001$. Specifically, we will focus on six cases (including $\delta n_{\rm Im}=0.0005,0.0015$ and four more cases that the arrows points to) and compare the results from QNMs and the full dipole simulation, where an excellent agreement is shown for the Purcell factors (cf.~Fig.~\ref{fig:2N}). Note that the eigenfrequency of QNM~1 for $\delta n_{\rm Im}=0.001$ is $\omega_{0}-i\gamma_{0}\approx2.50543552\times10^{15} - i3.768337\times10^{11}$ (${\rm rad/s}$), i.e., $\omega_{0}$ and $\gamma_{0}$ are the real part and the opposite imaginary part of this pole eigenfrequency. With the increase of $\delta n_{\rm Im}$, the decay rates  of these QNMs increase.}\label{fig:1N}
    \centering
    \includegraphics[width=0.65\columnwidth]{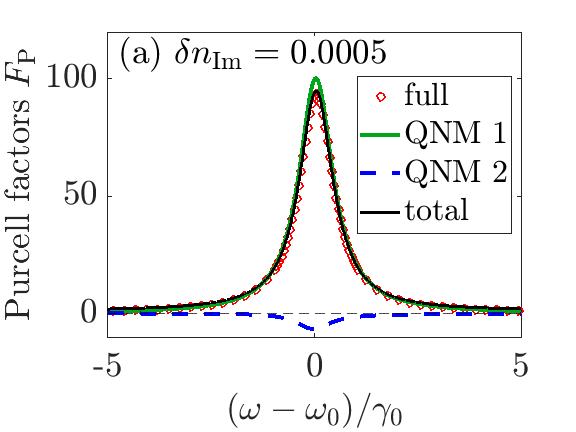}
    \includegraphics[width=0.65\columnwidth]{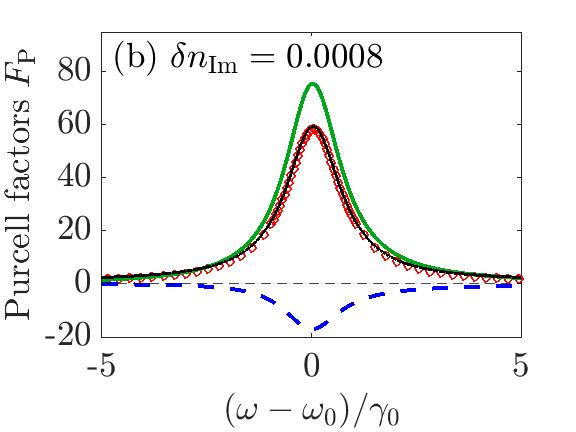}
    \includegraphics[width=0.65\columnwidth]{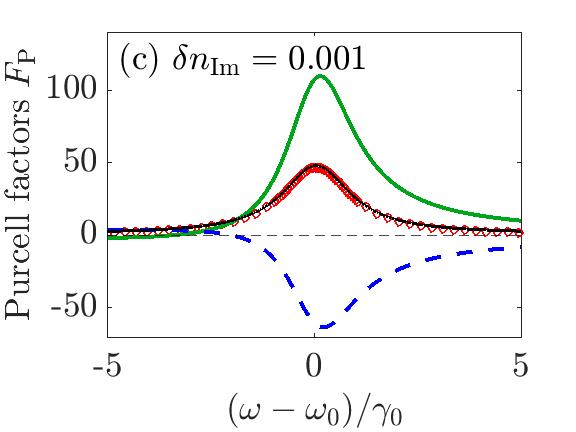}
    \includegraphics[width=0.65\columnwidth]{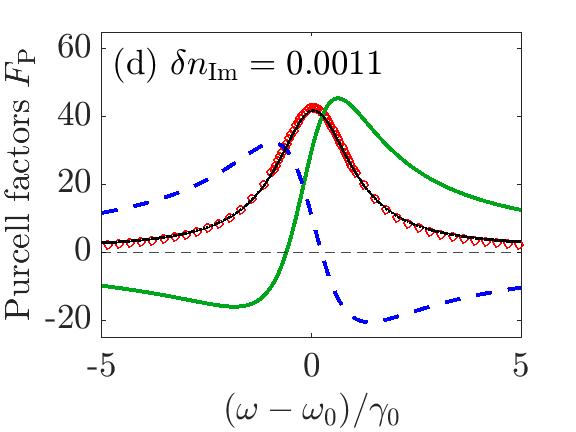}
    \includegraphics[width=0.65\columnwidth]{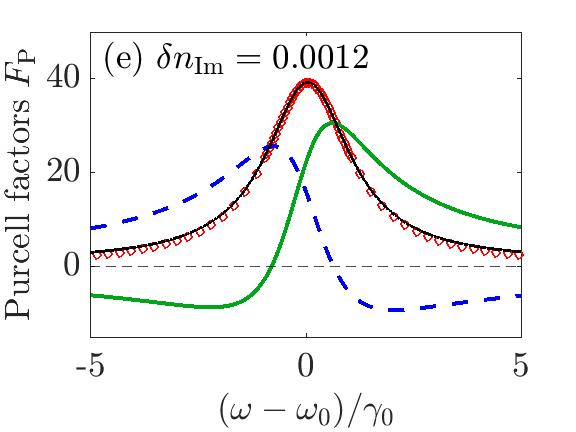}
    \includegraphics[width=0.65\columnwidth]{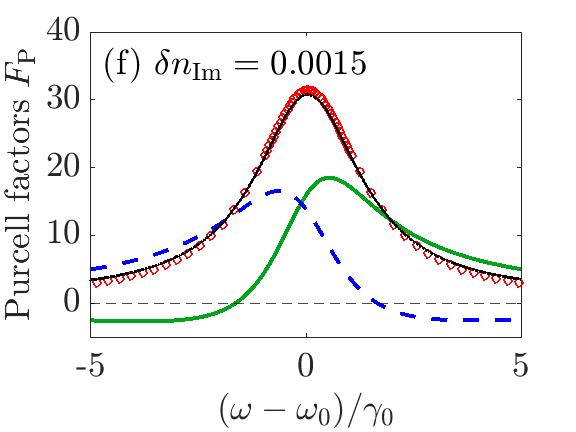}
    \caption{Purcell factors for six cases investigated with the analytical two QNM expansion, including (a) $\delta n_{\rm Im}=0.0005$, (b) $\delta n_{\rm Im}=0.0008$, (c) $\delta n_{\rm Im}=0.001$, (d) $\delta n_{\rm Im}=0.0011$, (e) $\delta n_{\rm Im}=0.0012$, and (f) $\delta n_{\rm Im}=0.0015$, while keeping $\delta n_{\rm Re}=0.001$. The Purcell factors are for a linear $\hat{\boldsymbol{\phi}}$ dipole at $\phi=0$ (i.e., a linear $\hat{\bf y}$ dipole at $\phi=0$) and $10~$nm away from the outer ring surface.
    $\omega_{0}$ and $\gamma_{0}$ are the real part and the opposite imaginary part of the pole eigenfrequency for QNM~1 with $\delta n_{\rm Im}=0.001$.
   }\label{fig:2N}
\end{figure*}

In order to test our general theory even further,
and  motivated by the recent experiments
in Ref.~\cite{chen_revealing_2020},
we next study QNMs for EP-like behaviour
from index-modulated ring resonators. In these experiments, the authors reported a novel 
reversible
chiral emission for linearly polarized dipoles coupled near the ring resonators, using both optical and acoustic systems.
One of the novel conclusions was that the emitter becomes disentangled from the system eigenmodes, and coupled to a ``missing dimension'', such as the Jordan vector.
Here, we study a similar structure, and show that the same features naturally emerge and are fully expected using a two QNM picture, which is again due to the special properties of the QNM phases and how they couple and hybridize.

Specifically,
we consider a microring resonator, similar to those in Refs.~\onlinecite{chen_revealing_2020,martin-cano_chiral_2019}, with a outer radius of $R_{\rm out}=1271~$nm and a inner radius of $r_{\rm in}=1109$~nm (width of $L=R_{\rm out}-r_{\rm in}=162$~nm) (cf.~Fig.~\ref{fig:1N}(a)). 
The two blue arrows in Fig.~\ref{fig:1N}(a) show the direction of CW and CCW fields. In our calculations, we  investigate a 2D ring resonator and coupled it with an in-plane linear dipole. This geometry yields
a TE-like mode.
As shown in Fig.~\ref{fig:1N}(c), the refractive index of the ring is modulated periodically (along CCW direction), which has the following form:
 \begin{align}
 \begin{split}\label{period_n}
    n_{1}&=n_{0}+\delta n_{\rm Re}+i\delta n_{\rm Im},~[l\pi/m<\phi<(l+1/4)\pi/m], \\
    n_{2}&=n_{0}+i\delta n_{\rm Im},~[(l+1/4)\pi/m<\phi<(l+1/2)\pi/m],\\
    n_{3}&=n_{0},~[(l+1/2)\pi/m<\phi<(l+3/4)\pi/m],\\
    n_{4}&=n_{0}+\delta n_{\rm Re},~[(l+3/4)\pi/m<\phi<(l+1)\pi/m],
\end{split}
 \end{align}
where $l=0,1,2 \cdots 2m-1$, and $m$ is the azimuthal mode number (we considering a TE mode with $q=1$, and $m=21$). There are $2m$ periods and $8m$ sections in total.
In the following, we will keep $n_{0}=3.0$ and $\delta n_{\rm Re}=0.001$, while $\delta_{\rm Im}$ is changed in the range of $0.0005\sim0.0015$. 

With the dipole-excitation QNM technique, we use a linear $\hat{\boldsymbol{\phi}}$ dipole placed at $\phi=\phi_1=0$ (which is also a linear $\hat{\bf y}$ dipole at $\phi=0$) and $10~$nm away from the ring surface (the bottom dipole $\mathbf{d}_{1}=d_{0}\hat{\boldsymbol{\phi}}=d_{0}\hat{\bf y}$ in the Fig.~\ref{fig:1N}(c) schematically showing its position and polarization).
Two QNMs (QNM~1 and QNM~2) are found, whose eigenfrequencies ($\tilde{\omega}_{1}=\omega_{1}-i\gamma_{1}$ and $\tilde{\omega}_{2}=\omega_{2}-i\gamma_{2}$) are shown in Fig.~\ref{fig:1N}(e) (red and green stars). We also investigate the eigenfrequecies from the approximate COMSOL eigenfrequency solver (black circles), which agree well with those obtained from
the dipole-excited QNMs technique (mainly as we are dealing with high $Q$ resonators). Note in this section, QNM~1 (QNM~2) and $\tilde{\omega}_{1}$ ($\tilde{\omega}_{1}$) are for coupled modes of the index-modulated ring resonators. Specifically, the eigenfrequency of QNM~1 for $\delta n_{\rm Im}=0.001$ is $\omega_{0}-i\gamma_{0}\approx2.50543552\times10^{15} - i3.768337\times10^{11}$ (${\rm rad/s}$), i.e., $\omega_{0}$ and $\gamma_{0}$ are the real part and the opposite imaginary part of this pole eigenfrequency (different from previous sections).

The field distributions ${\rm Re}[\tilde{f}_s]$, 
with $\tilde{\bf f}=\tilde{f}_{x}\hat{\bf x}+\tilde{f}_{y}\hat{\bf y}=\tilde{f}_{s}\hat{\bf s}+\tilde{f}_{\phi}\hat{\boldsymbol{\phi}}$, of QNM~1 and QNM~2 for the ring structures, with $\delta n_{\rm Re}=\delta n_{\rm Im}=0.001$, are shown in Fig.~\ref{fig:1N}(b) and (d). 
The black square in Fig.~\ref{fig:1N}(b) is enlarged and shown in Fig.~\ref{fig:1N}(c), where the detailed periodic refractive index is displayed. In addition to the general ($\hat{\bf x},\hat{\bf y}$) basis, we also utilize a polar ($\hat{\bf s},\hat{\boldsymbol{\phi}}$) basis. Their unit vectors are related from $\hat{\bf s}=\hat{\bf x}\cos{\phi}+\hat{\bf y}\sin{\phi}$ and $\hat{\boldsymbol{\phi}}=-\hat{\bf x}\sin(\phi)+\hat{\bf y}\cos{\phi}$ (the positive direction of polar angle $\phi$ is along CCW direction). 
As mentioned before, with the dipole QNM technique, we use a linear $\hat{\boldsymbol{\phi}}$ dipole at $\phi=0$ (the bottom one $\mathbf{d}_{1}$ in Fig.~\ref{fig:1N}(c)). Once we obtain the QNMs, then we can easily couple them to any in-plane linear dipoles with any polarization; we schematically give two example dipoles in Fig.~\ref{fig:1N} (c), one is a linear $\hat{\bf s}$ dipole at $\phi=\phi_{2}=\pi/4m$ (the center dipole, $\mathbf{d}_{2}$), and the other one is a linear $\hat{\boldsymbol{\phi}}$ dipole at $\phi=\phi_{3}=3\pi/4m$ (the top dipole, $\mathbf{d}_{3}$).

\begin{figure*}
    \centering
    \includegraphics[width=1.8\columnwidth]{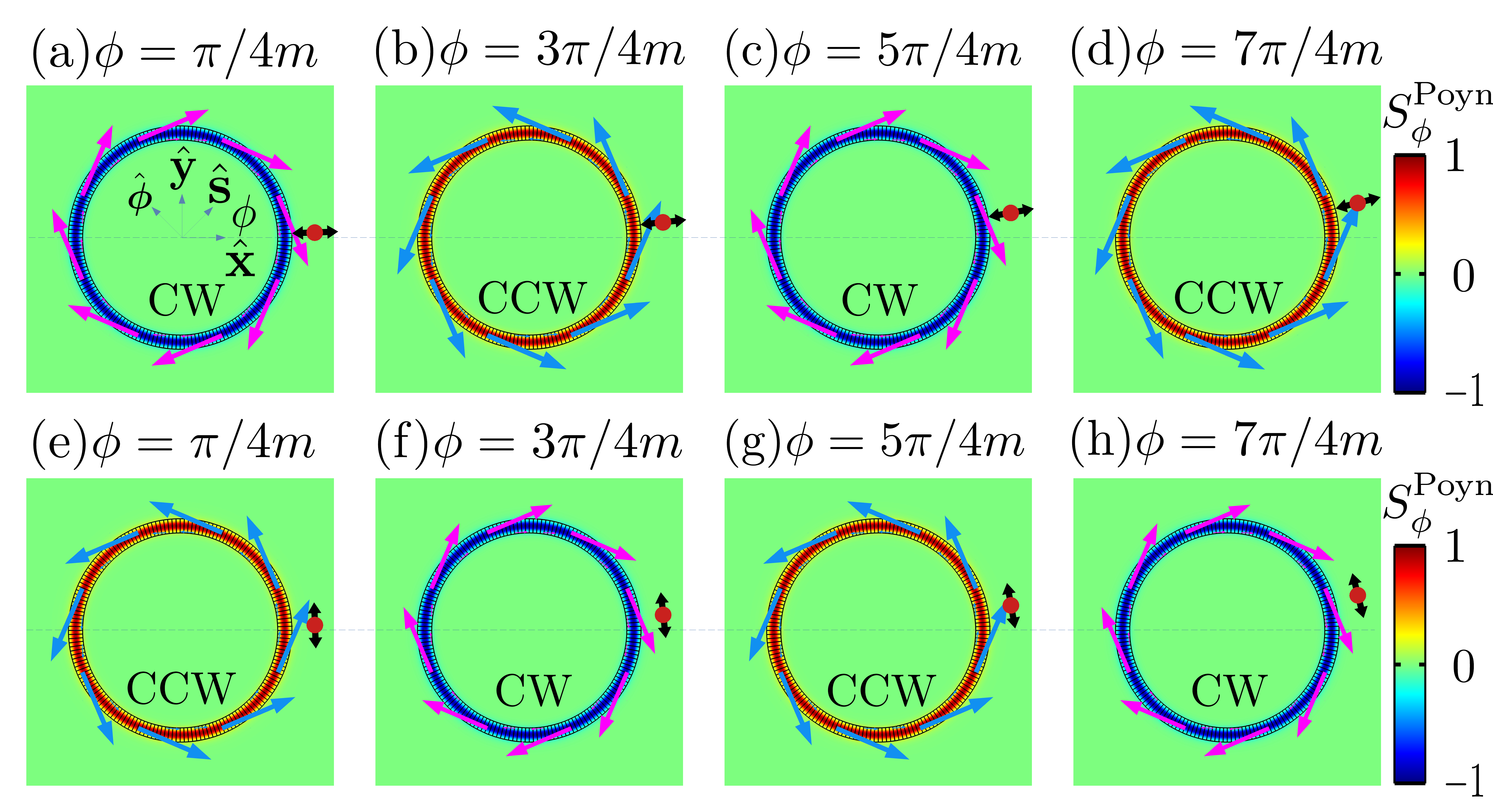}
    \includegraphics[width=1.8\columnwidth]{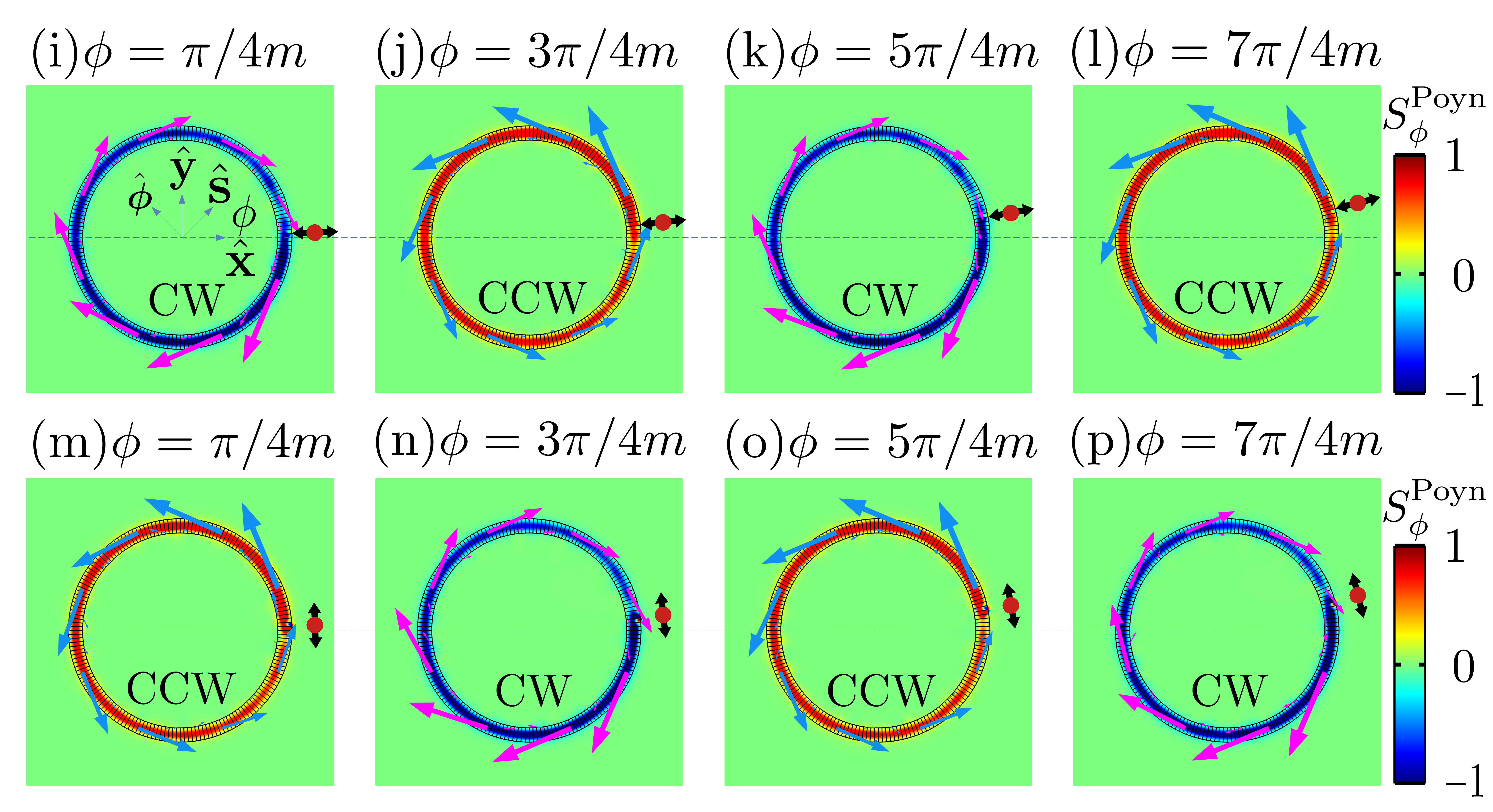}
       \caption{Poynting vectors ($\mathbf{S}^{\rm Poyn}=S_{x}^{\rm Poyn}\hat{\bf x}+S_y^{\rm Poyn}\hat{\bf{y}}=S_{s}^{\rm Poyn}\hat{\bf s}+S_{\phi}^{\rm Poyn}\hat{\boldsymbol{\phi}}$) from (a-h) QNMs approaches and (i-p) full dipole method for (a-d) ((i-l)) $\hat{\bf s}$ and (e-h) ((m-p)) $\hat{\boldsymbol{\phi}}$ dipoles at $\phi=\pi/4m$, $\phi=3\pi/4m$, $\phi=5\pi/4m$, and $\phi=7\pi/4m$ (the radial distance between the dipole and the ring surface is $10~$nm ($\delta n_{\rm Re}=\delta n_{\rm Im}=0.001$). The red dot with double black arrows schematically describe the position and polarization of the linear dipoles. The Poynting vectors are labelled as the magenta (for CW direction) and blue (for CCW direction) arrows. The distributions are normalized to $S^{\rm Poyn}_{\phi}$ (for those from full dipole method, they are with arb. units). These power flows are calculated at a example real frequency $\omega=2.50543\times10^{15}$~(${\rm rad/s}$) (close to QNM resonance). In the ring region, one will find $S^{\rm Poyn}_{\phi}<0$ for CW propagation and $S^{\rm Poyn}_{\phi}>0$ for CCW propagation; the net power flow from QNMs approaches are in excellent agreement with the full dipole simulations. 
       }\label{Fig:34N}
\end{figure*}

\subsection{Purcell factors versus frequency for different dipole locations}

To confirm the accuracy of our two QNM description for the
index-modulated ring resonators, we calculate the Purcell factors analytically from the QNM properties (Eq.~\eqref{QNMpurcell}) for a $\hat{\boldsymbol{\phi}}$ dipole at $\phi=0$ (equal to $\hat{\bf y}$ at $\phi=0$, $\mathbf{d}_{1}$ in Fig.~\ref{fig:1N} (c)), and compare them with the full dipole method (Eq.~\eqref{Purcellfulldipole}). For all configurations investigated, an excellent agreement is shown as a function of frequency as demonstrated in Fig.~\ref{fig:2N}, where the results for six cases are investigated with the analytical two QNM expansion are studied, including $\delta n_{\rm Im}=0.0005,0.0008,0.001,0.0011,0.0012,0.0015$ while keeping $\delta n_{\rm Re}=0.001$. 
Their eigenfrequencies are marked in Fig.~\ref{fig:1N}(e). Although there are no material gain regions in these resonators, we also note the appearance of modal negative Purcell factors.
This level of agreement is unusual in the sense that
the eigenmodes in Ref.~\cite{chen_revealing_2020} were reported to be completely decoupled, which was used to argue  a chiral emission, an effect that is not expected nor obtained for a regular ring resonator. We will connect to the power flow emission below, and show that it is also well explained in terms of the underlying QNMs.


\subsection{Chiral power flow from linearly polarized dipoles}

To directly connect with the experimentally measured chirality in Ref.~\onlinecite{chen_revealing_2020} from linear dipole emitters, here we show chiral power flow using only the QNM propagators in a similar microring with refractive index modulation.
We can {\it easily} obtain the power flow by computing the
scattered fields and the Poynting vector from QNMs and the QNM propagator, and thus the calculations are basically instantaneous.

Once the QNMs are known,
the scattered field (in real frequency space) at any point ${\bf r}$
from a dipole at ${\bf r}_{\rm d}$ is simply
\begin{align}
\mathbf{E}^{\rm scatt}(\mathbf{r}_{\rm },\omega) &=\mathbf{G}(\mathbf{r}_{\rm },\mathbf{r}_{\rm d},\omega)\cdot\mathbf{d} \nonumber \\
=[A_{1}(\omega)&\tilde{\bf f}_1({\bf r}) \tilde{\bf f}_1({\bf r}_{\rm d})
+ A_{2}(\omega)\tilde{\bf f}_2({\bf r}) \tilde{\bf f}_2({\bf r}_{\rm d})]
\cdot d_{0}{\bf n_{\rm d}},
\end{align}
where ${\bf d}=d_0 {\bf n}_{\rm d}$,
and we will consider both
$\hat{\bf s}$ dipoles
and $\hat{\boldsymbol{\phi}}$ dipoles (i.e., ${\bf n}_{\rm d}=\hat{\bf s}$ or ${\bf n}_{\rm d}=\hat{\boldsymbol{\phi}}$). 
The magnetic scattered field is then
\begin{equation}
\begin{split}
\nabla\times\mathbf{E}^{\rm scatt}(\mathbf{r}_{\rm },\omega)
&=i\omega\mu_{0}\mathbf{H^{\rm scatt}}(\mathbf{r}_{\rm },\omega)
\end{split}
\end{equation}
and thus the Poynting vector is simply
\begin{equation}
\begin{split}\label{Eq:Spoyn}
\mathbf{S}^{\rm Poyn}(\mathbf{r}_{\rm },\omega)&=\frac{1}{2}{\rm Re}[\mathbf{E}^{\rm scatt}(\mathbf{r}_{\rm },\omega)\times\mathbf{H}^{\rm scatt~*}(\mathbf{r}_{\rm },\omega)]\\
&={S}_{x}^{\rm Poyn}(\mathbf{r}_{\rm },\omega)\hat{\bf x}+{S}_{y}^{\rm Poyn}(\mathbf{r}_{\rm },\omega)\hat{\bf y}\\
&={S}_{s}^{\rm Poyn}(\mathbf{r}_{\rm },\omega)\hat{\bf s}+{S}_{\phi}^{\rm Poyn}(\mathbf{r}_{\rm },\omega)\hat{\boldsymbol{\phi}},
\end{split}
\end{equation}
where $S_{\phi}^{\rm Poyn}(\mathbf{r}_{\rm },\omega)$ is the projection along $\hat{\boldsymbol{\phi}}$,  and one will find $S_{\phi}^{\rm Poyn}(\mathbf{r}_{\rm },\omega)<0$ ($>0$) when it goes along the CW (CCW) direction. 
Using the QNMs, we stress that we can compute the direction of the power flow and scattered fields analytically, for any dipole position and orientation. This further demonstrates the power of having an analytical expression for the Green function in terms of only QNM properties.

In this section, we focus on the case with $\delta n_{\rm Re}=\delta n_{\rm Im}=0.001$, which is closest to the EP as seen from Fig.~\ref{fig:1N}(e) and Fig.~\ref{fig:2N}(c).
Figure \ref{Fig:34N}(a-h) shows the chiral power flow (Eq.~\eqref{Eq:Spoyn}) from our QNMs model, for $\hat{\bf s}$ and $\hat{\boldsymbol{\phi}}$ dipole at several specific positions, including $\phi=\pi/4m$ (center of two lossy sections, see Fig.~\ref{fig:1N}(c)), $\phi=3\pi/4m$ (center of two lossless sections), $\phi=5\pi/4m$ (center of two lossy sections), and $\phi=7\pi/4m$ (center of two lossless sections) (the radial distance between the dipole and the ring surface is $10~$nm). The red dot with a double black arrows in Fig.~\ref{Fig:34N} schematically shows the position and polarization of these linear dipoles (though this is not to scale). Also note that these power flows presented in Fig.~\ref{Fig:34N} are calculated at a example real frequency $\omega=2.50543\times10^{15}$~(${\rm rad/s}$) (close to QNM resonance), and we stress that the chirality will remain the same in a range around $\omega_{0}\pm1.5\gamma_{0}$.

The first thing to observe, is that when a linear $\hat{\bf s}$ dipole is placed at $\phi=\pi/4m$ (Fig.~\ref{Fig:34N} (a)), the net power flow ($\mathbf{S}^{\rm Poyn}$, Eq.~\eqref{Eq:Spoyn}), indicate by the magenta arrows, goes along the CW direction. The  distribution shown is normalized to the projection $S_{\phi}^{\rm Poyn}$. We also find that $S_{\phi}^{\rm Poyn}<0$ in the ring region, which further confirms the net energy flow goes along the CW direction. In addition, the azimuthal period of such phenomenon is $\pi/m$, i.e., when a linear $\hat{\bf s}$ dipole is located at $\phi=\frac{\pi}{4m}+p\frac{\pi}{m}$ (center of the two lossy regions), where $p=0,1,2,...$, the net power will go along the CW direction, such as at $\phi=5\pi/4m$ shown in Fig.~\ref{Fig:34N} (c).

To further justify that our QNM picture is both correct
and rigorously accurate for describing this chiral emission,
we have confirmed these unusual properties with full dipole calculations shown in Fig.~\ref{Fig:34N} ((i) and (k)). This also supports the experimental findings in Ref.~\onlinecite{chen_revealing_2020}, where they measured CW propagation when putting a linear dipole at $\phi=\pi/4l$ (center of lossy region) (azimuthal mode number $l=1$ in their considered structure, and we are using azimuthal mode number $m=21$ for a similar ring structure). However, our interpretation is drastically different. Instead or arguing a decoupling from the eigenmodes and coupling to a missing dimension, we find that our net chiral power flows are fully explained from the underlying QNMs, which are the correct natural eigenmodes of such resonators\footnote{Assuming that one is not at a perfect EP, which we have discussed earlier is highly unlikely, and thus not a practical concern. }.

Moreover, we also  find that such unusual properties are not limited to $\hat{\bf s}$ dipoles at $\phi=\frac{\pi}{4m}+p\frac{\pi}{m}$. As shown in Fig.~\ref{Fig:34N}(b,d) (QNMs results) and
Fig.~\ref{Fig:34N} (j,l) (full dipole results), when a linear $\hat{\bf s}$ dipole is placed at $\phi=\frac{3\pi}{4m}+p\frac{\pi}{m}$ (center of the two lossless sections, where $p=0,1,2,...$), the net power flow is along the CCW direction.
In addition, as shown in Fig.~\ref{Fig:34N} (e-h) (QNMs results) and Fig.~\ref{Fig:34N} (m-p) (full dipole results), when a linear $\hat{\boldsymbol{\phi}}$ dipole is placed at $\phi=\frac{\pi}{4m}+p\frac{\pi}{m}$ ($\phi=\frac{3\pi}{4m}+p\frac{\pi}{m}$), the net power flow will go along the CCW (CW) direction.
Also note, 
for $\hat{\bf s}$ dipoles at $\phi=n\frac{\pi}{m}$ 
and $\phi=\frac{2\pi}{4m}+n\frac{\pi}{m}$
(cf.~Fig.~\ref{fig:1N}),  respectively, 
the power flow direction will change from CCW  to CW 
and CW to CCW, around resonance, i.e., at resonance, there is no net flow direction.
This is also true for $\hat{\boldsymbol{\phi}}$ dipoles; for $\hat{\boldsymbol{\phi}}$ at $\phi=n\frac{\pi}{m}$ ($\phi=\frac{2\pi}{4m}+n\frac{\pi}{m}$), the power flow direction will change from CW (CCW)  to CCW (CW).  
This gives one an external control to change the directionality by simply changing the frequency.

  Note that such chirality behaviour is not found in regular ring resonators (without a refractive modulation) for linear $\hat{\bf s}$ and $\hat{\boldsymbol{\phi}}$ dipoles, i.e., no net power flow is obtained.
However, by using a circular dipole close to or inside the general ring resonator, chirality will exist, as shown in Ref.~\onlinecite{martin-cano_chiral_2019}, where positional dependent chirality for right- or left-handed circular dipoles are demonstrated. 
This is also similar to how one excites unidirectional 
propagation in photonic crystal waveguides~\cite{young_polarization_2015,Sllner2015}.
However, in all these cases, local symmetry breaking is possible by using
a circularly polarized dipole.

\section{Conclusions}
\label{sec:sec6}

We have introduced a powerful and highly accurate QNM approach to coupled loss-gain 
resonators, and have 
presented a rigorous and intuitive CMT based on the photonic Green function, which allows one to solve the coupled system efficiently with just the bare QNM solutions from the individual resonators. 
We have also highlighted the failure of using a NM CMT approach when defining the general conditions for finding EPs.

For the SE response of embedded dipoles in these systems, we have carried out detailed calculations for coupled microdisk resonators;
as well as finding
 Lorentzian-like and Lorentzian-suqared like responses at the EP, consistent with other works, we have shown much richer Purcell factor lineshapes near EPs for various designs and  spatial dipole positions, showing excellent agreement with QNM CMT and  full dipole
 calculations. 
 In particular, we have shown how the
 Purcell factors can also be negative,
 in loss-gain media, even when the hybrid modes are both lossy ($\gamma_\mu>0$). This is caused by  a breakdown of Fermi's golden rule which incorrectly assumes that the SE rate is propositional to the (projected) LDOS)~\cite{EPQuantumPaper}.
In addition, we also showed how the Green function propagators (related to various experimental observables, such as the emitted spectrum) also take on rich non-Lorentzian features,
which depend on the values of the QNM phases.

In addition to the coupled loss and gain resonators, 
we also investigated EP-like resonances
formed from index-modulated ring resonators,
where unusual chiral emission from linearly polarized diploes
was recently observed~\cite{chen_revealing_2020}.
Once again, we showed that the full dipole Purcell factor response
is quantitatively well explained in terms of the
main two QNMs of this resonator. Moreover, 
when a linear $\hat{\bf s}$ dipole is located at $\phi=\pi/4m$, the Poynting vectors from QNMs propagators goes along the CW direction, which supports the experimental findings in Ref.~\onlinecite{chen_revealing_2020} for similar ring resonators.
Notably,
our explanation is in contrast with the view that the emitter does not couple to the system eigenmodes.
In addition, 
we also showed that such chirality is not limited to $\hat{\bf s}$ dipoles at $\phi=\frac{\pi}{4m}+p\frac{\pi}{m}$, and we also show the opposite chirality for $\hat{\bf s}$ dipoles at $\phi=\frac{3\pi}{4m}+p\frac{\pi}{m}$.  There is also similar chirality for linear $\hat{\boldsymbol{\phi}}$ dipoles at these positions. We stress, again, that these net power flows can be well explained and interpreted from the two underling QNMs and the corresponding Green function, where the QNM phases play a decisive and fundamental role on the light emission,
without having to invoke any unusual interpretation such as a missing dimension (Jordan vector).


Apart from providing a detailed and intuitive formalism for understanding the classical mode properties of these complex coupled resonator systems,
our QNM formalism forms the basis for a rigorous quantum optics approach in media with gain and loss, using new approaches with quantized QNMs~\cite{franke_quantization_2019,PhysRevResearch.2.033332,PhysRevResearch.2.033456}, which has already 
lead to a revision of the usual photonic Fermi's golden rule for coupled loss and gain resonators~\cite{EPQuantumPaper}, one
in which the net Purcell factor is always a positive quantity.

\section{Acknowledgements}
We  acknowledge funding from Queen's University,
the Canadian Foundation for Innovation, 
the Natural Sciences and Engineering Research Council of Canada, and CMC Microsystems for the provision of COMSOL Multiphysics.
We also acknowledge support from 
the Alexander von Humboldt Foundation through a Humboldt Research Award.
We thank Andreas Knorr
for discussions and  support, and 
Thomas Christopoulos for discussions
about PML properties in COMSOL.

\vspace{1cm}

\appendix

\section{Quasinormal mode (QNM)  normalization}
\label{sec:numerics}


There are several general
numerical approaches
to obtaining normalized QNMs~\cite{kristensen_modeling_2020}, including a dipole excitation technique in complex frequency~\cite{bai_efficient_2013-1}, 
PML normalization~\cite{sauvan_theory_2013}, 
finite domain normalization with a surface term~\cite{muljarov_brillouin-wigner_2010,Kristensen2015,muljarov_exact_2016},
 finite-difference time-domain methods~\cite{ge_design_2014},
and  Riesz-projection-based techniques~\cite{PhysRevA.98.043806}.
In the main text, we use the dipole technique in complex frequency space~\cite{bai_efficient_2013-1}, and more details are shown in the following subsection~\ref{AppendixA_sub1}. A brief description of  two alternative methods is also given in subsections~\ref{AppendixA_sub2}-\ref{AppendixA_sub3}; numerically, we find all three approaches give the same normalized QNMs for the
QNMs used in this work
(within numerical precision).

\subsection{QNM normalization and numerically exact Green function from a  dipole  source}\label{AppendixA_sub1}
 
As shown in main text, with CMT and the Green function theory, only the bare mode solutions (for a single lossy resonator or  gain resonator) are required. We employ an efficient dipole scattering approach to obtain the uncoupled QNMs in complex frequency~\cite{bai_efficient_2013-1}, where an out-of-plane line current (a point in 2D, red dot in Fig.~\ref{disk_sche}, for a TM mode; it's a in-plane point dipole for a TE mode) is placed close to the lossy resonator, or the gain resonator, or the index-modulated ring resonators. We can also use this approach for the coupled resonator problem, which we also do to check the accuracy of the analytical CMT for the hybrid modes (see Appendix~\ref{sec:directQNMs}).

The scattered field of this dipole at $\mathbf{r}_{0}$ is related to the 2D Green function, from 
\begin{equation}
 \mathbf{E^{\rm scatt}(\mathbf{r},\omega)}=\mathbf{G}^{\rm 2D}(\mathbf{r},\mathbf{r}_{0},\omega)\cdot\frac{\mathbf{d}^{\rm 2D}}{\epsilon_{0}}.  
\end{equation}
where the units of the scattered field $\mathbf{E^{\rm scatt}}$, 2D Green function $\mathbf{G}^{\rm 2D}$, and 2D dipole moment $\mathbf{d}^{\rm 2D}$ are, respectively, $\rm V/m$, ${\rm m}^{-2}$, and $\rm C$.

Expanding the 2D Green function with one QNM (dominating in the regime of interest), then
\begin{equation}\label{2DGreen}
\mathbf{G}^{\rm 2D}(\mathbf{r},\mathbf{r}_{0},\omega)=A_{\rm c}(\omega)\tilde{\mathbf{f}}_{\rm c}^{\rm 2D}(\mathbf{r})\tilde{\mathbf{f}}_{\rm c}^{\rm 2D}(\mathbf{r}_{0}),   
\end{equation}
so that
\begin{equation}\label{E_scatter_2D}
\mathbf{E}^{\rm scatt}(\mathbf{r},\omega)=\frac{1}{\epsilon_{0}}A(\omega)\tilde{\mathbf{f}}_{\rm c}^{\rm 2D}(\mathbf{r})\tilde{\mathbf{f}}_{\rm c}^{\rm 2D}(\mathbf{r}_{0})\cdot \mathbf{d}^{\rm 2D}.
\end{equation}
Multiplying Eq. \eqref{E_scatter_2D} with $\mathbf{d}^{\rm 2D}$ and using $\mathbf{r}=\mathbf{r}_{0}$, then 
\begin{equation}
\mathbf{d}^{\rm 2D}\cdot\tilde{\mathbf{f}}^{\rm 2D}_{\rm c}(\mathbf{r}_{0})=\sqrt{\frac{\epsilon_{0}\mathbf{d}^{\rm 2D}\cdot\mathbf{E}^{\rm scatt}(\mathbf{r}_{0},\omega)}{A(\omega)}}.
\end{equation}
Substituting this back to Eq.~\eqref{E_scatter_2D}, we obtain the 2D normalized QNM field as a function of space
\begin{align}
\begin{split}\label{2DQNM_norm}
\tilde{\mathbf{f}}^{\rm 2D}_{\rm c}(\mathbf{r})&=\sqrt{\frac{\epsilon_{0}}{A(\omega)\mathbf{d}^{\rm 2D}\cdot\mathbf{E}^{\rm scatt}(\mathbf{r}_{0},\omega)}}\mathbf{E}^{\rm scatt}(\mathbf{r},\omega),\\
&=\sqrt{\frac{2\epsilon_{0}(\tilde{\omega}_{\rm c}-\omega)}{\omega\mathbf{d}^{\rm 2D}\cdot\mathbf{E}^{\rm scatt}(\mathbf{r}_{0},\omega)}}\mathbf{E}^{\rm scatt}(\mathbf{r},\omega).
\end{split} 
\end{align}

The above QNM simulations are performed in the commercial COMSOL software~\cite{comsol}, where the frequency in Eq.~\eqref{2DQNM_norm} is set as $\omega=(1-10^{-7})\times\tilde{\omega}_{\rm c}$ (or $(1-10^{-6})\times\tilde{\omega}_{\rm c}$, $(1-10^{-8})\times\tilde{\omega}_{\rm c}$, adjust with the quality factors), very close to the pole frequency.
The computational domain (including PMLs) is 
around $804\sim814$ $\mu$m$^{2}$ (various gap distance for microdisk resonators; for microring resonators, it's around $72$~$\mu$m$^{2}$), where the maximum mesh element sizes are $0.1$ nm, $40$ nm and $75$ nm at the dipole point, inside and outside the 2D microdisks. To minimize boundary reflections, 
we used $15$ layers to form the PMLs with a total thickness of $1.5$ $\mu$m,
which was found to be well converged numerically.

In addition, once the normalized QNMs are available, the corresponding effective mode area (with units ${\rm m}^{2}$), which is a function of position, is obtained from
\begin{equation}
A_{\rm c}^{\rm eff}(\mathbf{r}_0)
=\frac{1}{\epsilon(\mathbf{r}_0){\rm Re}[{\bf f}^{2}_{\rm c,2D}(\mathbf{r}_0)]}.
\end{equation}

The decay rates for 2D dipoles are as follows:
 \begin{align}
 \begin{split}
 \Gamma_{\rm }^{\rm 2D}(\mathbf{r}_{0},\omega)&=\frac{2}{\hbar\epsilon_{0}}\mathbf{d}^{\rm 2D}\cdot{\rm Im}\{\mathbf{G}^{\rm 2D}(\mathbf{r}_{0},\mathbf{r}_{0},\omega)\}\cdot\mathbf{d}^{\rm 2D},\\
  \Gamma^{\rm 2D}_{0}(\mathbf{r}_{0},\omega)&=\frac{2}{\hbar\epsilon_{0}}\mathbf{d}^{\rm 2D}\cdot{\rm Im}\{\mathbf{G}^{\rm 2D}_{\rm B}(\mathbf{r}_{0},\mathbf{r}_{0},\omega)\}\cdot\mathbf{d}^{\rm 2D},
 \end{split}  
 \end{align}
where ${\rm Im}\{\mathbf{G}_{\rm B}(\mathbf{r}_{0},\mathbf{r}_{0},\omega)\}={\omega^2}/{4c^2}$ (${\omega^2}/{8c^2}$) for a 2D TM (TE) dipole, and  the units of $\Gamma^{\rm 2D}$  is $1/({\rm s}\cdot{\rm m})$.

The corresponding Purcell factor is simply
\cite{Anger2006,kristensen_modes_2014} 
 \begin{align}
 \begin{split}\label{eq:2Dpurcell}
     &F^{\rm 2D}(\mathbf{r}_0,\omega) =1+\frac{\Gamma^{\rm 2D}(\mathbf{r}_{0},\omega)}{\Gamma^{\rm 2D}_{0}(\mathbf{r}_{0},\omega)},
 \end{split}
 \end{align}
 and similar expressions are obtained for 3D systems.


\subsection{Perfectly matched later (PML) normalization}\label{AppendixA_sub2}



The PML normalization~\cite{sauvan_theory_2013,vial_quasimodal_2014,lalanne_light_2018} 
approach is another alternative way to get the normalized QNMs, which is given via 
(for dispersive and nonmagnetic materials)
\begin{align}
\begin{split}
\braket{\braket{\tilde{\mathbf{f}}_{\mu}(\mathbf{r})|\tilde{\mathbf{f}}_{\mu}(\mathbf{r})}}=\frac{1}{2}\int_{V-V_{\rm PML}}\bigg[\frac{\partial(\omega\epsilon(\mathbf{r},\omega))}{\partial\omega}\Big|_{\tilde{\omega}_{\mu}}\tilde{\mathbf{f}}_{\mu}(\mathbf{r})\cdot\tilde{\mathbf{f}}_{\mu}(\mathbf{r})\\
+\epsilon(\mathbf{r},\tilde{\omega}_{\mu})\tilde{\mathbf{f}}_{\mu}(\mathbf{r})\cdot\tilde{\mathbf{f}}_{\mu}(\mathbf{r})\bigg]dV\\
+\frac{1}{2\epsilon_{0}}\int_{V_{\rm PML}}\bigg[\epsilon_{0}\frac{\partial(\omega\epsilon_{\rm PML}(\mathbf{r},\omega))}{\partial\omega}\Big|_{\tilde{\omega}_{\mu}}\tilde{\mathbf{f}}_{\mu}(\mathbf{r})\cdot\tilde{\mathbf{f}}_{\mu}(\mathbf{r})\\
-\mu_{0}\frac{\partial(\omega\mu_{\rm PML}(\mathbf{r},\omega))}{\partial\omega}\Big|_{\tilde{\omega}_{\mu}}\tilde{\mathbf{h}}_{\mu}(\mathbf{r})\cdot\tilde{\mathbf{h}}_{\mu}(\mathbf{r})\bigg]dV=1,
\end{split}
\end{align}
where $\tilde{\mathbf{h}}_{\mu}(\mathbf{r})$ is the magnetic field of QNM, $V$ is the whole simulation region, and $V_{\rm PML}$ denotes the PML region. Outside the PML region, then the fields are zero.

For a nondispersive and nonmagnetic material, 
\begin{align}
\begin{split}
\braket{\braket{\tilde{\mathbf{f}}_{\mu}(\mathbf{r})|\tilde{\mathbf{f}}_{\mu}(\mathbf{r})}}=\int_{V-V_{\rm PML}}\bigg[
\epsilon(\mathbf{r})\tilde{\mathbf{f}}_{\mu}(\mathbf{r})\cdot\tilde{\mathbf{f}}_{\mu}(\mathbf{r})\bigg]dV\\
+\frac{1}{2\epsilon_{0}}\int_{V_{\rm PML}}\bigg[\epsilon_{0}\frac{\partial(\omega\epsilon_{\rm PML}(\mathbf{r},\omega))}{\partial\omega}\Big|_{\tilde{\omega}_{\mu}}\tilde{\mathbf{f}}_{\mu}(\mathbf{r})\cdot\tilde{\mathbf{f}}_{\mu}(\mathbf{r})\\
-\mu_{0}\frac{\partial(\omega\mu_{\rm PML}(\mathbf{r},\omega))}{\partial\omega}\Big|_{\tilde{\omega}_{\mu}}\tilde{\mathbf{h}}_{\mu}(\mathbf{r})\cdot\tilde{\mathbf{h}}_{\mu}(\mathbf{r})\bigg]dV=1,
\end{split}
\label{eq:PML}
\end{align}
and extra care is needed for the PML region (second term in Eq.~\eqref{eq:PML}), though this contribution can be very small for certain problems and geometries.
In general, there are several kinds of transformation performed in PMLs to minimize boundary reflections. The first one is using special permittivity $\epsilon_{\rm PML}(\mathbf{r},\omega)$ and permeability  $\mu_{\rm PML}(\mathbf{r},\omega)$ values, which is not always available with some commercial software~\cite{lalanne_light_2018}.
A second approach  uses a coordinate transformation, which is what we used here, with the built-in stretched-coordinate PML of COMSOL, 
where the coordinates are transferred from real space to the complex plane~\cite{berenger_perfectly_1994}.

To verify that the dipole normalization technique and PML normalization are consistent with each other, we performed  the norm calculation (Eq.~\eqref{eq:PML}) using the QNM fields from the dipole technique.
We obtained 
$\braket{\braket{\tilde{\mathbf{f}}^{\rm L}(\mathbf{r})|\tilde{\mathbf{f}}^{\rm L}(\mathbf{r})}}=1.0004 - 0.0011i\sim1$ for the  QNM of interest from the single lossy WGM resonator with $n_{\rm loss}=2+10^{-5}i$, and $\braket{\braket{\tilde{\mathbf{f}}^{\rm G}(\mathbf{r})|\tilde{\mathbf{f}}^{\rm G}(\mathbf{r})}}=0.9991 + 0.0006i\sim1$ for the QNM of interest from single gain WGM resonator with $n_{\rm gain}=2-5\times10^{-6}i$.
However, note, that for our chosen PML geometry and resonator, the contribution from the PML is 
practically negligible
for this problem of interest (high $Q$ resonance).

\begin{figure*}
    \centering
    \includegraphics[width=1.99\columnwidth]{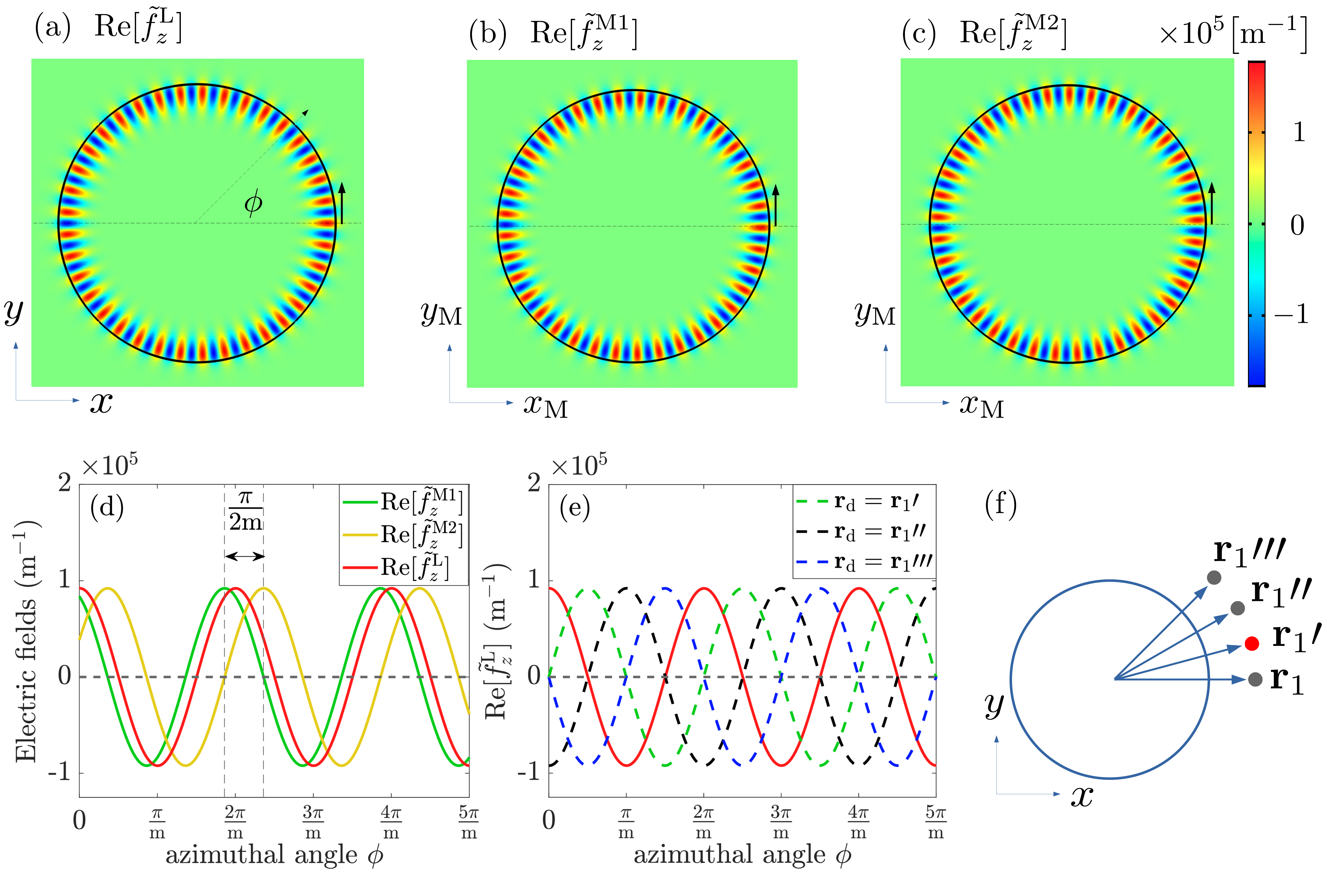}
    \caption{Spatial profile of the degenerate QNMs for a single lossy resonator. (a) QNM obtained with a general QNM approach (Eq.~\eqref{2DQNM_norm}) (it is the same as in Fig.~\ref{QNM_distribution} (a)), where only one of the degenerate standing wave modes is excited, which is with a $cos(m\phi)$ dependence, considering $\phi=0$ at the positive $x$-axis. (b-c)
    Normalized (Eq.~\eqref{eq:pmlnorm_M1M2}) degenerate  modes $1$ and $2$ from the eigenfrequency solver using COMSOL, where the boundary condition is with PML. The horizontal dashed lines in (a), (b), and (c) labeled the $x$-axis with their own coordinates.
    (d) The fields at the disk edge along vertical arrow (in (a), (b), and (c)).
    (e) Same as in (a) but when the dipole position is moved to a different position, as shown in (f).
   }
    \label{fig:degeneratemodes}
\end{figure*}

\subsection{Finite-domain normalization with a surface term}\label{AppendixA_sub3}

A third way to normalize QNMs is through a finite domain normalization with an outgoing surface 
term~\cite{muljarov_brillouin-wigner_2010,muljarov_exact_2016,Kristensen2015,kristensen_modeling_2020}, 
 is given by
\begin{align}
\braket{\braket{\tilde{\mathbf{f}}_{\mu}(\mathbf{r})|\tilde{\mathbf{f}}_{\mu}(\mathbf{r})}} &=\frac{1}{2\epsilon_{0}}\int_{V}\bigg[\epsilon_{0}\frac{\partial(\omega\epsilon(\mathbf{r},\omega))}{\partial\omega}\Big|_{\tilde{\omega}_{\mu}}\tilde{\mathbf{f}}_{\mu}(\mathbf{r})\cdot\tilde{\mathbf{f}}_{\mu}(\mathbf{r}) \nonumber \\
&-\mu_{0}\tilde{\mathbf{h}}_{\mu}(\mathbf{r})\cdot\tilde{\mathbf{h}}_{\mu}(\mathbf{r})\bigg]dV \nonumber \\
&+\frac{i}{2\epsilon_{0}\tilde{\omega}_{\mu}}\int_{\partial V}\bigg[\big(r\partial r\tilde{\mathbf{f}}_{\mu}(\mathbf{r})\big )\mathbf{\times}\tilde{\mathbf{h}}_{\mu}(\mathbf{r})\nonumber \\
&-\tilde{\mathbf{f}}_{\mu}(\mathbf{r})\mathbf{\times}\big(r\partial{r}\tilde{\mathbf{h}}_{\mu}(\mathbf{r})\big)\bigg]\cdot\hat{\mathbf{n}}dA=1,
\end{align}
or
\begin{align}
\braket{\braket{\tilde{\mathbf{f}}_{\mu}(\mathbf{r})|\tilde{\mathbf{f}}_{\mu}(\mathbf{r})}}& =\int_{V}\sigma(\mathbf{r},\tilde{\omega}_{\mu})\tilde{\mathbf{f}}_{\mu}(\mathbf{r})\cdot\tilde{\mathbf{f}}_{\mu}(\mathbf{r})d\mathbf{r} \nonumber \\
&+\frac{1}{2\tilde{k}_{\mu}^{2}}\int_{\partial V}\bigg[\tilde{\mathbf{f}}_{\mu}(\mathbf{r})\cdot\partial s\big[r\partial r\tilde{\mathbf{f}}_{\mu}(\mathbf{r})\big] \nonumber \\
&-r\big[\partial r\tilde{\mathbf{f}}_{\mu}(\mathbf{r})\big]\cdot\big[\partial s \tilde{\mathbf{f}}_{\mu}(\mathbf{r})\big]\bigg]dA=1,
\end{align}
where 
\begin{equation}
\sigma(\mathbf{r},\tilde{\omega}_{\mu})=\frac{1}{2\omega}\frac{\partial[\omega^2\epsilon(\mathbf{r},\omega)]}{\partial\omega}\Big|_{\tilde{\omega}_{\mu}},
\end{equation}
and the surface $\partial V$ is the surface of the finite domain $V$ that surrounding the entire resonators, and $\partial s$ is the derivative
in the normal direction of the  surface $\partial V$.
Now the integral region $V$ has a greater freedom of choice, but at the cost of adding a surface integral 
(though for 2D, it is a surface integral and line integral). 

For a nondispersive material, 
\begin{align}
\begin{split}\label{eq:surnorm_f_nondis}
\braket{\braket{\tilde{\mathbf{f}}_{\mu}(\mathbf{r})|\tilde{\mathbf{f}}_{\mu}(\mathbf{r})}}& =\int_{V}\epsilon(\mathbf{r})\tilde{\mathbf{f}}_{\mu}(\mathbf{r})\cdot\tilde{\mathbf{f}}_{\mu}(\mathbf{r})d\mathbf{r}\\
& +\frac{1}{2\tilde{k}_{\mu}^{2}}\int_{\partial V}\bigg[\tilde{\mathbf{f}}_{\mu}(\mathbf{r})\cdot\partial s\big[r\partial r\tilde{\mathbf{f}}_{\mu}(\mathbf{r})\big]\\
& -r\big[\partial r\tilde{\mathbf{f}}_{\mu}(\mathbf{r})\big]\cdot\big[\partial s \tilde{\mathbf{f}}_{\mu}(\mathbf{r})\big]\bigg]dA=1.
\end{split}
\end{align}

For completeness,  we have checked our normalized QNMs with this technique,
and obtain 
$\braket{\braket{\tilde{\mathbf{f}}^{\rm L}(\mathbf{r})|\tilde{\mathbf{f}}^{\rm L}(\mathbf{r})}}=1.0004 - 0.0011i\sim1$ for the single bare QNM from single lossy WGM resonartor with $n_{\rm loss}=2+10^{-5}i$, and $\braket{\braket{\tilde{\mathbf{f}}^{\rm G}(\mathbf{r})|\tilde{\mathbf{f}}^{\rm G}(\mathbf{r})}}=0.9991 + 0.0006i\sim1$ for the single bare QNM from single gain WGM resonartor with $n_{\rm gain}=2-5\times10^{-6}i$.
Again, we note for this problem, the contribution from the line integral is negligible, thus we obtain basically the same answer as the PML normalization.
As a further check, 
we also tried above norm with two smaller domain with area of $579~\mu$m$^{2}$ and $497~\mu$m$^{2}$, and obtain exactly the same numerical results as above.

\section{Degenerate QNMs for microdisk Resonators}\label{sec:degenerate_WGMs}

The  WGM resonator naturally supports degenerate modes because of circular symmetry.
For a specific azimuthal mode number $m$, the degenerate counterpropagating modes are~\cite{mazzei_controlled_2007,teraoka_resonance_2009,cognee_cooperative_2019}:
$E_{\rm cw}(\mathbf{r},\phi)=E(\mathbf{r})e^{(-im \phi)}$ with clockwise (cw) direction, and $E_{\rm ccw}(\mathbf{r},\phi)=E(\mathbf{r})e^{(im \phi)}$ with counter clockwise (ccw) direction; for a TM mode, we only have   the $z$ component for the electric fields. 
A linear combinations of these fields result in standing waves~\cite{mazzei_controlled_2007,teraoka_resonance_2009,cognee_cooperative_2019}, such as a symmetric standing mode ${E}_{\rm s}={E}_{\rm cw}+ {E}_{\rm ccw} \propto \cos{(m \phi)}$, and an antisymmetric standing mode ${E}_{\rm as}={E}_{\rm cw}- {E}_{\rm ccw} \propto \sin{(m \phi)}$.
These standing waves could of course also be normalized, and can be expanded to obtain the QNM Green function.

Using the direct eigenfrequency solver (where the boundary condition is PML) in COMSOL, one can obtain a pair of unnormalized standing modes distribution $\tilde{\mathbf{f}}^{\rm M1/M2}_{\rm un}$. This is not used generally as a QNM solver since it is not so robust as a nonlinear eigenvalue solver, but the results appear to be reasonable for the high $Q$ modes of interest in this work.

For example, one can use PML normalization to obtain the
relevant QNM from
\begin{equation}\label{eq:pmlnorm_M1M2}
\tilde{\mathbf{f}}^{\rm Mi}(\mathbf{r})=\frac{\tilde{\mathbf{f}}^{\rm Mi}_{\rm un}(\mathbf{r})}{\sqrt{\braket{\braket{\tilde{\mathbf{f}}^{\rm Mi}_{\rm un}(\mathbf{r})|\tilde{\mathbf{f}}^{\rm Mi}_{\rm un}(\mathbf{r})}}_{\rm PML}}},
\end{equation}
where $\braket{\braket{\tilde{\mathbf{f}}^{\rm Mi}_{\rm un}(\mathbf{r})|\tilde{\mathbf{f}}^{\rm Mi}_{\rm un}(\mathbf{r})}}_{\rm PML}$ takes the form of Eq.~\ref{eq:PML}.
For a single lossy WGM resonator with refractive index $n_{\rm loss}=2+10^{-5}i$, the normalized distribution ${\rm Re}[\tilde{f}_{z}^{\rm M1/M2}]$ (Eq.~\ref{eq:pmlnorm_M1M2}) is shown in Fig.~\ref{fig:degeneratemodes}(b) and (c).
To better check their differences, we show them along the edge of the disk (equator) along the small vertical arrow (see Fig.~\ref{fig:degeneratemodes}(b) and (c)) direction in Fig.~\ref{fig:degeneratemodes}(d) (green solid and orange solid curves).
One can find their antinode positions are off by $\frac{\pi}{2m}$ ($m=37$), corresponding to the difference between $\cos{m \phi}$ and $\sin{m \phi}$. However, their absolute phases are offset, since these are related to where 
 $\phi=0$ is defined in coordinate system ($x_{\rm M}$,~$y_{\rm M}$),  with the eigenfrequency solver. 
 
In contrast, in the dipole excitation technique~\cite{bai_efficient_2013-1}, only one of the two degenerate standing wave modes is excited, as shown in Fig.~\ref{fig:degeneratemodes}(a) (which is the same as Fig.~\ref{QNM_distribution} (a) in main text), where the dipole is placed at $\mathbf{r}_{\rm d}=\mathbf{r}_{1}$. The field ${\rm Re}[\tilde{f}_{z}^{\rm L}]$ along the equator is also shown in Fig.~\ref{fig:degeneratemodes}(d) and (e) (red solid curve), which is with the form $cos(m\phi)$ considering $\phi=0$ corresponds to the positive $x$-axis.
Moreover, with the dipole technique,   one can control the phases accurately by setting the dipole at different locations, as they naturally excite the same standing wave defined from that position. For example, if the dipole is placed at a position corresponding to   $\phi\prime=\pi/2m$ ($\mathbf{r}_{\rm d}=\mathbf{r}_{1}\prime$), or $\phi\prime\prime=2\pi/2m$ ($\mathbf{r}_{\rm d}=\mathbf{r}_{1}\prime\prime$), or $\phi\prime\prime\prime=3\pi/2m$ ($\mathbf{r}_{\rm d}=\mathbf{r}_{1}\prime\prime\prime$) (the distance between the dipoles and the tangent of the disk surface  remains at $10~$nm) (see schematic in Fig.~\ref{fig:degeneratemodes} (f)),
then the same QNM is obtained with a simple shift in phase; these are shown in Fig.~\ref{fig:degeneratemodes} (e) (green dashed, black dashed and blue dashed curves), with the phase terms $\cos(m\phi-m\phi\prime)=\cos(m\phi-\pi/2)$, $\cos(m\phi-m\phi\prime\prime)=\cos(m\phi-2\pi/2)$, and $\cos(m\phi-m\phi\prime\prime\prime)=\cos(m\phi-3\pi/2)$ (considering $\phi=0$ is defined at the positive $x$-axis).

\begin{figure}[hb]
    \centering
    \includegraphics[width=0.92\columnwidth]{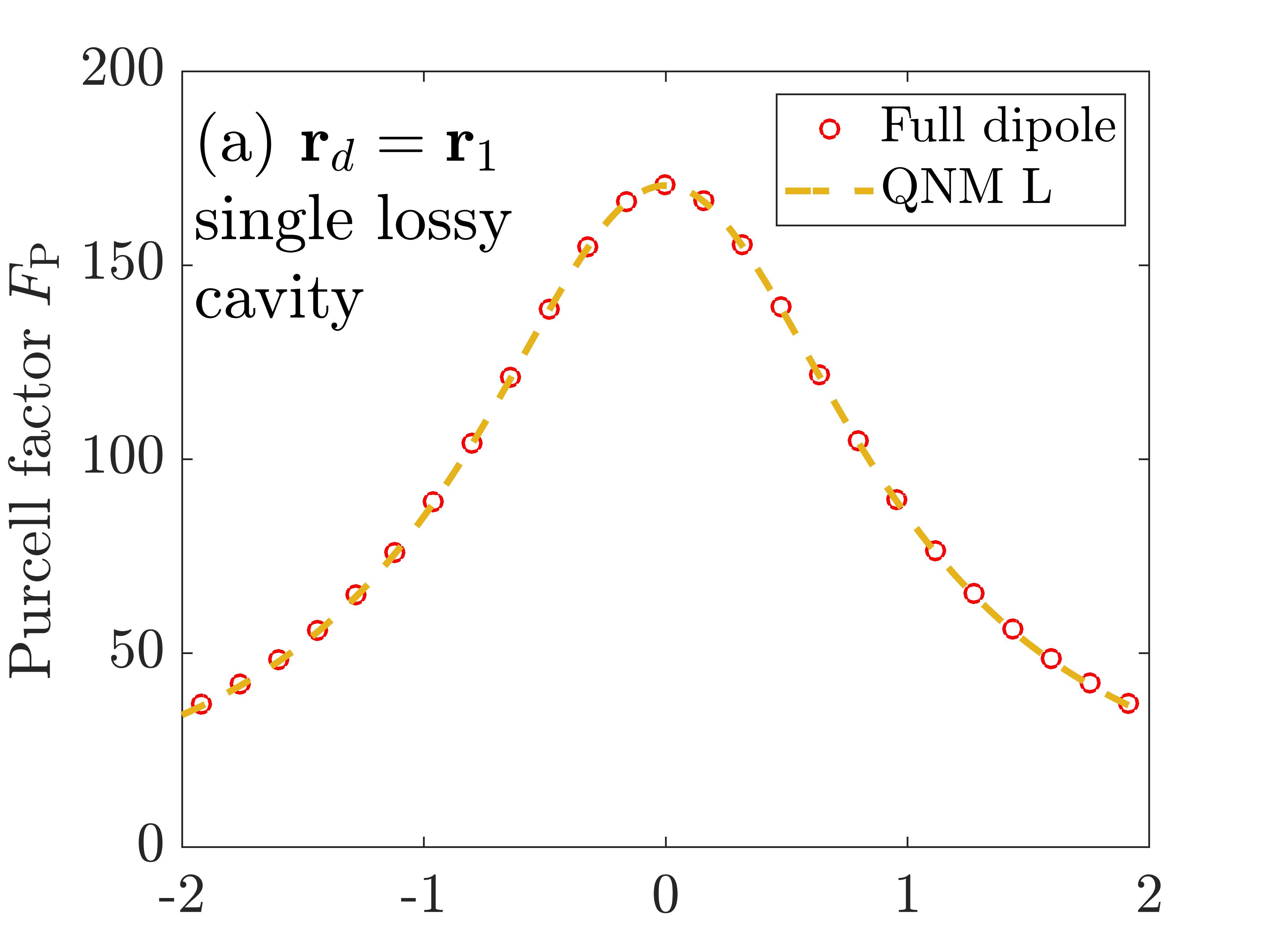}
    \includegraphics[width=0.92\columnwidth]{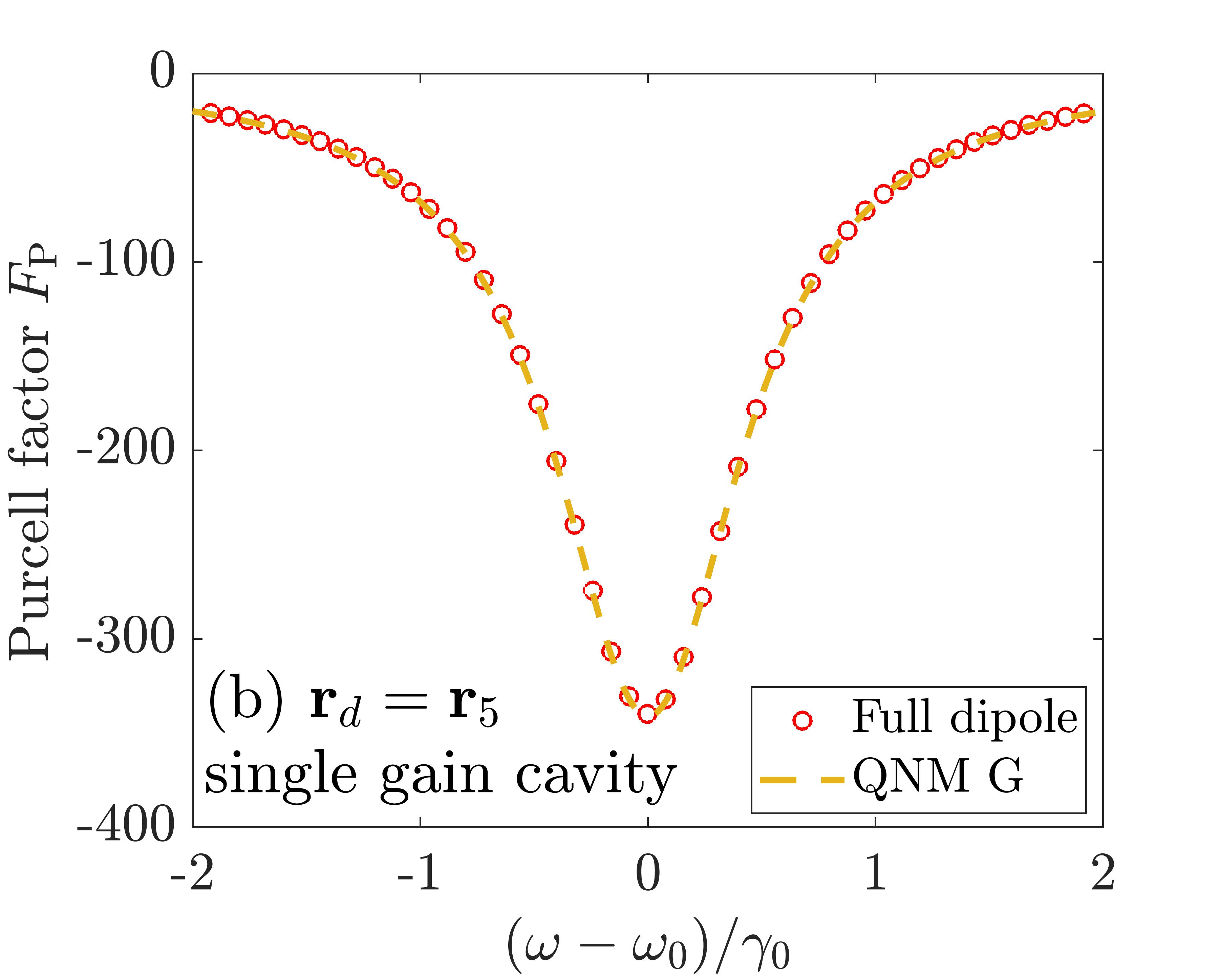}
    \caption{Purcell factors for (a) single lossy cavity with $n_{\rm loss}=2+10^{-5}i$ and (b) single gain cavity with $n_{\rm gain}=2-5\times10^{-6}i$ from single QNMs (Eq.\eqref{QNMpurcell} or Eq.\eqref{eq:2Dpurcell}) and full dipole method (Eq.\eqref{Purcellfulldipole}), where nice agreements are obtained. The dipole is put at $10~$nm away from single lossy cavity or gain cavity. The Purcell factors from single lossy cavity are net-positive. While the Purcell factors from single gain cavity are net-negative.
}\label{fig4a_app}
\end{figure}

\section{Full dipole simulations}\label{sec:full_simulation}

To check the validity of the Purcell factors from the QNMs (for microdisks and microrings) and also the CMT  Green function solutions (for microdisks), we compare these directly with the numerical results from a full dipole method (namely, with no modal approximations), which is obtained as follows:
\begin{equation}\label{Purcellfulldipole}
    F_{\rm P}^{\rm num,2D}(\mathbf{r}_{0},\omega)=\frac{\int_{ L_{\rm c}}\hat{\mathbf{n}}\cdot {\bf S}_{\rm dipole,total}(\mathbf{r},\omega)d{L_{\rm c}} }{\int_{ L_{\rm c}}\hat{\mathbf{n}}\cdot {\bf S}_{\rm dipole,background}(\mathbf{r},\omega)d{L_{\rm c}} },
\end{equation}
where $ L_{\rm c}$ is a small circle (with radius $1$ nm) surrounding point current (for TM mode; for 2D TE modes, it's a in-plane point dipole) and $\hat{\mathbf{n}}$ is a unit vector normal to $L_{\rm c}$, pointing outward.
The vector ${\bf S}(\mathbf{r},\omega)$ is the Poynting vector at this small circle and the subscript `total' and `background' represent the case with and without the resonator.

\begin{figure}[hb]
    \centering
    \includegraphics[width=0.99\columnwidth]{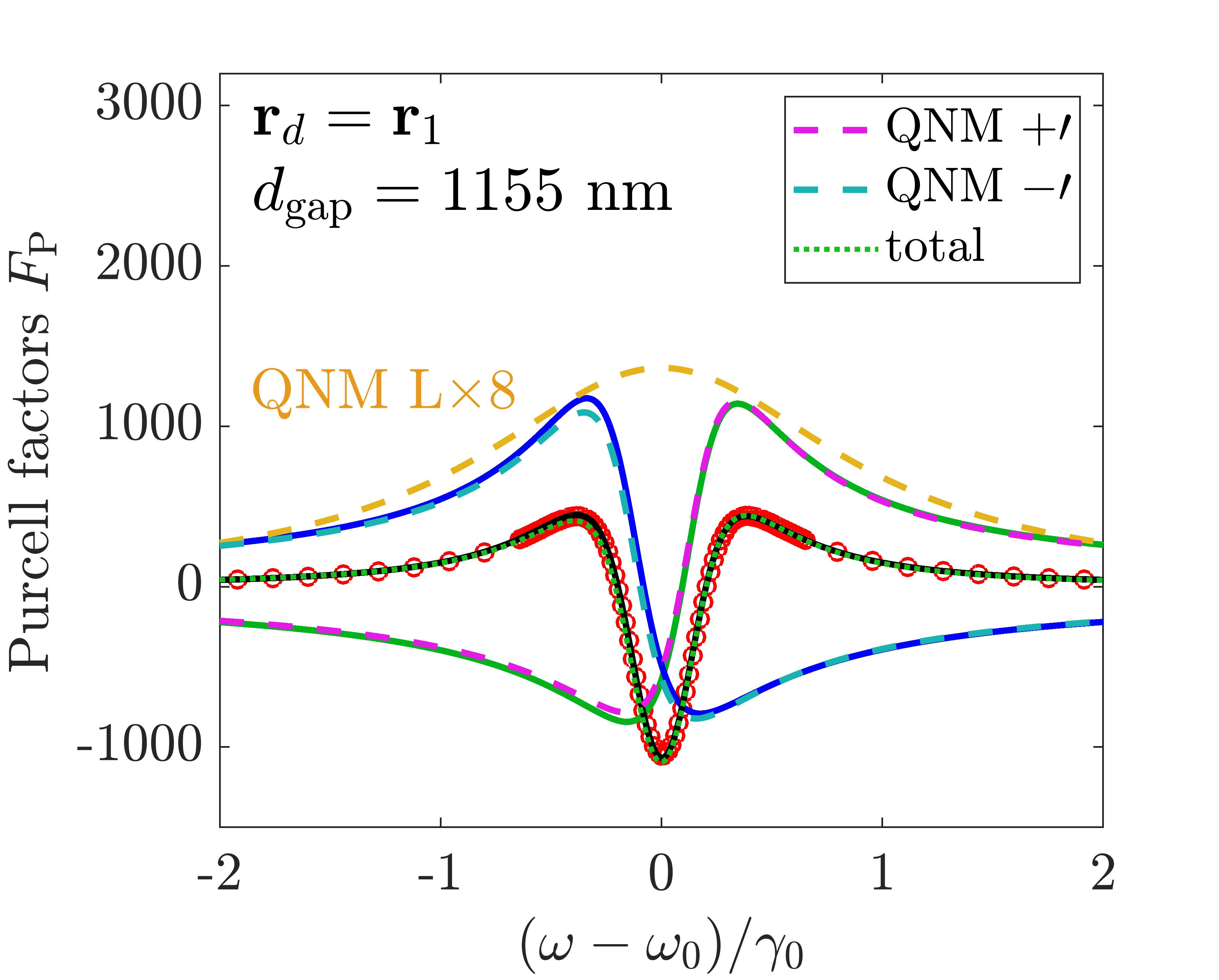}
    \caption{Purcell factors from the direct QNM approach, for the coupled resonators example shown in Fig.~\ref{fig4}(a) of the main text.
    For comparison, all original curves (analytical CMT results and full dipole results) on Fig.~\ref{fig4}(a) remain the same. With the direct QNM approach (when the dipole is located at $\mathbf{r}_{\rm d}=\mathbf{r}_{1}$), there are two dominant modes, QNM $+\prime$ and QNM $-\prime$. Their separate contributions to the total Purcell factors are shown as a dashed magenta curve and a dashed cyan curve, and the total one is shown as a green dotted curve. They show good agreements with CMT results and full dipole results, i.e., the contribution from QNM $+$ (green solid curve,  see  Fig.~\ref{fig4}(a)) and QNM $+\prime$ (dashed magenta curve) agrees very well, as does the contribution from QNM $-$ (blue solid curve) and QNM $-\prime$ (dashed cyan curve);
    the CMT and two direct QNM totals are practically indistinguishable  (black solid curve, green dotted curve;  and red circles show the full dipole results). 
}\label{fig4a_ap2}
\end{figure}

\section{Checking the accuracy of the QNMs for a single lossy resonator or a single gain resonator}\label{sec:single_modes}

\begin{figure*}[th]
    \centering
    \includegraphics[width=0.95\columnwidth]{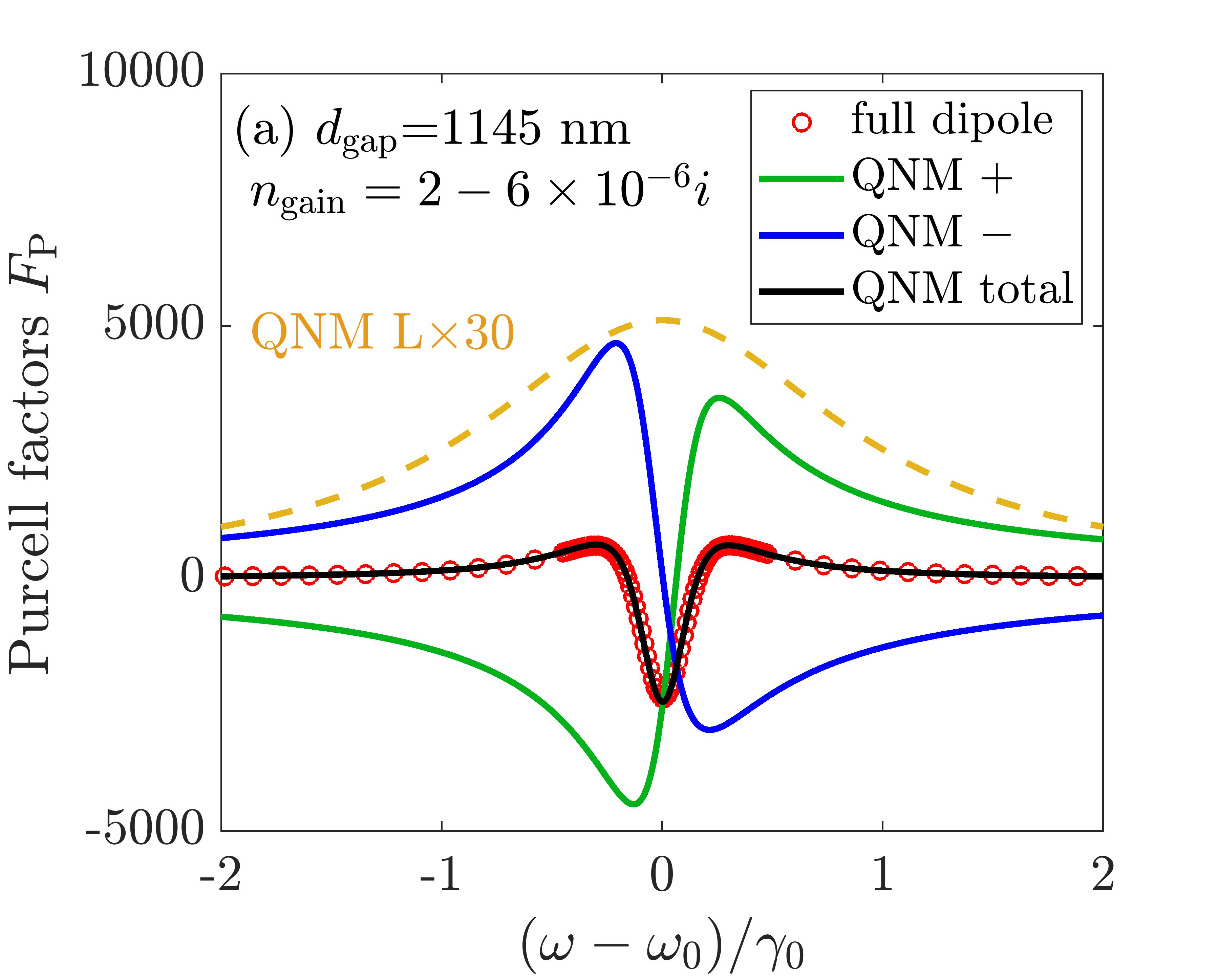}
    \includegraphics[width=0.95\columnwidth]{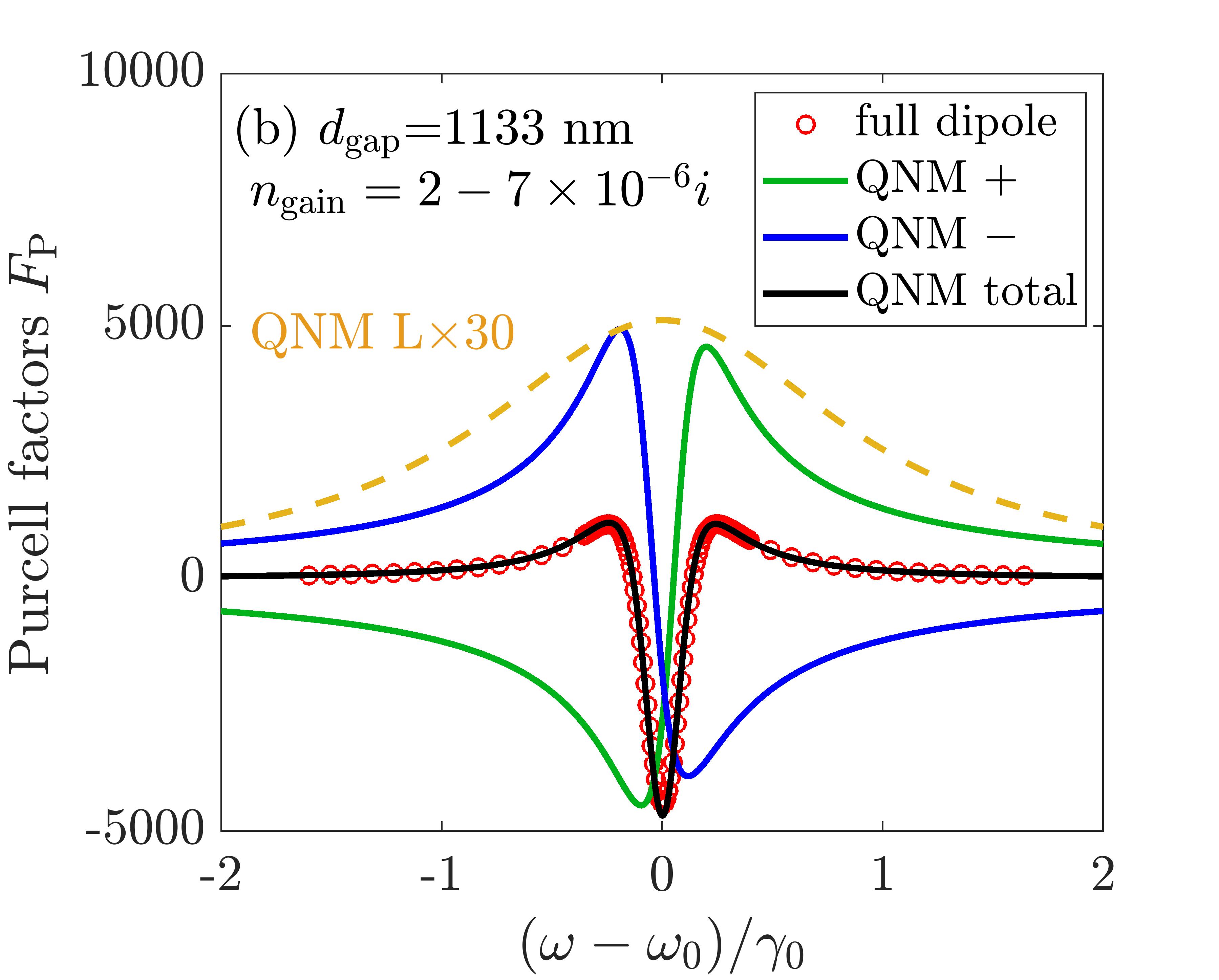}
    \caption{Purcell factors for coupled loss-gain resonators with  (a) $n_{\rm gain}=2-6\times10^{-6}i$ and (b) $n_{\rm gain}=2-7\times10^{-6}i$, while $n_{\rm loss}=2+10^{-5}i$ is same as in the main text for both cases. The dipole is placed at $\mathbf{r}_{1}$. The nice agreements with full dipole results are obtained. 
    }\label{fig6}
\end{figure*}

In order to compare with the coupled resonator results, and to confirm the accuracy of a single QNM approach for the single resonators, we show the results for single lossy cavity with $n_{\rm loss}=2+10^{-5}i$ or single gain cavity with $n_{\rm gain}=2-5\times10^{-6}i$. Using the dipole technique described above (Eq.~\eqref{2DQNM_norm}), the QNMs $\tilde{f}_{z}^{\rm L}$ and $\tilde{f}_{z}^{\rm G}$ are shown in Fig.~\ref{QNM_distribution} (a) and (b) separately. The dipole is put at $10~$nm away from single lossy cavity or the gain cavity.

The corresponding Purcell factors are shown in Fig.~\ref{fig4a_app}, which show quantitative agreement with full dipole method.
The  Purcell factors from single lossy cavity are net-positive and those with single gain cavity are net-negative.
Note these results are also shown as dashed curves in Fig.~\ref{fig4} and Fig.~\ref{fig4_prime} in the main text, but multiplied by constant for better comparison with the results from coupled system.

\section{Direct QNM approach for coupled loss resonators and gain resonators}\label{sec:directQNMs}

Naturally, one can also solve the coupled system with a QNM approach directly, instead of using coupled modes theory after the bare solutions are known.
In Fig.~\ref{fig4a_ap2}, we show their comparison for the specific example shown in Fig.~\ref{fig4}(a) of the main text. 
For comparison, all original curves (analytical CMT results and full dipole results) on Fig.~\ref{fig4}(a) remain the same.
With the dipole-excitation QNM approach (where the dipole is at $\mathbf{r}_{\rm d}=\mathbf{r}_{1}$), there are two dominant modes,  QNM $+\prime$ and QNM $-\prime$  (the hybrid modes). Their separate contributions to the Purcell factors are shown as magenta dashed curve and cyan dashed curve, and the total one is shown as green dotted curve. There are very  good agreements with the CMT results and the direct QNM results (QNM $+$ versus QNM $+\prime$, QNM $-$ versus QNM $-\prime$, and their total contributions), as well as 
quantitatively good agreement between the 
direct QNM results and the 
full dipole simulations (green dotted curve and red circles).\\

\section{Additional loss-gain resonator examples with different gain coefficients}\label{sec:morelossgain}

Finally,  we also show the Purcell factors for two additional loss-gain resonator examples, for different amounts of gain. The refractive index for the lossy resonator is fixed at $n_{\rm loss}=2+10^{-5}i$,  as in the main text. When $n_{\rm gain}=2-6\times10^{-6}i$, the Purcell factors with a gap distance $d_{\rm gap}=1145~$nm (close to the lossy EP) for a dipole at $\mathbf{r}_{1}$ ($10$~nm away from the lossy cavity) are shown in Fig.~\ref{fig6}(a), which show very good agreement with  the full dipole method. Again, we see that negative Purcell factors are obtained over a wide frequency range. The separate contributions from $\tilde{\mathbf{f}}^{+}$ and $\tilde{\mathbf{f}}^{-}$ are also given. For better comparison, the Purcell factors with single lossy cavity are shown as an orange dashed curve (the dipole is at $\mathbf{r}_{1}$), which is net positive and multiplied by $30$ for clarity.

Similarly, the corresponding results for the case with $n_{\rm gain}=2-7\times10^{-6}i$ are shown in Fig.~\ref{fig6}(b), where the gap distance is $d_{\rm gap}=1133~$nm. 
Excellent agreement with full dipole results are also obtained.
The absolute values of the negative Purcell factors increase here mainly because it is closer to the EP and the gap distance is smaller.


\bibliography{refs}

\end{document}